\documentclass[11pt,a4paper]{article}   	
\pdfoutput=1

\usepackage{geometry}                	

\usepackage{graphicx}		
\usepackage[table]{xcolor}

\usepackage{xspace}
\usepackage{subfig}

\usepackage{amsmath} 
\usepackage{ifthen}  
\usepackage{amssymb}
\usepackage{amsfonts}
\usepackage{upgreek} 

\usepackage{float}

\usepackage{array}
\newcolumntype{L}[1]{>{\raggedright\let\newline\\\arraybackslash\hspace{0pt}}m{#1}}
\newcolumntype{C}[1]{>{\centering\let\newline\\\arraybackslash\hspace{0pt}}m{#1}}
\newcolumntype{R}[1]{>{\raggedleft\let\newline\\\arraybackslash\hspace{0pt}}m{#1}}

\usepackage{hyperref}
\usepackage{cancel}
\newboolean{uprightparticles}
\setboolean{uprightparticles}{false}

\newcommand{\decay}[2]{\ensuremath{#1\!\to #2}\xspace}         

\def\PB      {\ensuremath{B}\xspace}                 
\def\B       {\ensuremath{\PB}\xspace}
\def\Bd      {\ensuremath{\B^0}\xspace}
\def\Bbar    {\kern 0.18em\overline{\kern -0.18em \PB}{}\xspace}
\def\Bdb     {\ensuremath{\Bbar^0}\xspace}
\def\PK      {\ensuremath{K}\xspace}                 
\def\kaon  {\ensuremath{\PK}\xspace}
\def\Kbar  {\kern 0.2em\overline{\kern -0.2em \PK}{}\xspace}
\def\Kp    {\ensuremath{\kaon^+}\xspace}
\def\Km    {\ensuremath{\kaon^-}\xspace}
\def\Kstarz  {\ensuremath{\kaon^{*0}}\xspace}
\def\Kstarzb {\ensuremath{\Kbar^{*0}}\xspace}
\def\Pmu         {\ensuremath{\mu}\xspace}                 
\def\mumu       {\ensuremath{\Pmu^+\Pmu^-}\xspace}
\def\Ppi         {\ensuremath{\pi}\xspace}                 
\def\pion  {\ensuremath{\Ppi}\xspace}
\def\pip   {\ensuremath{\pion^+}\xspace}
\def\pim   {\ensuremath{\pion^-}\xspace}
\def\Ps      {\ensuremath{s}\xspace}                 
\def\squark    {\ensuremath{\Ps}\xspace}
\def\Bs      {\ensuremath{\B^0_\squark}\xspace}

\def\invfb   {\ensuremath{\mbox{\,fb}^{-1}}\xspace}
\newcommand{\tev}{\ensuremath{\mathrm{\,Te\kern -0.1em V}}\xspace}
\newcommand{\gev}{\ensuremath{\mathrm{\,Ge\kern -0.1em V}}\xspace}
\newcommand{\mev}{\ensuremath{\mathrm{\,Me\kern -0.1em V}}\xspace}
\newcommand{\kev}{\ensuremath{\mathrm{\,ke\kern -0.1em V}}\xspace}
\newcommand{\ev}{\ensuremath{\mathrm{\,e\kern -0.1em V}}\xspace}
\newcommand{\gevc}{\ensuremath{{\mathrm{\,Ge\kern -0.1em V\!/}c}}\xspace}
\newcommand{\mevc}{\ensuremath{{\mathrm{\,Me\kern -0.1em V\!/}c}}\xspace}
\newcommand{\gevcc}{\ensuremath{{\mathrm{\,Ge\kern -0.1em V\!/}c^2}}\xspace}
\newcommand{\gevgevcccc}{\ensuremath{{\mathrm{\,Ge\kern -0.1em V^2\!/}c^4}}\xspace}
\newcommand{\mevcc}{\ensuremath{{\mathrm{\,Me\kern -0.1em V\!/}c^2}}\xspace}

\def\ctl       {\ensuremath{\cos{\theta_l}}\xspace}
\def\ctk       {\ensuremath{\cos{\theta_K}}\xspace}

\def\CP                {\ensuremath{C\!P}\xspace}

\newcommand{\thetal}{\ensuremath{\theta_l}\xspace}
\newcommand{\thetak}{\ensuremath{\theta_K}\xspace}
\newcommand{\aparl}{\ensuremath{{\cal A}_\parallel^{\rm L}}\xspace}
\newcommand{\aparr}{\ensuremath{{\cal A}_\parallel^{\rm R}}\xspace}
\newcommand{\aperpl}{\ensuremath{{\cal A}_\perp^{\rm L}}\xspace}
\newcommand{\aperpr}{\ensuremath{{\cal A}_\perp^{\rm R}}\xspace}
\newcommand{\azl}{\ensuremath{{\cal A}_0^{\rm L}}\xspace}
\newcommand{\azr}{\ensuremath{{\cal A}_0^{\rm R}}\xspace}
\newcommand{\asl}{\ensuremath{{\cal A}_{\rm S}^{\rm L}}\xspace}
\newcommand{\asr}{\ensuremath{{\cal A}_{\rm S}^{\rm R}}\xspace}
\newcommand{\fs}{\ensuremath{{F}_{\rm S}}\xspace}
\newcommand{\rmd}{\ensuremath{{\rm d}}\xspace}
\newcommand{\rmdcos}{\ensuremath{{\rm dcos}}\xspace}

\newcommand{\aparlst}{\ensuremath{{\cal A}_\parallel^{\rm L*}}\xspace}
\newcommand{\aparrst}{\ensuremath{{\cal A}_\parallel^{\rm R*}}\xspace}
\newcommand{\aperplst}{\ensuremath{{\cal A}_\perp^{\rm L*}}\xspace}
\newcommand{\aperprst}{\ensuremath{{\cal A}_\perp^{\rm R*}}\xspace}
\newcommand{\azlst}{\ensuremath{{\cal A}_0^{\rm L*}}\xspace}
\newcommand{\azrst}{\ensuremath{{\cal A}_0^{\rm R*}}\xspace}

\makeatletter
\newcommand{\parenbar}{\mathpalette\p@renb@r}
\def\p@renb@r#1#2{\vbox{%
  \ifx#1\scriptscriptstyle \dimen@.7em\dimen@ii.2em\else
  \ifx#1\scriptstyle \dimen@.8em\dimen@ii.25em\else
  \dimen@1em\dimen@ii.4em\fi\fi \offinterlineskip
  \ialign{\hfill##\hfill\cr
    \vbox{\hrule width\dimen@ii}\cr
    \noalign{\vskip-.3ex}%
    \hbox to\dimen@{$\mathchar300\hfil\mathchar301$}\cr
    \noalign{\vskip-.3ex}%
    $#1#2$\cr}}}
\makeatother

\newboolean{articletitles}
\setboolean{articletitles}{true}

\usepackage{cite}
\usepackage{mciteplus}

\newboolean{inbibliography}

\begin{document}

\def\thefootnote{\arabic{footnote}}
\hfill{\tt CERN-TH-2017-178, MITP/17-053}

\vspace*{1.0cm}
\begin{center}
{\huge\bfseries \boldmath
  Direct determination of Wilson coefficients using $\decay{\Bd}{\Kstarz\mumu}$ decays\\[1.0 cm]
}
{\Large
T.~Hurth\footnote{Email:~\href{mailto:tobias.hurth@cern.ch}{tobias.hurth@cern.ch}}$^{,a}$, 
C.~Langenbruch\footnote{Email:~\href{mailto:christoph.langenbruch@cern.ch}{christoph.langenbruch@cern.ch}}$^{,b}$, 
F.~Mahmoudi\footnote{Also Institut Universitaire de France, 103 boulevard Saint-Michel, 75005 Paris, France; Email:~\href{mailto:nazila@cern.ch}{nazila@cern.ch}}$^{,c,d}$
}\\[0.4 cm] 
{\small
  $^a$PRISMA Cluster of Excellence and Institute for Physics (THEP)\\
  Johannes Gutenberg University, D-55099 Mainz, Germany\\[0.2cm]
  $^b$I. Physikalisches Institut B,\\
  RWTH Aachen, D-52074 Aachen, Germany\\[0.2cm]
  $^c$Univ Lyon, Univ Lyon 1, CNRS/IN2P3, Institut de Physique Nucl\'eaire de Lyon UMR5822, F-69622 Villeurbanne, France\\[0.2cm]
  $^d$CERN, Theoretical Physics Department, CH-1211 Geneva 23, Switzerland
} \\[0.5 cm]
\small
\end{center}

\vspace*{2.0cm}

\begin{abstract} 
  A method to directly determine the Wilson coefficients for rare $b\to s$ transitions using $\decay{\Bd}{\Kstarz\mumu}$ decays in an unbinned maximum likelihood fit is presented.
  The method has several advantages compared to the conventional 
  determination of the Wilson coefficients from angular observables that are determined in bins of $q^2$, the square of the mass of the dimuon system.
  The method uses all experimental information in a more efficient way and 
  automatically accounts for experimental correlations. 
  Performing pseudoexperiments, we show the improved sensitivity of the proposed method for the Wilson coefficients.   
  We also demonstrate that it will be possible to use the method 
  with the combined Run~1 and~2 data sample taken by the LHCb experiment. 
\end{abstract} 

\clearpage

\section{Introduction} 
\label{sec:introduction}
Rare flavour changing neutral current~(FCNC) decays constitute sensitive probes for New Physics~(NP) since they are forbidden at tree-level in the Standard Model~(SM) and can only occur at loop order. 
New heavy particles can appear in competing diagrams and affect both the branching fractions as well as angular distributions of rare processes. 

The rare $\decay{b}{s\mumu}$ decay $\decay{\Bd}{\Kstarz(\to\Kp\pim)\mumu}$\,\footnote{Charge conjugation is implied throughout this paper unless otherwise noted.} exhibits a particularly rich phenomenology since it allows access to many angular observables that are sensitive to NP contributions. 
The final state of the decay is completely determined by the three decay angles $\vec{\Omega}=(\cos\thetal, \cos\thetak, \phi)$, 
the square of the invariant mass of the dimuon system, $q^2$, and the decay flavour that can be inferred from the kaon charge.  
Angular observables are typically determined by performing angular fits in bins of $q^2$ that are then compared with $q^2$-binned SM predictions.

The angular distributions of the decay have been studied by the BaBar, Belle, CDF, CMS and LHCb collaborations~\cite{Aubert:2006vb,Lees:2015ymt,Wei:2009zv,Aaltonen:2011ja,Chatrchyan:2013cda,Khachatryan:2015isa,LHCb-PAPER-2015-051,Wehle:2016yoi,ATLAS:2017dlm,CMS:2017ivg}. 
The LHCb collaboration has performed the first full angular analysis using the full data sample from Run 1 of the LHC, corresponding to an integrated luminosity of $3\invfb$~\cite{LHCb-PAPER-2015-051}. 
The resulting complete set of angular observables and their correlations constitutes the most precise measurement of these observables to date. 
For this analysis, some tension has emerged with the SM predictions~\cite{Straub:2015ica,Horgan:2013hoa,Horgan:2015vla,Descotes-Genon:2014uoa}, which is particularly visible in the angular observable $P_5^\prime$. 
For this observable, for definitions see Ref.~\cite{Descotes-Genon:2013vna},
uncertainties from the hadronic $\decay{\Bd}{\Kstarz}$ form factors are designed to cancel at leading order. 
The local deviations in this observable correspond to $2.8$ and $3.0$ standard deviations ($\sigma$) for the $q^2$ bins $4<q^2<6\gevgevcccc$ and $6<q^2<8\gevgevcccc$.\,\footnote{We note that the significances depend on assumptions on the size of $\Lambda_{\rm QCD}/m_b$ power corrections.} 
This confirms a tension seen in an earlier analysis by LHCb that used only $1\invfb$ of data~\cite{LHCb-PAPER-2013-037}. 
The recent measurements of $P_5^\prime$ by Belle and ATLAS~\cite{Wehle:2016yoi,ATLAS:2017dlm} are also in good agreement with the LHCb result and show tensions with the SM. 
The analysis by the CMS collaboration is compatible with both the SM and the LHCb result~\cite{CMS:2017ivg}. 

Together with the branching fractions of $\decay{b}{s\mu\mu}$ decays reported in Refs.~\cite{Aaij:2016flj,Aaij:2015esa,Aaij:2014pli,LHCb-PAPER-2016-045} that tend to lie below SM predictions and the tensions in tests of Lepton universality~\cite{Aaij:2014ora,Aaij:2017vbb}, 
the angular distributions of $\decay{\Bd}{\Kstarz\mumu}$ constitute the so-called ``flavour anomalies'' in rare decays~\cite{Blake:2017wjz}. 

Several theory groups performed global fits of the available data on rare $\decay{b}{s}$ decays, 
including the $q^2$-binned data on the decay $\decay{\Bd}{\Kstarz\mumu}$, see for example Refs.~\cite{Beaujean:2013soa,Descotes-Genon:2015uva,Hurth:2016fbr,Hurth:2017hxg,Altmannshofer:2014rta,Altmannshofer:2017fio,Capdevila:2017bsm,Geng:2017svp}. 
The fits determine the Wilson coefficients, the effective coupling strengths of the contributing local operators. 
The global fits find that the description of the data can be improved by a shift of the effective vector-coupling ${\cal C}_9$ of around $\Delta{\cal C}_9\sim -1$ from its SM prediction. 
This deviation is found to have a significance of around $3$--$5\,\sigma$, depending on the combination of varied Wilson coefficients, the used experimental input and the handling of theory nuisance parameters. 
Nuisance parameters that can impact the significance of the deviation are the hadronic form factor parameters and uncertainties on subleading $\Lambda_{\rm QCD}/m_b$ suppressed corrections of QCD factorisation.
The deviations in the angular observables can be interpreted as signs of NP, most notably new heavy $Z^\prime$ gauge bosons~\cite{Gauld:2013qba,Buras:2013qja,Altmannshofer:2013foa,Altmannshofer:2014cfa,Crivellin:2015mga,Sierra:2015fma,Crivellin:2015lwa,Celis:2015ara,Falkowski:2015zwa,Chiang:2016qov,Crivellin:2016ejn,Bhatia:2017tgo,Kamenik:2017tnu,Ellis:2017nrp,Bonilla:2017lsq} 
or leptoquarks~\cite{Hiller:2014yaa,Biswas:2014gga,Buras:2014fpa,Gripaios:2014tna,Bauer:2015knc,Varzielas:2015iva,Becirevic:2017jtw,Barbieri:2015yvd,Fajfer:2015ycq,Alonso:2015sja,Calibbi:2015kma,Becirevic:2016oho,Hiller:2016kry,Cai:2017wry,Chauhan:2017ndd}. 
However, the $\decay{\Bd}{\Kstarz\mumu}$ angular observables can also be affected by 
contributions from SM $c\bar{c}$-loop processes~\cite{Jager:2012uw,Lyon:2014hpa,Ciuchini:2015qxb,Chobanova:2017ghn}, which are part of the $\Lambda_{\rm QCD}/m_b$ corrections. 
A large effort from both theory and experiment is currently ongoing to disentangle these effects. 
A more efficient method to exploit the experimental information on semileptonic rare $b\to s$ decays is therefore highly desirable. 

We propose a new approach to the determination of the Wilson coefficients using $\decay{\Bd}{\Kstarz\mumu}$ decays. 
The proposed method uses all available experimental data, the decay angles, $q^2$, and the $\Bd$ decay flavour 
to determine the Wilson coefficients in a $q^2$-unbinned maximum likelihood fit. 
Furthermore, the invariant mass of the $\Kp\pim$ system is included to improve the control of contributions by the S-wave, where $\Kp$ and $\pim$ are in a state of relative angular momentum zero. 
Unlike previous $q^2$-unbinned approaches that fit a parameterisation of the $\Kstarz$ spin amplitudes~\cite{Egede:2008uy,Egede:2010zh,Egede:2015kha,LHCb-PAPER-2015-051}, we propose to instead fit the Wilson coefficients and nuisance parameters directly. 
The decay $\decay{\Bd}{\Kstarz\mumu}$ can be described using QCD factorisation at low $q^2$~\cite{Beneke:2001at} and an operator product expansion in $1/m_b$ at high $q^2$~\cite{Grinstein:2004vb}. 
Several open-source software packages implement these calculations and provide the four-differential decay rate $\rmd^4\Gamma(\decay{\Bd}{\Kstarz\mumu})/\left(\rmdcos\thetal\,\rmdcos\thetak\,\rmd\phi\,\rmd q^2\right)$~\cite{Bobeth:2010wg,EOS,Beaujean:2013soa,Mahmoudi:2007vz,Mahmoudi:2008tp,Mahmoudi:2009zz,flavio}. 
We extend both the EOS software~\cite{Bobeth:2010wg,EOS,Beaujean:2013soa} and SuperIso~\cite{Mahmoudi:2007vz,Mahmoudi:2008tp,Mahmoudi:2009zz} to include the S-wave contribution 
resulting in the five-differential decay rate $\rmd^5\Gamma(\decay{\Bd}{\Kstarz\mumu})/\left(\rmdcos\thetal\,\rmdcos\thetak\,\rmd\phi\,\rmd q^2\,\rmd m_{K\pi}^2\right)$ we use as probability density function. 
In the following we will use SuperIso for the generation of pseudoexperiments and EOS for their fit. 

The proposed method has several advantages over the conventional method of first determining angular observables in bins of $q^2$ and then performing the determination of the Wilson coefficients in a second step:
\begin{enumerate}
\item The direct fit uses all available experimental information resulting in a more efficient exploitation of the available data and a 
  more precise determination of the Wilson coefficients.
  This leads to higher sensitivity to possible NP contributions. 
\item The full statistics of the data is available in a single fit 
which leads to improved fit stability and reduces the potential need to perform computationally expensive coverage correction. 
\item Theory nuisance parameters, in particular form factor parameters that require non-perturbative calculation, can be better constrained using the full information on $q^2$. 
\item Finally, experimental correlations, including possible non-linear correlations, are automatically accounted for. 
\end{enumerate}

The paper is structured as follows: In Sec.~\ref{sec:decay} we introduce the four-differential decay rate of the decay $\decay{\Bd}{\Kstarz\mumu}$ and
expand it to include the mass of the $\Kp\pim$ system to constrain contributions from the S-wave. 
Section~\ref{sec:directfit} describes the proposed procedure to determine the Wilson coefficients from a direct fit of $\decay{\Bd}{\Kstarz\mumu}$ decays without binning in $q^2$. 
Section~\ref{sec:validation} details the validation of the procedure and the evaluation of its performance using pseudoexperiments. 
In Sec.~\ref{sec:conclusions} we conclude. 

\section[The decay $\decay{\Bd}{\Kstarz\mumu}$]{The decay \boldmath{$\decay{\Bd}{\Kstarz\mumu}$}} 
\label{sec:decay}
\subsection{Four-differential decay rate}
\label{sec:decayrate}
The four-differential decay rates for the rare decay $\decay{\Bdb}{\Kstarzb\mumu}$ and its conjugate $\decay{\Bd}{\Kstarz\mumu}$ are given by 
\begin{align}
  \frac{\rmd^4\Gamma(\decay{\Bdb}{\Kstarzb\mumu})}{\rmd\vec{\Omega}\,\rmd q^2} &= \sum_i I_i(q^2) f_i(\vec{\Omega})\nonumber\\
  \frac{\rmd^4\Gamma(\decay{\Bd}{\Kstarz\mumu})}{\rmd\vec{\Omega}\,\rmd q^2} &= \sum_i \bar{I}_i(q^2) f_i(\vec{\Omega}),\label{eq:pwave}
\end{align}
where $I_i(q^2)$ and $\bar{I}_i(q^2)$ denote the $q^2$-dependent angular observables given by bilinear combinations of $\Kstarz$ spin amplitudes~\cite{Kruger:1999xa,Altmannshofer:2008dz}. 
Both $\parenbar{I_i}(q^2)$ 
and the angular terms $f_i(\ctl,\ctk,\phi)$ are given in Tab.~\ref{tab:fi}. 
\begin{table}
  \centering
  \renewcommand{\arraystretch}{1.5}
  \begin{tabular}{lll}\hline
    i & \multicolumn{1}{c}{$I_i(q^2)$} & \multicolumn{1}{c}{$f_i(\vec{\Omega})$}\\\hline\hline
   1s & $\frac{3}{4}\left[|\aparl|^2+|\aperpl|^2+|\aparr|^2+|\aperpr|^2\right]$ & $\sin^2\thetak$\\
   1c & $|\azl|^2+|\azr|^2$ & $\cos^2\thetak$\\
   2s & $\frac{1}{4}\left[|\aparl|^2+|\aperpl|^2+|\aparr|^2+|\aperpr|^2\right]$ & $\sin^2\thetak\cos 2\thetal$ \\
   2c & $-|\azl|^2-|\azr|^2$ & $\cos^2\thetak\cos 2\thetal$\\
   3 & $\frac{1}{2}\left[|\aperpl|^2-|\aparl|^2+|\aperpr|^2-|\aparr|^2\right]$ & $\sin^2\thetak\sin^2\thetal\cos 2\phi$\\
   4 & $\sqrt{\frac{1}{2}}{\rm Re}\left(\azl\aparlst + \azr\aparrst\right)$& $ \sin 2\thetak \sin 2\thetal \cos\phi$\\
   5 & $\sqrt{2} {\rm Re}\left(\azl\aperplst-\azr\aperprst\right)$& $\sin 2\thetak\sin\thetal\cos\phi$\\
   6s & $2{\rm Re}\left(\aparl\aperplst-\aparr\aperprst\right)$ & $\sin^2\thetak\cos\thetal$\\
   7 & $\sqrt{2}{\rm Im}\left(\azl\aparlst-\azr\aparrst\right)$ & $\sin 2\thetak\sin\thetal\sin\phi$\\
   8 & $\sqrt{\frac{1}{2}}{\rm Im}\left(\azl\aperplst+\azr\aperprst\right)$ & $\sin 2\thetak\sin 2\thetal\sin\phi$\\
   9 & ${\rm Im}\left(\aparlst\aperpl+\aparrst\aperpr\right)$ & $\sin^2\thetak\sin^2\thetal\sin 2\phi$\\ \hline
   10 & $\frac{1}{3}\left[|\asl|^2+|\asr|^2\right]$ & $1$\\
   11 & $\sqrt{\frac{4}{3}}{\rm Re}\left(\asl\azlst + \asr\azrst\right)$ & $\cos\thetak$\\
   12 & $-\frac{1}{3}\left[|\asl|^2+|\asr|^2\right]$ & $\cos 2\thetal$\\
   13 & $-\sqrt{\frac{4}{3}}{\rm Re}\left(\asl\azlst+\asr\azrst\right)$ & $\cos\thetak\cos 2\thetal$\\
   14 & $\sqrt{\frac{2}{3}}{\rm Re}\left(\asl\aparlst+\asr\aparrst\right)$ & $\sin\thetak\sin 2\thetal\cos\phi$\\
   15 & $\sqrt{\frac{8}{3}}{\rm Re}\left(\asl\aperplst-\asr\aperprst\right)$ & $\sin\thetak\sin\thetal\cos\phi$\\
   16 & $\sqrt{\frac{8}{3}}{\rm Im}\left(\asl\aparlst-\asr\aparrst\right)$ & $\sin\thetak\sin\thetal\sin\phi$\\
   17 & $\sqrt{\frac{2}{3}}{\rm Im}\left(\asl\aperplst+\asr\aperprst\right)$ & $\sin\thetak\sin 2\thetal\sin\phi$\\
  \hline\end{tabular}
  \caption{Dependence of the angular observables $I_i(q^2)$ on the transversity amplitudes ${\cal A}_{0,\parallel,\perp}^{\rm L,R}$ and the corresponding angular terms $f_i(\ctl,\ctk,\phi)$.
    The angular observables $\bar{I}_i(q^2)$ are given by complex conjugation of all weak phases ${\cal A}\to \bar{\cal A}$.\label{tab:fi}}
\end{table}
The conventional approach of angular analysis relies on integration of $\parenbar{I_i}(q^2)$ over a $q^2$ bin $q^2_{\rm min}<q^2<q^2_{\rm max}$ and determination of the angular observables
\begin{align}
    S_i(q^2_{\rm min},q^2_{\rm max}) &= \frac{\int_{q^2_{\rm min}}^{q^2_{\rm max}} I_i(q^2)+\bar{I}_i(q^2) \,\rmd q^2}{\int_{q^2_{\rm min}}^{q^2_{\rm max}} \frac{\rmd\Gamma(q^2)}{\rmd q^2} + \frac{\rmd\bar{\Gamma}(q^2)}{\rmd q^2} \,\rm dq^2}\nonumber\\
    A_i(q^2_{\rm min},q^2_{\rm max}) &= \frac{\int_{q^2_{\rm min}}^{q^2_{\rm max}} I_i(q^2)-\bar{I}_i(q^2) \,\rmd q^2}{\int_{q^2_{\rm min}}^{q^2_{\rm max}} \frac{\rmd\Gamma(q^2)}{\rmd q^2} + \frac{\rmd\bar{\Gamma}(q^2)}{\rmd q^2} \,\rmd q^2},
\end{align}
where the \CP-averaged (\CP-violating) angular observables $S_i$ ($A_i$) have been defined according to Ref.~\cite{Altmannshofer:2008dz}. 
Neglecting lepton mass effects the \CP-averaged $S_i$ reduce to the longitudinal polarisation fraction $F_{\rm L}=S_{1c}$, 
the forward-backward asymmetry $A_{\rm FB}=\frac{3}{4}S_{6s}$ and the remaining $S_{3,4,5,7,8,9}$. 
Additional ratios of $S_i$ have been proposed as observables, for which form factor uncertainties cancel at leading order~\cite{Kruger:2005ep,DescotesGenon:2012zf}.
Examples are
\begin{align*}
  P_{1} &= \frac{2S_3}{1-F_{\rm L}},\\
  P_{2} &= \frac{2}{3}\frac{A_{\rm FB}}{1-F_{\rm L}},\\
  P_{3} &= \frac{-S_9}{1-F_{\rm L}}~\text{and}\\
  P_{4,5,6,8}^\prime &= \frac{S_{4,5,7,8}}{\sqrt{F_{\rm L}(1-F_{\rm L})}}.
\end{align*}
In this work, we use the $q^2$-dependent amplitudes ${\cal A}_{0,\parallel,\perp,t}^{\rm L,R}$ 
that are given by 
\begin{align}
    {\cal A}_{\perp}^{\rm L(R)} &= {\cal N}\sqrt{2\lambda}\biggl\{ \bigl[ ({\cal C}_9^{\rm eff} + {\cal C}_9^{\prime\rm eff}) \mp ({\cal C}_{10}^{\rm eff} + {\cal C}_{10}^{\prime\rm eff})\bigr] \frac{{V(q^2)}}{m_B+m_{K^*}} 
    + \frac{2m_b}{q^2} ({\cal C}_7^{\rm eff} + {\cal C}_7^{\prime\rm eff}){T_1(q^2)}\biggr\}\nonumber\\
    {\cal A}_{\parallel}^{\rm L(R)} &= -{\cal N}\sqrt{2}(m_B^2-m_{K^*}^2) \biggl\{ \bigl[ ({\cal C}_9^{\rm eff} - {\cal C}_9^{\prime\rm eff}) \mp ({\cal C}_{10}^{\rm eff} - {\cal C}_{10}^{\prime\rm eff})\bigr] \frac{{A_1(q^2)}}{m_B-m_{K^*}}\nonumber \\
    &+ \frac{2m_b}{q^2} ({\cal C}_7^{\rm eff} - {\cal C}_7^{\prime\rm eff}){T_2(q^2)}\biggr\}\nonumber\\
    {\cal A}_{0}^{\rm L(R)} &= -\frac{{\cal N}}{2m_{K^*}\sqrt{q^2}}\biggl\{ \bigl[ ({\cal C}_9^{\rm eff} - {\cal C}_9^{\prime\rm eff}) \mp ({\cal C}_{10}^{\rm eff} - {\cal C}_{10}^{\prime\rm eff})\bigr]\nonumber\\ 
    & \times\bigl[(m_B^2-m_{K^*}^2-q^2)(m_B+m_{K^*}){A_1(q^2)} - \lambda \frac{{A_2(q^2)}}{m_B+m_{K^*}} \bigr]\nonumber\\
    & + 2m_b ({\cal C}_7^{\rm eff} - {\cal C}_7^{\prime\rm eff})\bigl[ (m_B^2+3m_{K^*}-q^2){T_2(q^2)} - \frac{\lambda}{m_B^2-m_{K^*}^2}{T_3(q^2)}\bigr] \biggr\}\nonumber\\
    {\cal A}_t &= \frac{{\cal N}}{\sqrt{q^2}}\sqrt{\lambda}\biggl\{2({\cal C}_{10}^{\rm eff}-{\cal C}_{10}^{\prime\rm eff})+\frac{q^2}{m_\mu}({\cal C}_P-{\cal C}_P^\prime)\biggr\}A_0(q^2), 
\end{align}
where ${\cal N}$ denotes a normalisation factor given by
\begin{align*}
  {\cal N}&=G_F \alpha_{\rm em} |V_{\rm tb}V_{\rm ts}| \sqrt{\frac{q^2 \sqrt{\lambda} \beta_\ell}{3\cdot 1024 \pi^5 m_B^3}},
\end{align*}
and $\lambda$ is given by $\lambda= m_B^4+m_{K^*}^4 + q^4 - 2\left(m_B^2m_{K^*}^2+m_{K^*}^2q^2+m_B^2q^2\right)$~\cite{Altmannshofer:2008dz}. 
The symbols $A_{0,1,2}(q^2)$, $V(q^2)$ and $T_{1,2,3}(q^2)$ denote the $q^2$-dependent hadronic form factors. 

The form factors require non-perturbative calculations and are determined using light cone sum rules (LCSR)~\cite{Ball:2004rg,Khodjamirian:2010vf,Straub:2015ica} or lattice calculations~\cite{Horgan:2013hoa,Horgan:2015vla}. 
This paper uses the full form factor approach with the results from Ref.~\cite{Straub:2015ica} for the decay $\decay{\Bd}{\Kstarz\mumu}$, 
determined from a combination from LCSR~\cite{Straub:2015ica} and lattice calculations~\cite{Horgan:2013hoa,Horgan:2015vla}. 
We also include the correlations between the form factor parameters. 

\subsection{S-wave contribution}
\label{sec:swave}
Besides contributions to the final state $\decay{\Bd}{\Kp\pim\mumu}$ from the decay of the vector-meson $\Kstarz$ (P-wave) the $\Kp\pim$ system in the final state can also be in an S-wave configuration that can originate either from a non-resonant decay or from the decay of scalar resonances. 
This results in two additional complex amplitudes ${\cal A}_{S0,St}^{\rm L,R}$ that contribute to the decay and that affect the distributions in the decay angles and $q^2$. 
As a result, the four-differential decay rate in Eq.~\ref{eq:pwave} needs to be modified according to
\begin{align}
  \frac{\rmd^4\Gamma(\decay{\Bdb}{\Km\pip\mumu})}{\rmd\vec{\Omega}\,\rmd q^2} &=
  \left(1-\fs\right) \frac{\rmd^4\Gamma(\decay{\Bdb}{\Kstarzb\mumu})}{\rmd\vec{\Omega}\,\rmd q^2}\bigg|_{\rm P-wave}\nonumber\\
  &+ \frac{3}{16\pi} \fs \sin^2\thetal + \frac{9}{32\pi}\left(I_{11}(q^2)+I_{13}(q^2)\cos 2\thetal\right)\cos\thetak \nonumber\\
  &+ \frac{9}{32\pi}\left(I_{14}(q^2)\sin 2\thetal + I_{15}(q^2)\sin\thetal\right)\sin\thetak\cos\phi\nonumber\\
  &+ \frac{9}{32\pi}\left(I_{16}(q^2)\sin \thetal + I_{17}(q^2)\sin 2\thetal\right)\sin\thetak\sin\phi,\label{eq:spwave}
\end{align}
where the $I_{10-17}(q^2)$ are again given in Tab.~\ref{tab:fi} and the four-differential decay rate for the process $\decay{\Bd}{\Kp\pim\mumu}$ is given by the replacements $I_i(q^2)\to \bar{I}_i(q^2)$. The fraction of S-wave is denoted by \fs, which is defined as
\begin{align*}
\fs &= \frac{|\asl|^2+|\asr|^2}{|\azl|^2+|\azr|^2 + |\aparl|^2+|\aparr|^2 + |\aperpl|^2+|\aperpr|^2 + |\asl|^2+|\asr|^2}. 
\end{align*}
The S-wave amplitudes are given by~\cite{Becirevic:2012dp,Lu:2011jm}
\begin{align}
{\cal A}_{S0}^{\rm L(R)} &= {\cal N}_0 \sqrt{\frac{\lambda_{K^*_0}}{q^2}} \biggl\{\bigl[ ({\cal C}_9 - {\cal C}_9^\prime) \mp ({\cal C}_{10} - {\cal C}_{10}^\prime) \bigr] f_+(q^2) + ({\cal C}_7-{\cal C}_7^\prime) 2m_b \frac{f_T(q^2)}{m_B+m_{K^*_0}} \biggr\}\nonumber\\
{\cal A}_{St} &= {\cal N}_0  \frac{1}{\sqrt{q^2}} 2 ({\cal C}_{10} - {\cal C}_{10}^\prime) (m_B^2-m_{K^*_0}^2) f_0(q^2),
\end{align}
where ${\cal N}_0$ denotes a normalisation factor given by
\begin{align*}
  {\cal N}_0&=G_F \alpha_{\rm em} |V_{\rm tb}V_{\rm ts}| \sqrt{\frac{q^2 \sqrt{\lambda_{K^*_0}} \beta_\ell}{3\cdot 1024 \pi^5 m_B^3}},
\end{align*}
and $\lambda_{K^*_0}$ is given by $\lambda_{K^*_0}= m_B^4+m_{K^*_0}^4 + q^4 - 2\left(m_B^2m_{K^*_0}^2+m_{K^*_0}^2q^2+m_B^2q^2\right)$. 
The symbols $f_+(q^2)$, $f_T(q^2)$ and $f_0(q^2)$ denote the $q^2$-dependent hadronic form factors. 
In the large energy limit they reduce to a single soft form factor $\xi_\parallel(q^2)=f_+(q^2)=m_Bf_T(q^2)/(m_B+m_{K^*_0})=m_Bf_0(q^2)/(2E)$. 
Following Ref.~\cite{Lu:2011jm} we use the soft form factor approach with $\xi_\parallel(q^2)=0.22\pm0.03$ for the S-wave contribution. 

\subsection[$m_{K\pi}$ distribution]{\boldmath{$m_{K\pi}$} distribution}
\label{sec:mkpi}
\begin{figure}
  \centering
  \includegraphics[width=9cm]{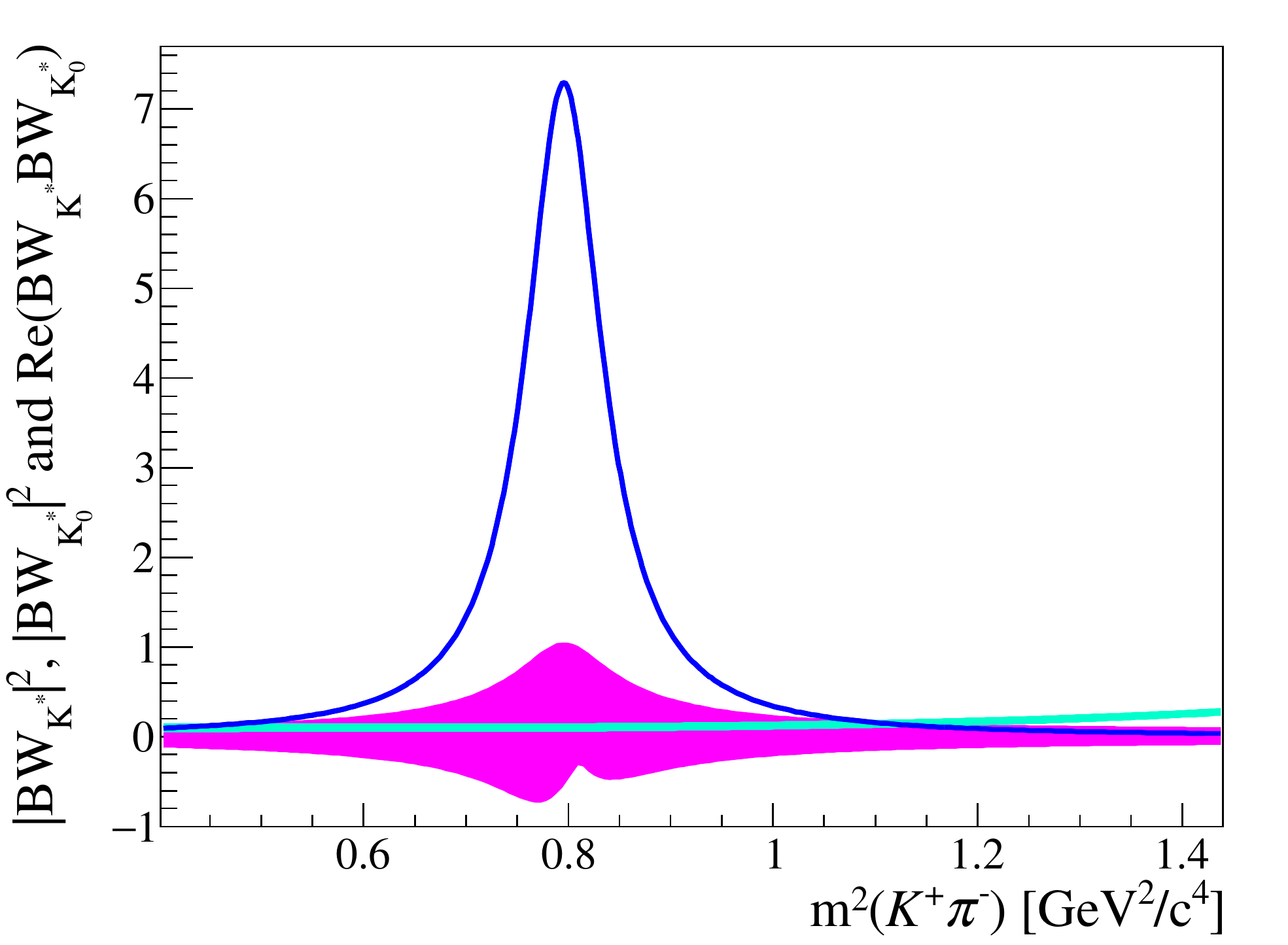}
  \caption{(Blue) $|{\cal BW}_{K^*}(m_{K\pi}^2)|^2$, (teal) $|{\cal BW}_{K^*_0}(m_{K\pi}^2)|^2$ and (magenta) the interference term ${\rm Re}({\cal BW}_{K^*}^\dagger(m_{K\pi}^2){\cal BW}_{K^*_0}(m_{K\pi}^2))$ when varying $g_\kappa$ and the relative phase between P-wave and S-wave as discussed in the text.     
    \label{fig:mkpi}}
\end{figure}

To statistically separate the contributions of the S-wave from the P-wave, the mass of the $\Kp\pim$ system is extremely valuable. 
The $\Kp\pim$ mass is included following Refs.~\cite{Lu:2011jm,Becirevic:2012dp}. 
The different 
amplitudes are affected as follows:
\begin{align}
{\cal A}_{0,\parallel,\perp,t}^{\rm L,R}(q^2,m_{K\pi}^2) &= {\cal A}_{0,\parallel,\perp}^{\rm L,R}(q^2)\times {\cal BW}_{K^*}(m_{K\pi}^2)\nonumber\\  
{\cal A}_{S0,St}^{\rm L,R}(q^2,m_{K\pi}^2) &= {\cal A}_{S0,St}^{\rm L,R}(q^2)\times {\cal BW}_{K^*_0}(m_{K\pi}^2),\label{eq:mkpidependence}
\end{align}
where the $m_{K\pi}$ dependent terms are given by
\begin{align}
{\cal BW}_{K^*}(m_{K\pi}^2)  &= \frac{\sqrt{m_{K^*}\Gamma_{K^*}/\pi}}{m_{K^*}^2-m_{K\pi}^2-im_{K^*}\Gamma_{K^*}}\nonumber\\
{\cal BW}_{K^*_0}(m_{K\pi}^2)  &= {\cal N}_m \left[-\frac{g_\kappa}{\left(m_\kappa-i\Gamma_\kappa/2\right)^2-m_{K\pi}^2}+\frac{1}{\left(m_{K^*_0}-i\Gamma_{K^*_0}/2\right)^2-m_{K\pi}^2}\right],
\end{align}
and the normalisation factor ${\cal N}_m$ is determined by the normalisation condition\linebreak $\int_0^\infty |{\cal BW}_{K^*_0}(m_{K\pi}^2)|^2 \rmd m_{K\pi}^2 = 1$. 
The symbol $K_0^*$ refers to the $K_0^*(1430)$ with mass $m_{K^*_0}=(1425\pm 50)\mevcc$ and width $\Gamma_{K^*_0}=(270\pm 80)\mevcc$. The symbol $\kappa$ refers to the $K^*_0(800)$ state with $m_\kappa=(658\pm 13)\mevcc$ and $\Gamma_\kappa=(557\pm 24)\mevcc$~\cite{DescotesGenon:2006uk}. 
For simplicity, the masses and widths of the $K^*_0$ and $\kappa$ will be fixed to their central values in the following. 
The complex coefficient $g_\kappa$ determines the relative magnitude and phase of the two contributions. 
We allow this parameter to vary in the range $0<|g_\kappa|<0.2$, following Ref.~\cite{Becirevic:2012dp}, and $0<\arg(g_\kappa)<2\pi$. 
Figure~\ref{fig:mkpi} gives the resulting $m_{K\pi}$ distribution of the P-wave and S-wave contributions in the SM, as well as their interference depending on the parameter $g_\kappa$. 
Good separation between the P-wave and S-wave component is observed. 

\begin{figure}
  \centering
\includegraphics[width=7cm]{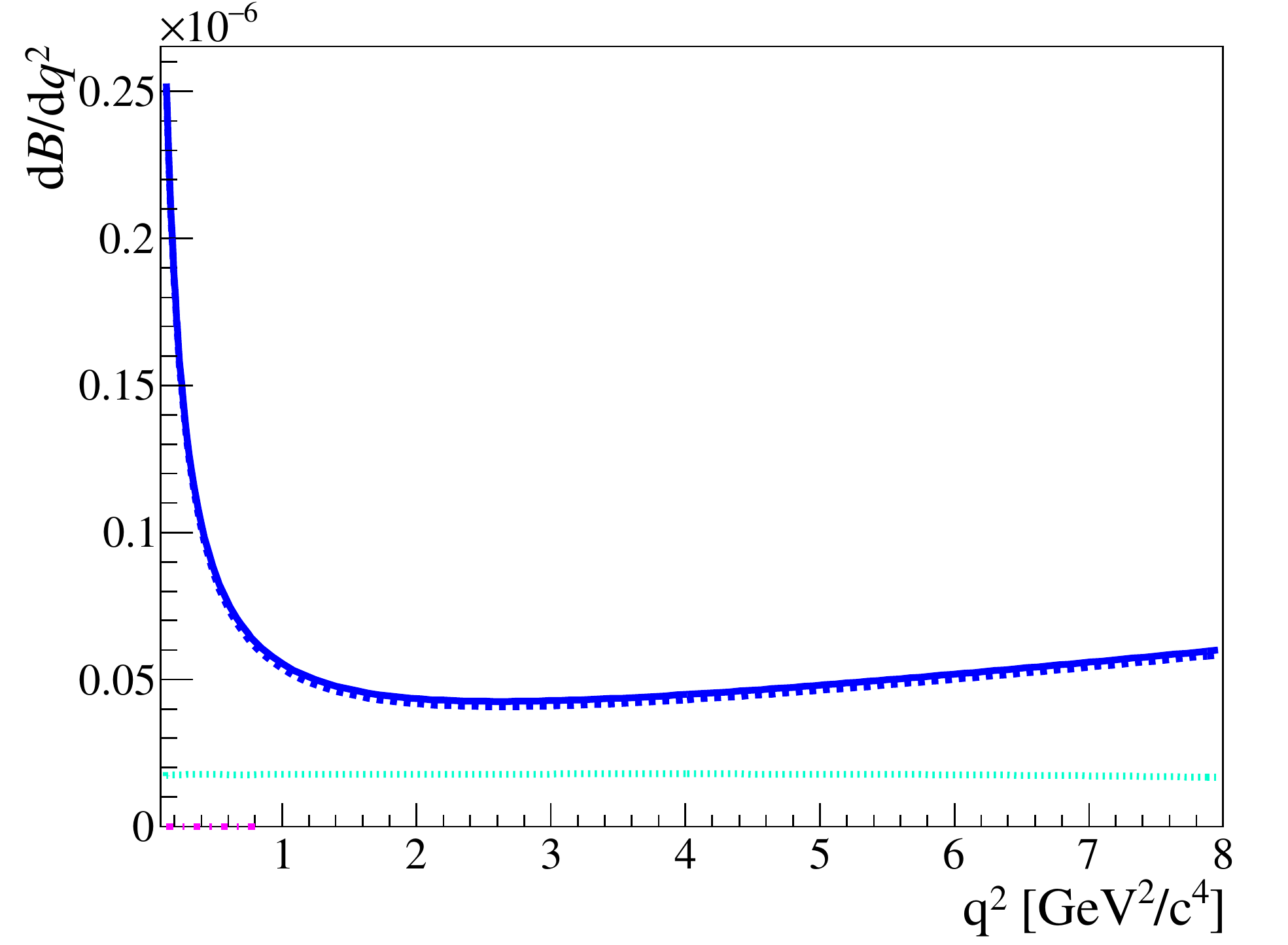}
\includegraphics[width=7cm]{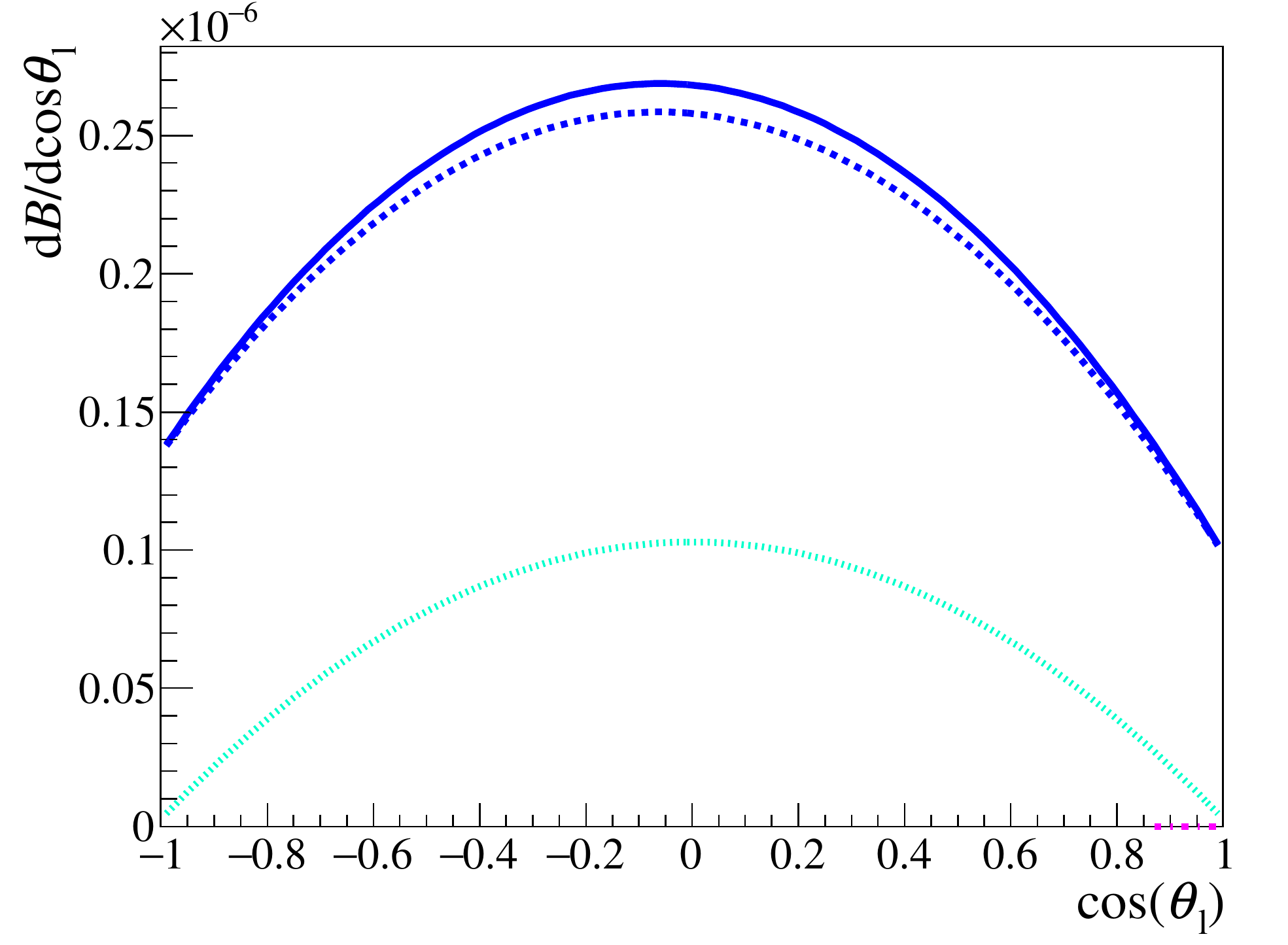}\\
\includegraphics[width=7cm]{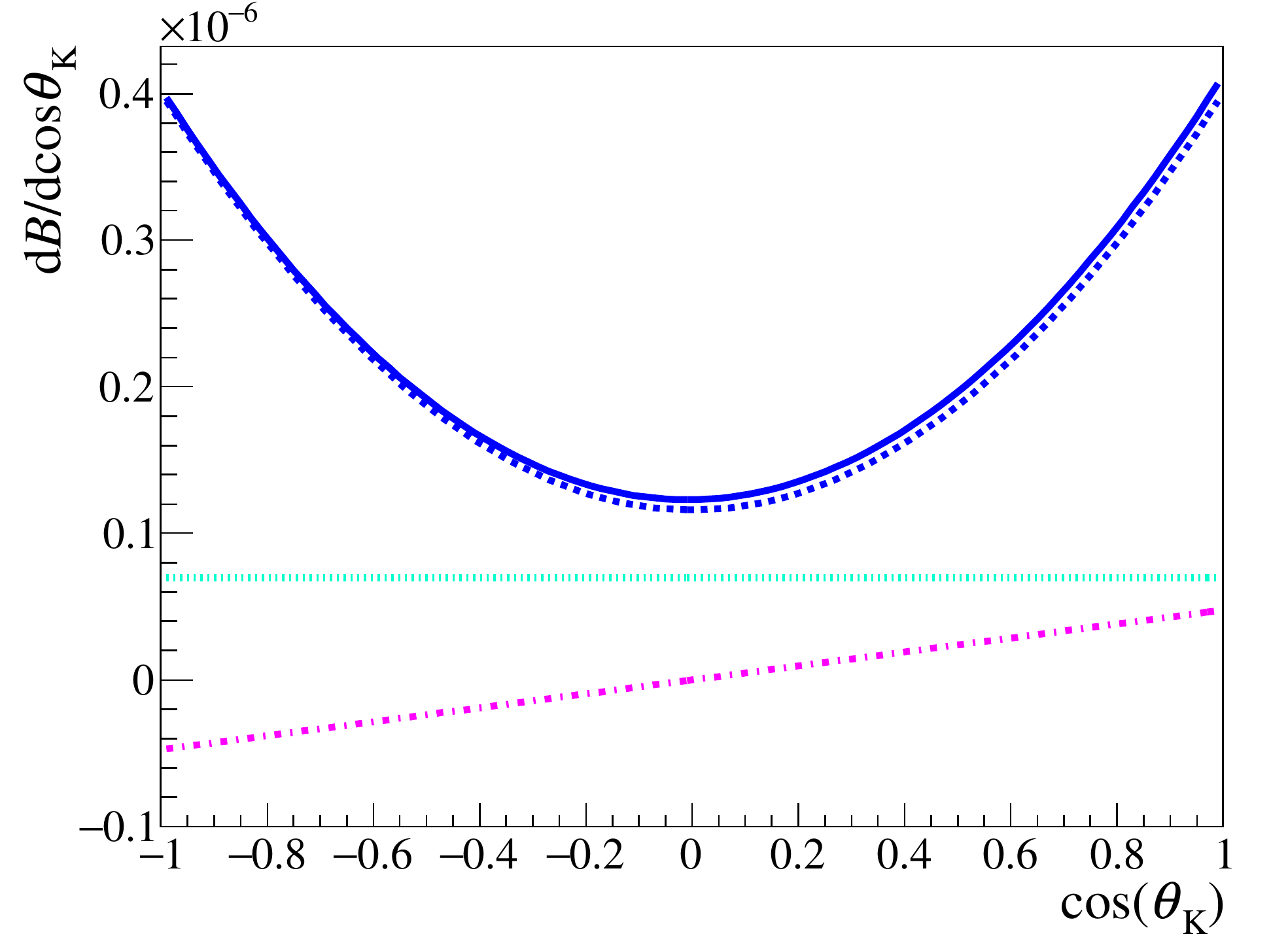}
\includegraphics[width=7cm]{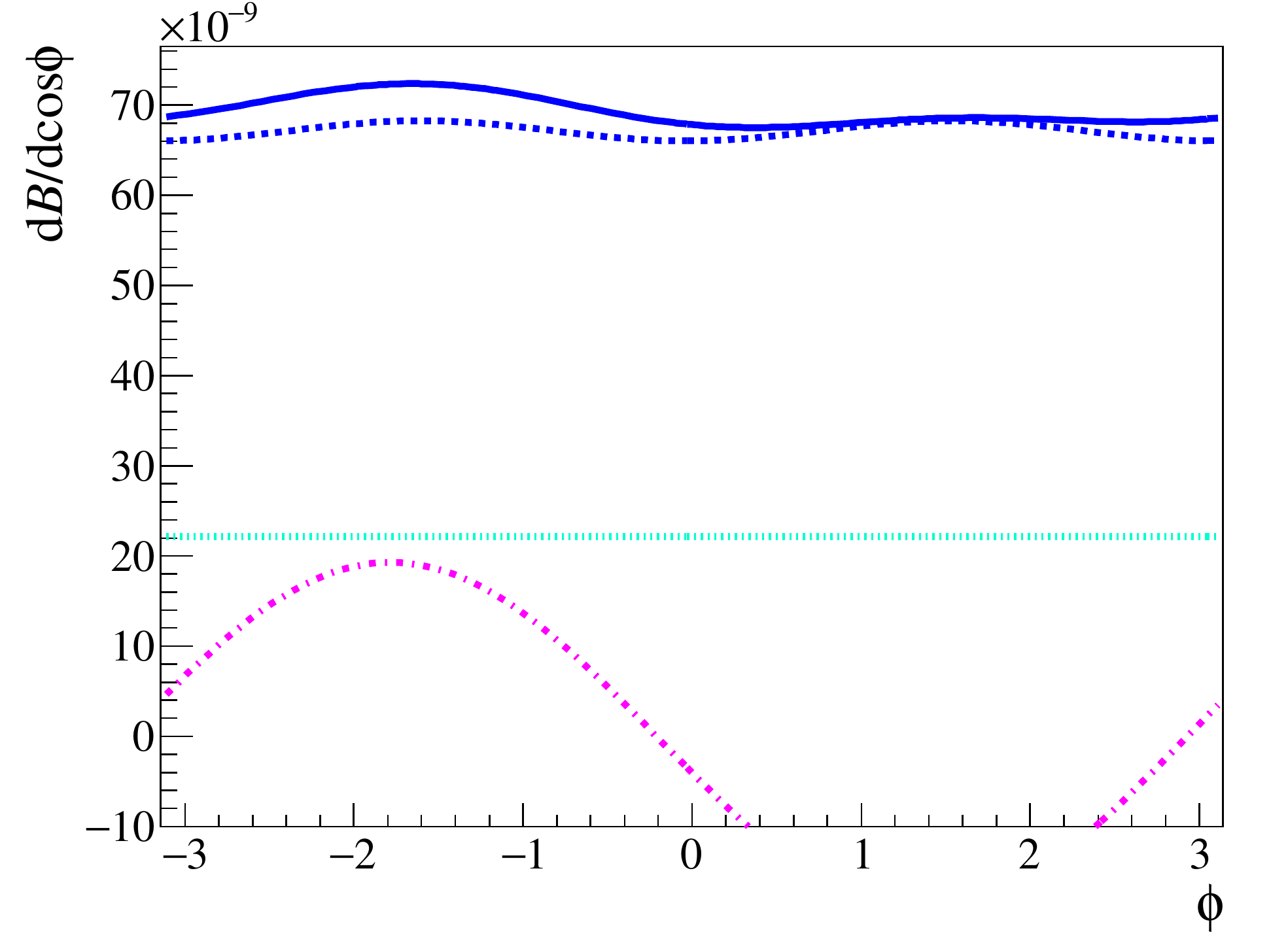}\\
\includegraphics[width=7cm]{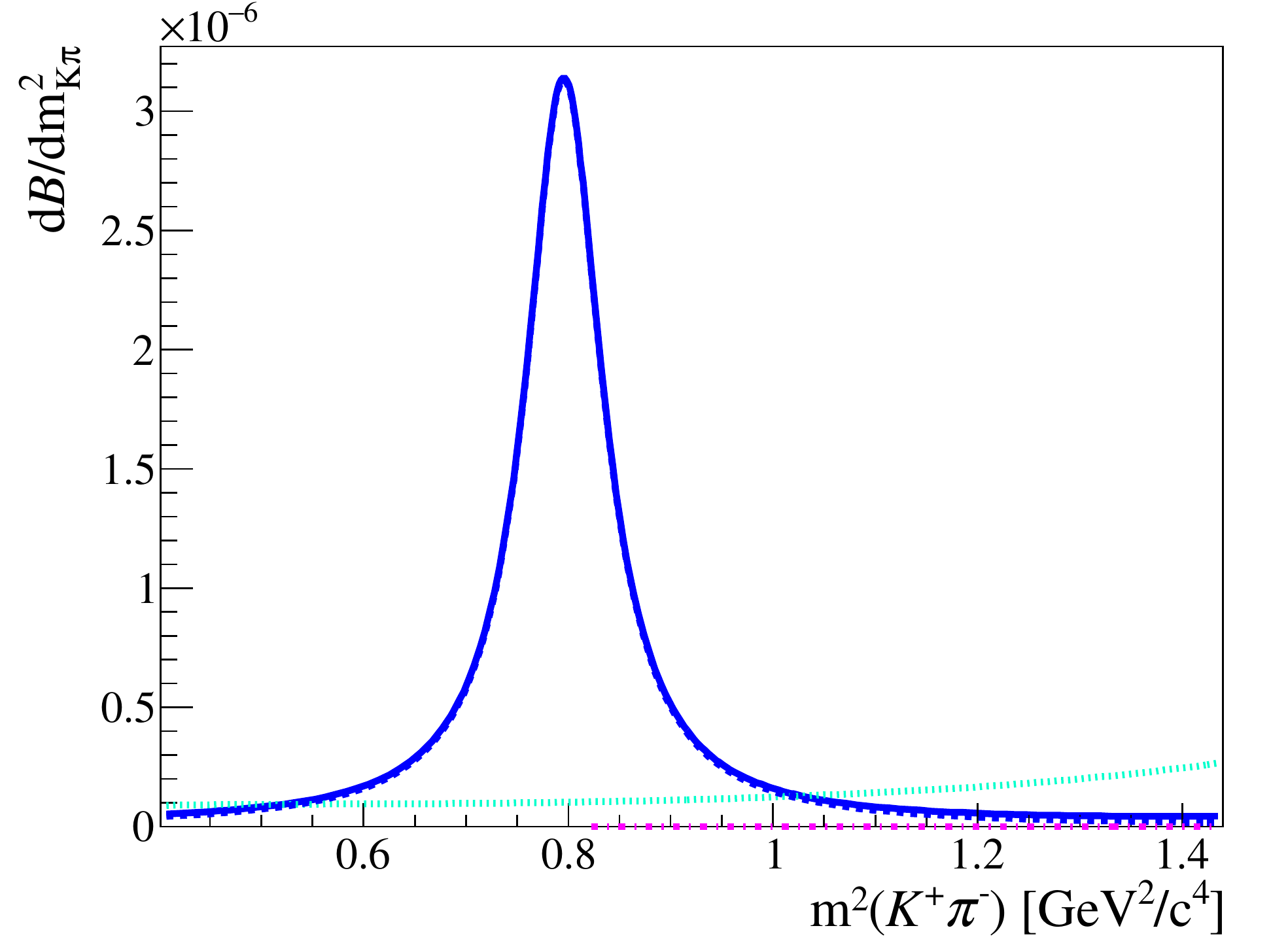}
\caption{
  Differential $\decay{\Bd}{\Kstarz\mumu}$ branching fraction depending on $q^2$, the three decay angles $\cos\thetal$, $\cos\thetak$ and $\phi$ and $m_{K\pi}^2$ in the $q^2$ range $0.1<q^2<8.0\gevgevcccc$. 
  The blue solid line denotes the full P+S-wave prediction, the blue dashed line the P-wave component, the teal dotted line the S-wave component
  and the magenta dash-dotted line the P-wave/S-wave interference. 
  Both the S-wave component and the interference are scaled by a factor 10 to improve readability. 
  \label{fig:projections_lowq2}}
\end{figure}

\begin{figure}
  \centering
\includegraphics[width=7cm]{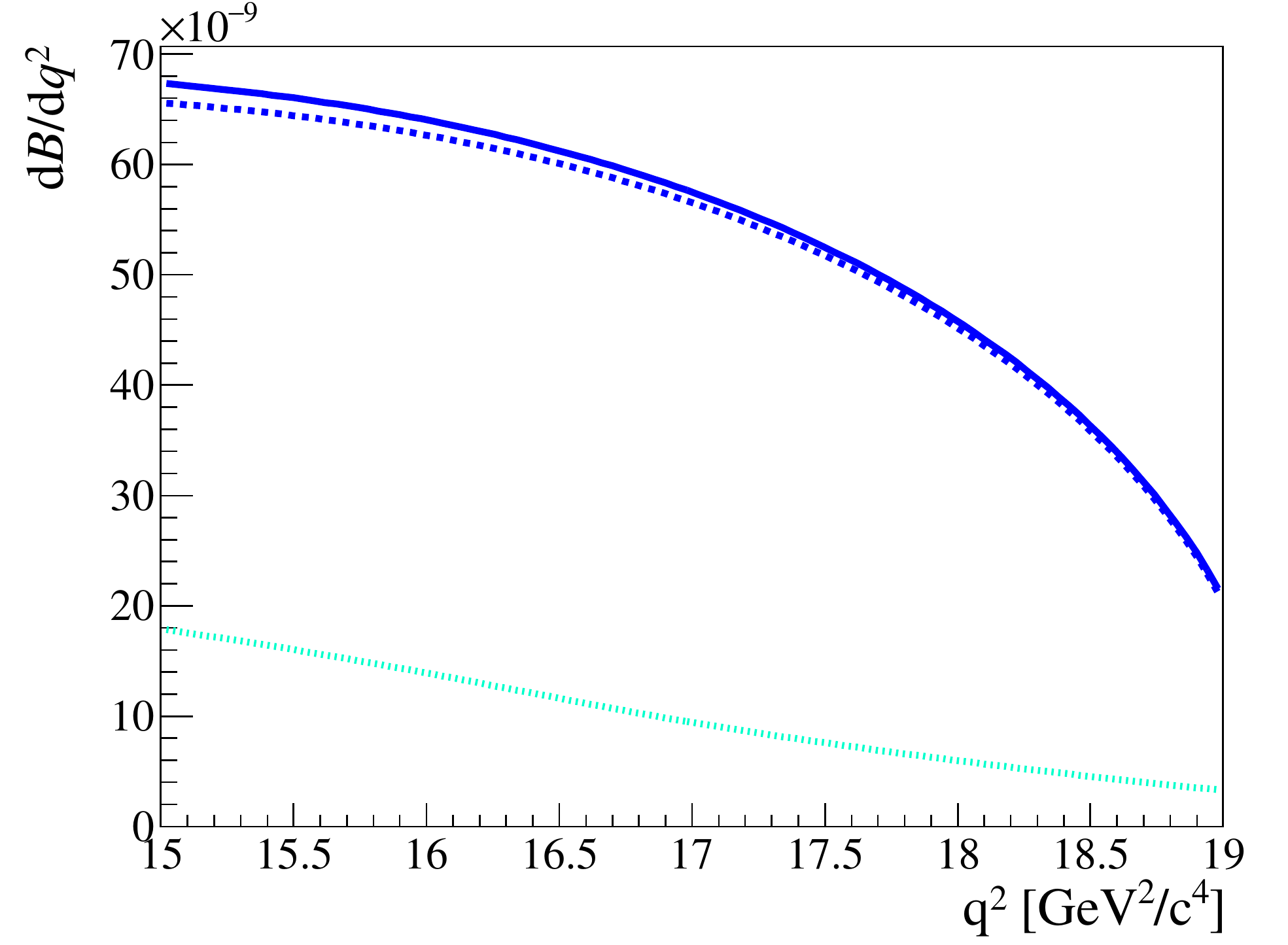}
\includegraphics[width=7cm]{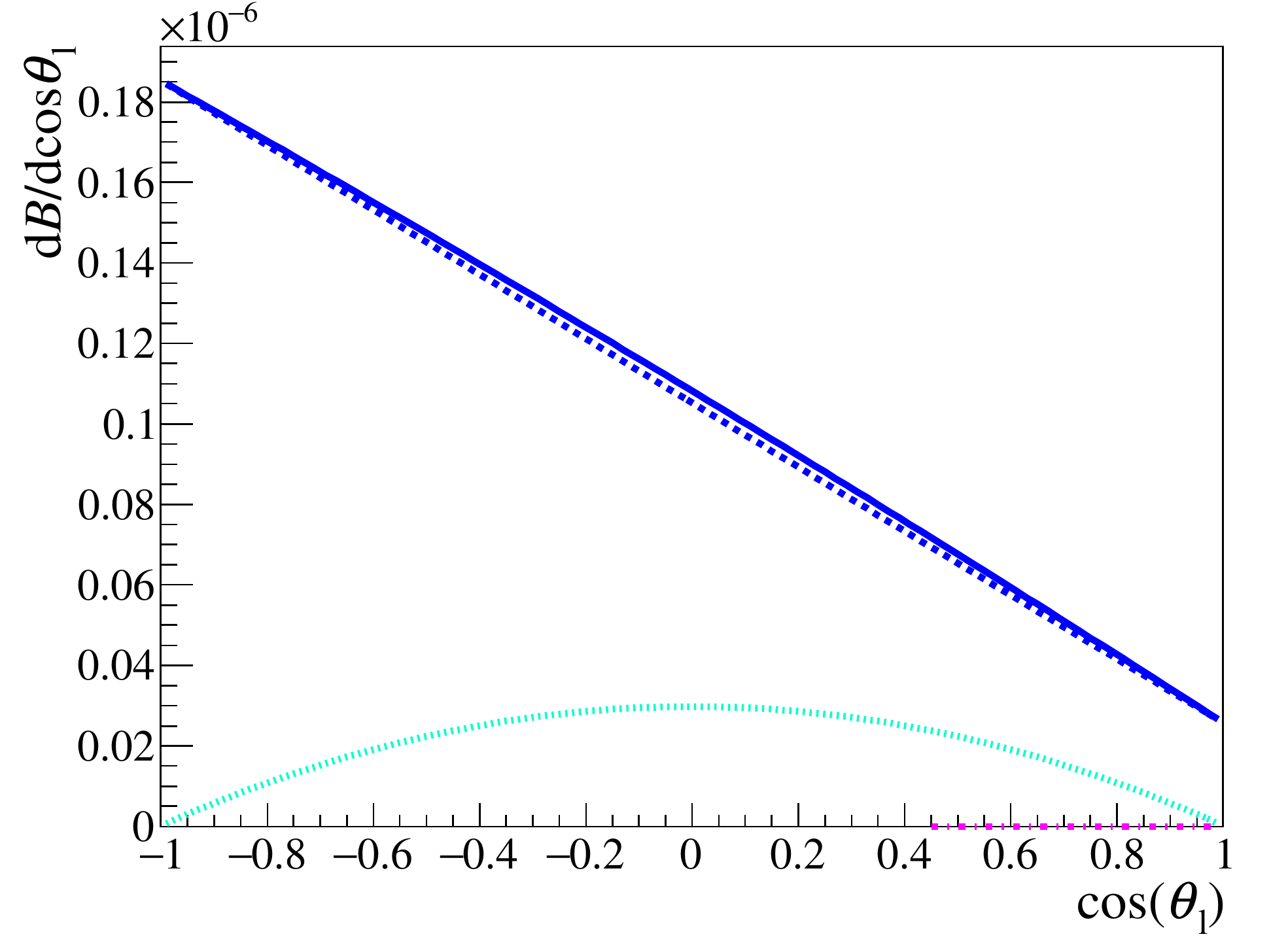}\\
\includegraphics[width=7cm]{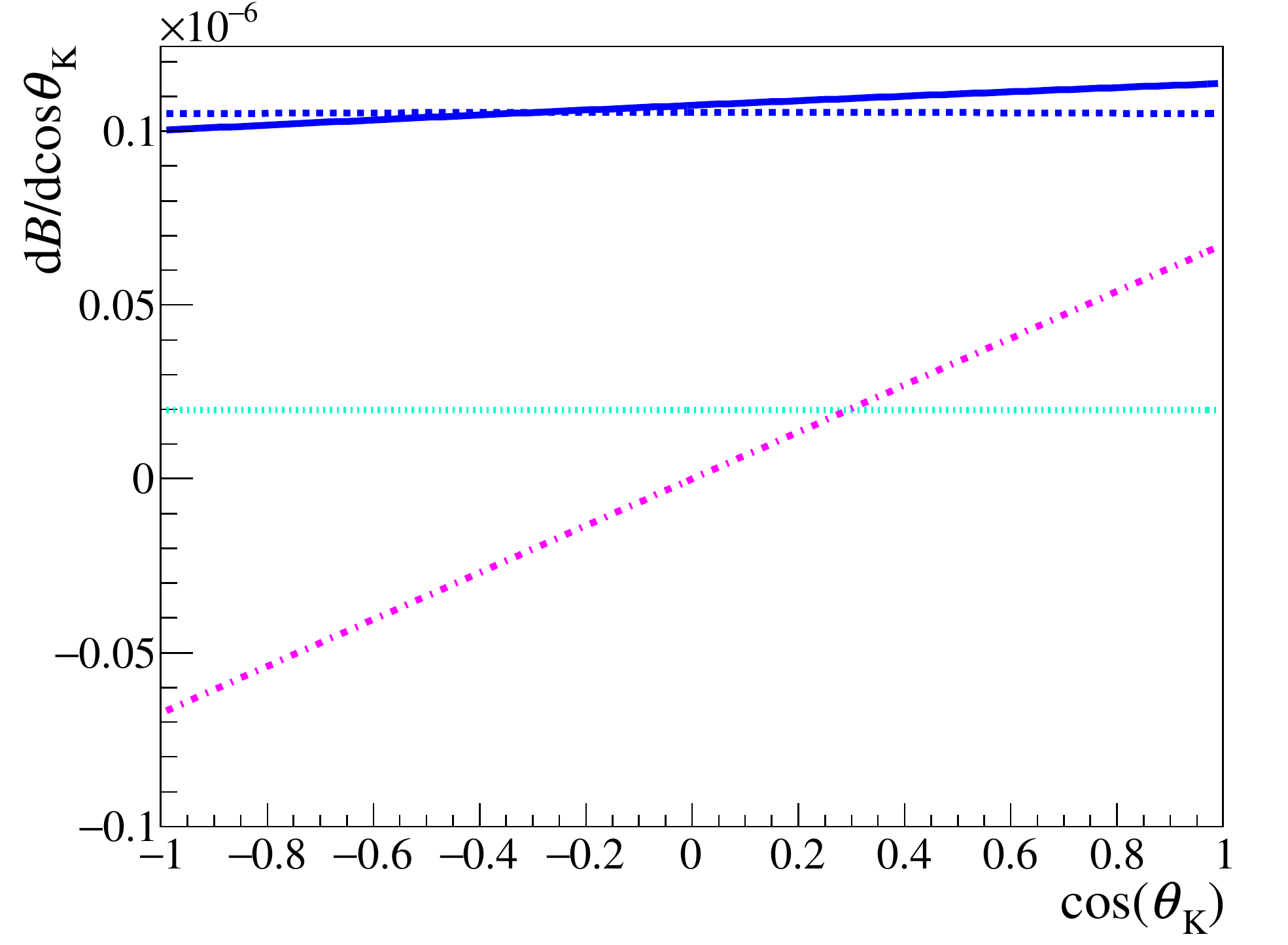}
\includegraphics[width=7cm]{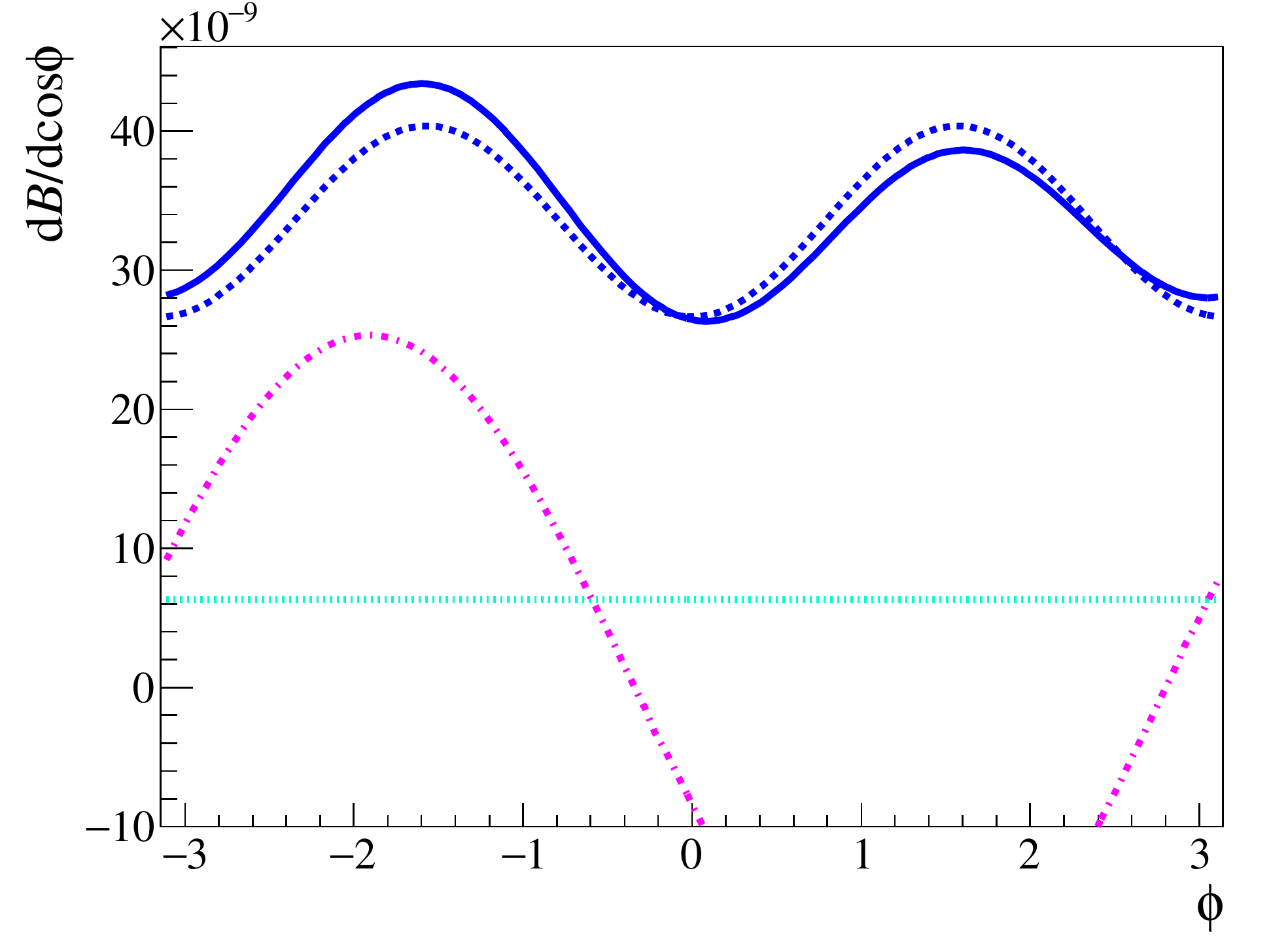}\\
\includegraphics[width=7cm]{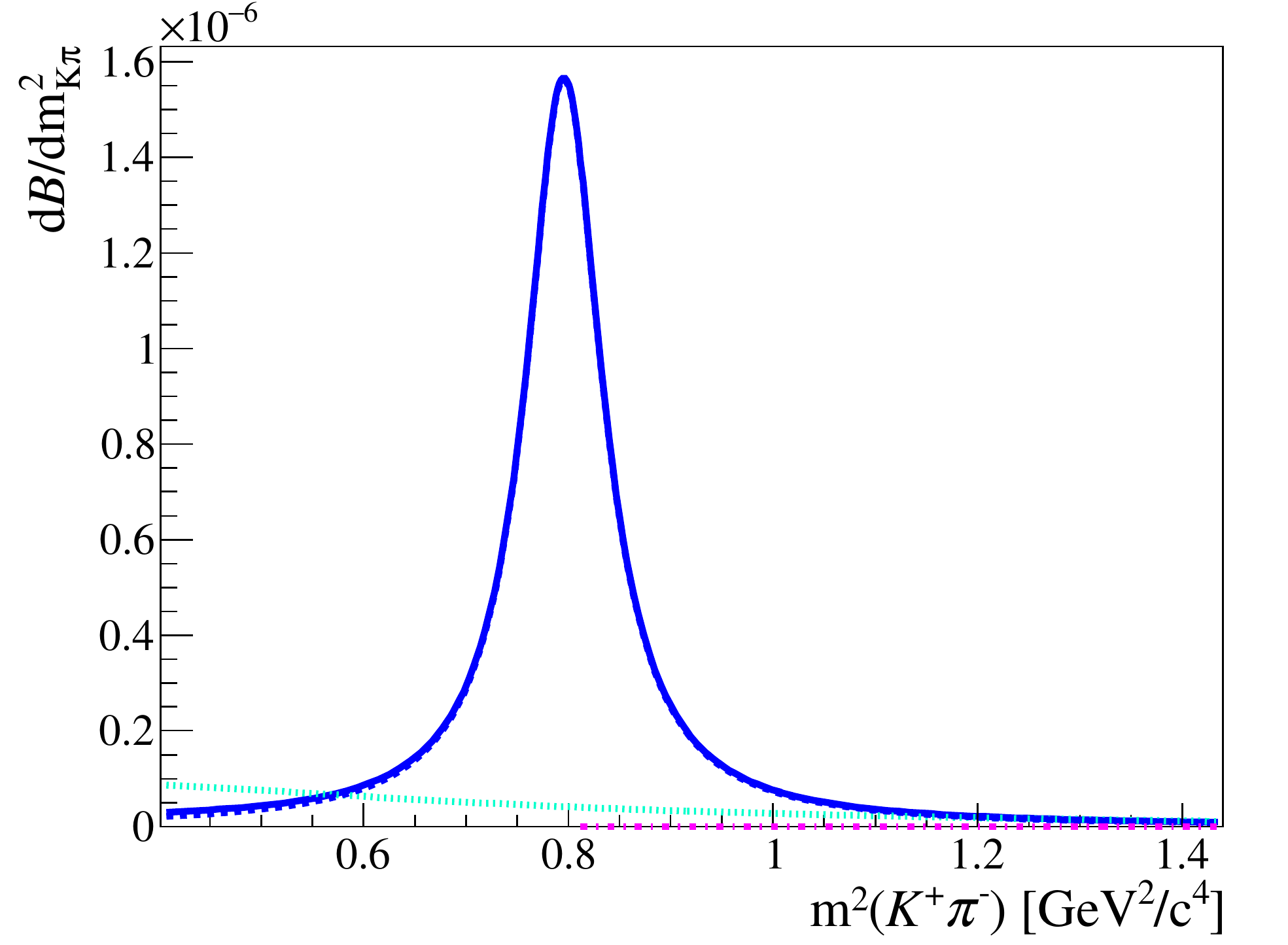}
\caption{
  Differential $\decay{\Bd}{\Kstarz\mumu}$ branching fraction depending on $q^2$, the three decay angles $\cos\thetal$, $\cos\thetak$ and $\phi$ and $m_{K\pi}^2$ in the $q^2$ range $15.0<q^2<19.0\gevgevcccc$. 
  The blue solid line denotes the full P+S-wave prediction, the blue dashed line the P-wave component, the teal dotted line the S-wave component
  and the magenta dash-dotted line the P-wave/S-wave interference. 
  Both the S-wave component as well as the interference are scaled by a factor 10 to improve readability.
  The $q^2$ distribution is only given for illustration, since the prediction relies on integration over a large $q^2$ range and
  does not predict any of the present and seen charmonium structures in this $q^2$ region. 
  \label{fig:projections_highq2}}
\end{figure}

The five differential decay rate
$\rmd^5\Gamma(\decay{\Bd}{\Kp\pim\mumu})/\left(\rmdcos\thetal\,\rmdcos \thetak\,\rmd \phi\,\rmd q^2\,\rmd m_{K\pi}^2\right)$
including both the P-wave and the S-wave contributions  
is given by Eq.~\ref{eq:spwave} when including the $m_{K\pi}$ dependence for the decay amplitudes ${\cal A}_{0,\parallel,\perp,t}^{\rm L,R}$ and ${\cal A}_{S0,St}^{\rm L,R}$ as detailed in Eq.~\ref{eq:mkpidependence}. 
For illustration, projections of the differential branching fraction on $q^2$, the decay angles, and $m_{K\pi}^2$ are given in Fig.~\ref{fig:projections_lowq2} for low $q^2$ and in Fig.~\ref{fig:projections_highq2} for high $q^2$.  

\clearpage

\section{Direct determination of Wilson coefficients} 
\label{sec:directfit}
\subsection{Detailed description of the method}
\label{sec:method}
An unbinned maximum likelihood fit is performed to determine the Wilson coefficients, using as input the three decay angles, $q^2$, $m_{K\pi}$ and the reconstructed $\Bd$ mass, as well as the $\Bd$ decay flavour determined by the kaon charge. 
As probability density function (PDF) for the signal decay $\decay{\Bd}{\Kp\pim\mumu}$, Eq.~\ref{eq:spwave} is used after including the $m_{K\pi}$ dependence according to Eq.~\ref{eq:mkpidependence}. 
The normalised signal PDF 
in the low $q^2$ region $0.1<q^2<8.0\gevgevcccc$ is then given by
\begin{align}
  {\cal S}(\vec{\Omega},q^2,m_{K\pi}^2| {\cal C}_{7,9,10}, \vec{\lambda})  &= \frac{\frac{{\rm d} \Gamma^5 (\decay{\parenbar{B}^0}{\Kp\pim\mumu})}{\rmd\vec{\Omega}\,\rmd q^2\,\rmd m_{K\pi}^2}}
  {\int\frac{{\rm d} \Gamma^5 (\decay{\parenbar{B}^0}{\Kp\pim\mumu})}{\rmd\vec{\Omega}\,\rmd q^2\,\rmd m_{K\pi}^2} \rmd{\vec{\Omega}}\,\rmd q^2\,\rmd m_{K\pi}^2}\nonumber\\
  &= \frac{\sum_i I_i(q^2,m_{K\pi}^2) f_i(\vec{\Omega})}{\sum_i \int I_i(q^2,m_{K\pi}^2)\rmd q^2\,\rmd m_{K\pi}^2\times \int f_i(\vec{\Omega})  \rmd\vec{\Omega}},\label{eq:fullpdf}
\end{align}
where $\vec{\lambda}$ denotes the nuisance parameters, including the form factor parameters and parameters describing subleading corrections, as well as quark masses, CKM parameters and S-wave parameters that will be discussed in detail in Sec.~\ref{sec:nuisances}. 
Where constraints from theory on the nuisance parameters are available, they are included in the fit using Gaussian constraints. 
The signal PDF is implemented using the EOS~\cite{Bobeth:2010wg,EOS,Beaujean:2013soa} and SuperIso~\cite{Mahmoudi:2007vz,Mahmoudi:2008tp,Mahmoudi:2009zz} software packages that are extended to include the S-wave contribution. 
Since the operator product expansion at low recoil 
relies on quark hadron duality it is only valid when integrating over a large $q^2$ range.
For the high $q^2$ region $15<q^2<19\gevgevcccc$ we therefore use the $q^2$-binned prediction in the fit. 

The low and high $q^2$ regions are fitted simultaneously using an extended unbinned maximum likelihood fit. 
The signal yields at low and high $q^2$ can be used to constrain the branching fraction of the decay, 
however in this paper we concentrate on the differential distribution and do not relate the determined yields to the total branching fraction. 
We use the Minuit minimiser to determine the Wilson coefficients and the nuisance parameters. 
The Hesse algorithm is used for the determination of the covariance matrix. 

\subsection{Theory nuisance parameters}
\label{sec:nuisances}
The theory nuisance parameters are summarised in Tab.~\ref{tab:nuisances}. 
The central values for the CKM parameters used in the generation of the pseudoexperiments are taken from Ref.~\cite{Bona:2006ah}. 
In the fit we allow their variation inside $\pm 3\sigma$ with $\sigma$ denoting their uncertainty. 
Their uncertainties are furthermore included as Gaussian constraints in the fit. 

The mass parameters $m_c$ and $m_b$ are also allowed to vary $\pm 3\sigma$ around their central values~\cite{PDG2012}. 
The uncertainties are also included as Gaussian constraints in the fit. 
For simplicity the top mass $m_t$ and the scale $\mu=4.2\gev$ are fixed. 

For the form factor parameters we use the results from the combined fit to lattice and LCSR calculations from Refs.~\cite{Straub:2015ica,Horgan:2013hoa,Horgan:2015vla}. 
The central values of the form factor parameters are given in Tab.~\ref{tab:nuisances}. 
For the uncertainties we use the full covariance matrix reported in Ref.~\cite{Straub:2015ica} and include it in the fit using a multivariate Gaussian constraint. 

For the S-wave contribution the soft form factor $\xi_\parallel(q^2)=0.22\pm0.03$ from Ref.~\cite{Lu:2011jm} is used. 
In the fit we apply a scale factor $S(\xi_\parallel)$ to the central value and include the relative uncertainty as Gaussian constraint. 
Further nuisances concerning the S-wave contribution are the fraction and relative phase of the $\kappa$ which are allowed to float as discussed in Sec.~\ref{sec:mkpi}. 
Finally, the S-wave contribution is allowed an overall relative phase $\delta_S$ with respect to the P-wave that is unconstrained in the fit. 

With respect to the non-factorisable subleading $\Lambda_{\rm QCD}/m_b$ corrections at low $q^2$ we follow Refs.~\cite{Hurth:2016fbr,Descotes-Genon:2014uoa} and include them as multiplicative factors $\left(1+a_i +b_i (q^2/6\gev^2)\right)$ with $i=0,\parallel,\perp$ and $a_i,b_i\in \mathbb{C}$ to the corresponding hadronic terms. 
At high $q^2$ we multiply the full transversity amplitudes with the factor $\left(1+c_i\right)$ with $c_i\in \mathbb{C}$. 
For the subleading corrections we use Gaussian constraints of $\pm 0.1$ around zero for 
the real and imaginary parts of $a_i$ and $c_i$. 
The real and imaginary parts of $b_i$ use Gaussian constraints of $\pm 0.25$ around zero. 
We note that the size of these Gaussian constraints are currently assumptions. 
Larger values of the (non-factorisable) power corrections cannot be ruled out at present, 
however a new approach using analyticity properties of the physical amplitudes may allow for real estimates of such contributions~\cite{Bobeth:2017vxj}. 
Such a parameterisation of the power corrections based on analyticity methods can also be implemented in the direct fit approach, which should be explored in future work.

\begin{table}
  \renewcommand{\arraystretch}{1.15}
  \centering
\begin{minipage}[t]{0.4\linewidth}
  \begin{tabular}[t]{L{2cm}R{3.5cm}}\hline
  Parameter & Value \\\hline\hline
\multicolumn{2}{c}{CKM parameters}\\\hline
$A$   & $0.807 \pm 0.02$ \\
$\lambda$ & $0.22535 \pm 0.00065$\\
$\bar{\rho}$ & $0.128 \pm 0.055$\\
$\bar{\eta}$ & $0.375 \pm 0.060$\\
\hline \multicolumn{2}{c}{Quark masses and scales}\\\hline
$m_c$ & $(1.275 \pm 0.025)\gev$\\
$m_b$ & $(4.18 \pm 0.03)\gev$\\
$m_t$ & $173.3\gev$\\
$\mu$ & $4.2\gev$\\
\hline \multicolumn{2}{c}{Subleading corrections}\\\hline
${\rm Re}(a_{0,\parallel,\perp})$ & $0\pm 0.1$\\
${\rm Im}(a_{0,\parallel,\perp})$ & $0\pm 0.1$\\
${\rm Re}(b_{0,\parallel,\perp})$ & $0\pm 0.25$\\
${\rm Im}(b_{0,\parallel,\perp})$ & $0\pm 0.25$\\
${\rm Re}(c_{0,\parallel,\perp})$ & $0\pm 0.1$\\
${\rm Im}(c_{0,\parallel,\perp})$ & $0\pm 0.1$\\ 
\hline \multicolumn{2}{c}{S-wave parameters}\\\hline
$\xi_\parallel$ & $0.22\pm 0.03$\\
$\delta_S$ & $\pi~(\in [0,+2\pi])$\\
$|g_{\kappa}|$ & $0.1~(\in [0,0.2])$\\
${\rm arg}(g_{\kappa})$ & $\pi/2~(\in [0,+2\pi])$\\
\hline\end{tabular}
\end{minipage}
\hspace{0.1\linewidth}
\begin{minipage}[t]{0.4\linewidth}
\begin{tabular}[t]{L{2cm}R{3.5cm}}\hline
  Parameter & Value \\\hline\hline
\multicolumn{2}{c}{Form factor parameters}\\\hline
$\alpha^{A_0}_0$ & $0.37\pm 0.03$\\
$\alpha^{A_0}_1$ & $-1.37\pm 0.26$\\
$\alpha^{A_0}_2$ & $0.13\pm 1.63$\\
$\alpha^{A_1}_0$ & $0.30\pm 0.03$\\
$\alpha^{A_1}_1$ & $0.39\pm 0.19$\\
$\alpha^{A_1}_2$ & $1.19\pm 1.03$\\
$\alpha^{A_{12}}_1$ & $0.53\pm 0.13$\\
$\alpha^{A_{12}}_2$ & $0.48\pm 0.66$\\                          
$\alpha^{V}_0$ & $0.38\pm 0.03$\\
$\alpha^{V}_1$ & $-1.17\pm 0.26$\\
$\alpha^{V}_2$ & $2.42\pm 1.53$\\                          
$\alpha^{T_1}_0$ & $0.31\pm 0.03$\\
$\alpha^{T_1}_1$ & $-1.01\pm 0.19$\\
$\alpha^{T_1}_2$ & $1.53\pm 1.64$\\
$\alpha^{T_2}_1$ & $0.50\pm 0.17$\\
$\alpha^{T_2}_2$ & $1.61\pm 0.80$\\
$\alpha^{T_{23}}_0$ & $0.67\pm 0.06$\\
$\alpha^{T_{23}}_1$ & $1.32\pm 0.22$\\
$\alpha^{T_{23}}_2$ & $3.82\pm 2.20$\\ 
\hline\end{tabular}
\end{minipage}
\caption{Nuisance parameters from theory used in the pseudoexperiments. 
  The given uncertainties indicate the Gaussian constraints used in the fit. 
  The form factor parameters are constrained in the fit using the full covariance matrix reported in Ref.~\cite{Straub:2015ica}.\label{tab:nuisances}}
\end{table}

\subsection{Mass distributions and backgrounds}
\label{sec:backgrounds}
For a realistic description of backgrounds, we use $f_{\rm sig}=N_{\rm sig}/(N_{\rm sig}+N_{\rm bkg})=0.6$, corresponding to the signal fraction found by LHCb integrated over the full $q^2$ region $[0.1,8.0]\gevgevcccc \cup [15.0,19.0]\gevgevcccc$. 
Both signal and background yields are allowed to vary and are determined in the fit. 
To separate the signal from the background contribution, which is predominantly combinatorial in nature, the reconstructed $\Bd$ mass is used:
\begin{align}
  {\cal P}(\vec{\Omega},q^2,m_{K\pi}^2,m_{K\pi\mu\mu}) = & 
  f_{\rm sig}\times {\cal S}(m_{K\pi\mu\mu})\times {\cal S}(\vec{\Omega},q^2,m_{K\pi}^2)\nonumber\\ 
  & + (1-f_{\rm sig})\times {\cal B}(m_{K\pi\mu\mu})\times {\cal B}(m_{K\pi}^2)\times {\cal B}(\vec{\Omega})\times {\cal B}(q^2). 
\end{align}
The $m_{K\pi\mu\mu}$ distribution of the signal is modeled using a double Crystal Ball shape, as published in Ref.~\cite{LHCb-PAPER-2015-051}. 
The signal mass parameters are fixed in the fit, as is also the case in Ref.~\cite{LHCb-PAPER-2015-051}. 
The combinatorial background is modelled using an Exponential function, the slope is allowed to vary in the fit. 
The angular and $m_{K\pi}^2$ and $q^2$ distributions of the background are generated flat in our pseudoexperiments.
In the fit the distributions are modeled using first order polynomials. 
For simplicity, the background distributions are assumed to factorise. 
It should be noted that instead of parameterising the background contribution,
it is also possible to statistically subtract it using the {\it sPlot} technique~\cite{Pivk:2004ty}. 

\subsection{Detector effects}
\label{sec:acceptance}
The reconstruction and selection of the signal decay $\decay{\Bd}{\Kstarz\mumu}$ leads to a distortion of the angular distribution, as well as $q^2$ and potentially also $m_{K\pi}$.
This acceptance effect can be accounted for in the signal PDF using the efficiency $\epsilon(\vec{\Omega}, q^2, m_{K\pi}^2)$, resulting in
\begin{align}
  {\cal S}(\vec{\Omega},q^2,m_{K\pi}^2| {\cal C}_{7,9,10}, \vec{\lambda}) 
  &= \frac{\epsilon(\vec{\Omega}, q^2, m_{K\pi}^2) \sum_i I_i(q^2,m_{K\pi}^2) f_i(\vec{\Omega})}{\sum_i \int \epsilon(\vec{\Omega}, q^2, m_{K\pi}^2) I_i(q^2,m_{K\pi}^2) f_i(\vec{\Omega}) \rmd\vec{\Omega}\,\rmd q^2\,\rmd m_{K\pi}^2}\nonumber\\
  &= \frac{\epsilon(\vec{\Omega}, q^2, m_{K\pi}^2) \sum_i I_i(q^2,m_{K\pi}^2) f_i(\vec{\Omega})}{\sum_i \int I_i(q^2,m_{K\pi}^2) \xi_i(q^2,m_{K\pi}^2) \rmd q^2\,\rmd m_{K\pi}^2},
\end{align}
where the angular integration results in the $q^2$ and $m_{K\pi}$ dependent terms $\xi_i(q^2,m_{K\pi}^2)$. 
The efficiency $\epsilon(\vec{\Omega}, q^2, m_{K\pi}^2)$ can be parameterised using the Legendre polynomial technique used in Ref.~\cite{LHCb-PAPER-2015-051}, 
the multiplicative factor in the numerator can be neglected in the minimisation of the negative logarithmic likelihood. 
For simplicity, all pseudoexperiments in this paper are performed with flat acceptance. 

It should be noted that, in principle, 
the probability density function given in Eq.~\ref{eq:fullpdf} needs to be convoluted with the detector
resolution in the decay angles, as well as $q^2$ and $m_{K\pi}$. 
The variation of the probability density function with the decay angles is very slow compared to the angular resolution, and angular resolution effects are thus neglected. 
The natural width of the $K^{*0}$ is large compared to the experimental resolution and therefore the resolution in $m_{K\pi}$ is neglected as well. 
To study the effect of the resolution in $q^2$ we perform pseudoexperiments in which we smear the generated $q^2$-value with the dimuon mass resolution at LHCb published in Ref.~\cite{LHCb-PAPER-2016-045}. 
The pseudoexperiments are then fitted neglecting the $q^2$-resolution. 
We find no significant change in sensitivity or bias in the observables and thus neglect the resolution in $q^2$ in the following. 

\section{Validation and performance determination} 
\label{sec:validation}

\subsection{Pseudoexperiments}
\label{sec:toys}
To validate the proposed method and to evaluate its sensitivity to the Wilson coefficients and nuisance parameters we perform pseudoexperiments. 
To this end, ensembles of 500 simulated samples are generated using an accept/reject method and then fitted. 
For the validation of the method, the pull distributions are of central importance. 
For a parameter $x$, the pull of pseudoexperiment $i$ is calculated according to $p_i=(x_i^{\rm fit}-x^{\rm gen})/\sigma_i(x)^{fit}$. 
The pull distributions are expected to be compatible with Gaussians centered around zero with a width of one if the fit is unbiased and the uncertainties are evaluated correctly. 
The sensitivity of the method for a certain observable is taken from the width of the distribution of fit values, it corresponds to the expected fit uncertainty on the parameter. 

Each simulated sample contains $9.6\,{\rm k}$ signal candidates which is four times the size of the Run~1 data sample. 
This choice corresponds to the expected signal yield at LHCb after Run~2. 
The background is modeled as described in Sec.~\ref{sec:backgrounds}, the signal fraction integrated over $q^2$ and in the $5170<m_{K\pi\mu\mu}<5700\mevcc$ mass range is given by $f_{\rm sig}=0.6$. 
The SM values of the Wilson coefficients are used in the generation of the pseudoexperiments. 
An overview of the nuisance parameters from theory used in the generation is given in Tab.~\ref{tab:nuisances}. 
While the central values for the nuisance parameters are used in the generation, in the subsequent fit these parameters are allowed to float. 

\subsection{Fits of a single pseudoexperiment}
\label{sec:single}
\begin{figure}
  \centering
\includegraphics[width=7cm]{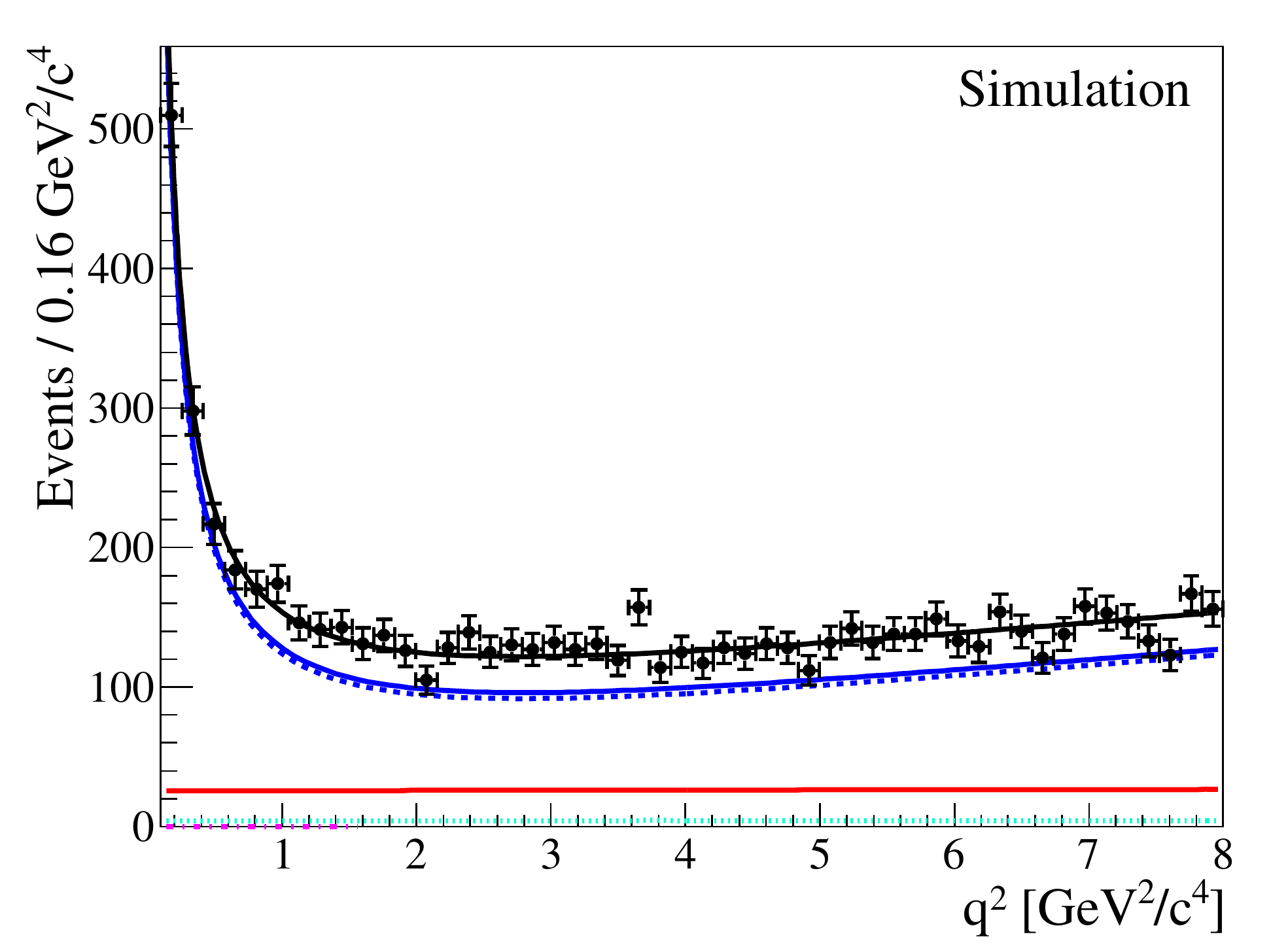}
\includegraphics[width=7cm]{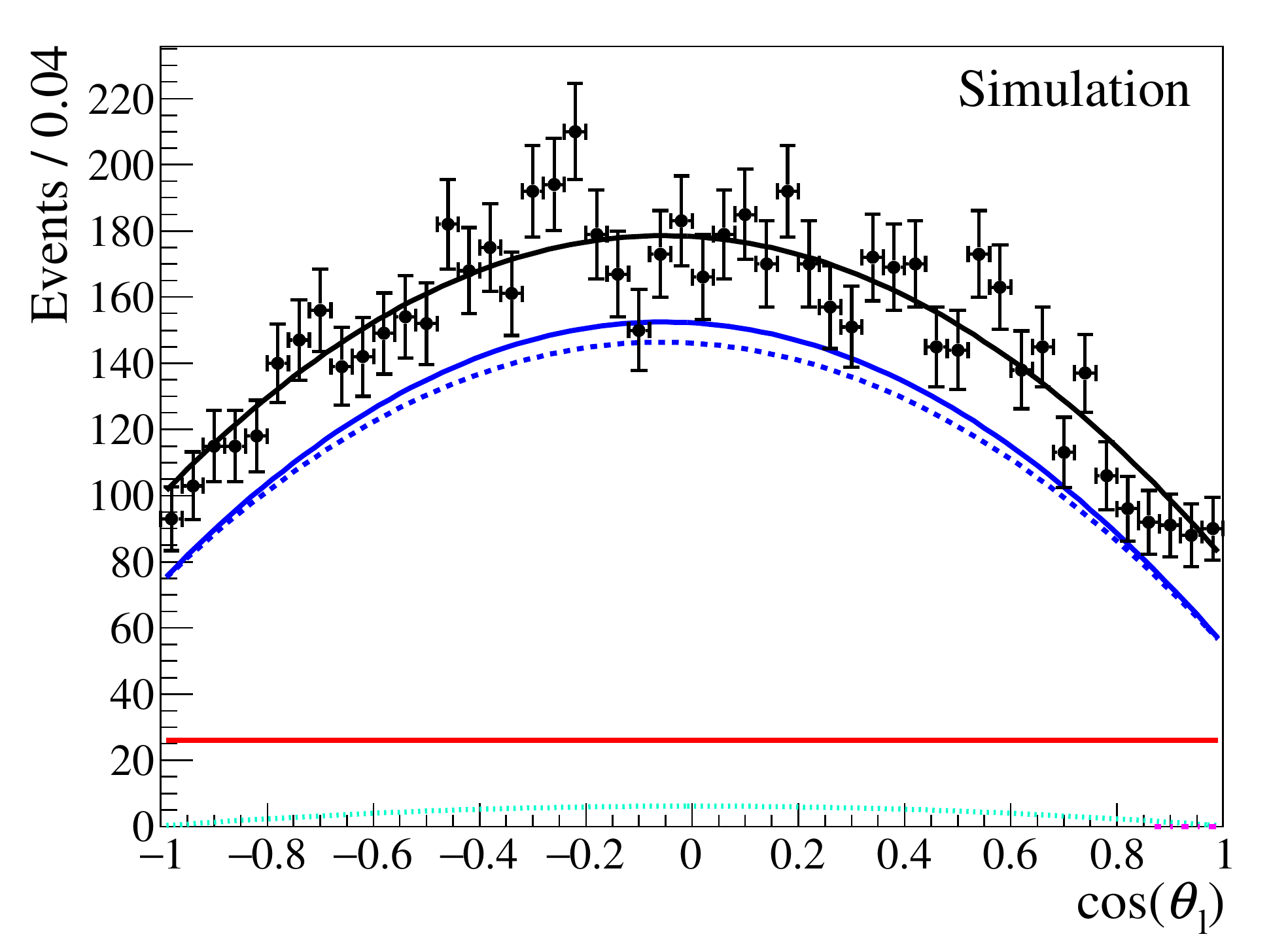}\\
\includegraphics[width=7cm]{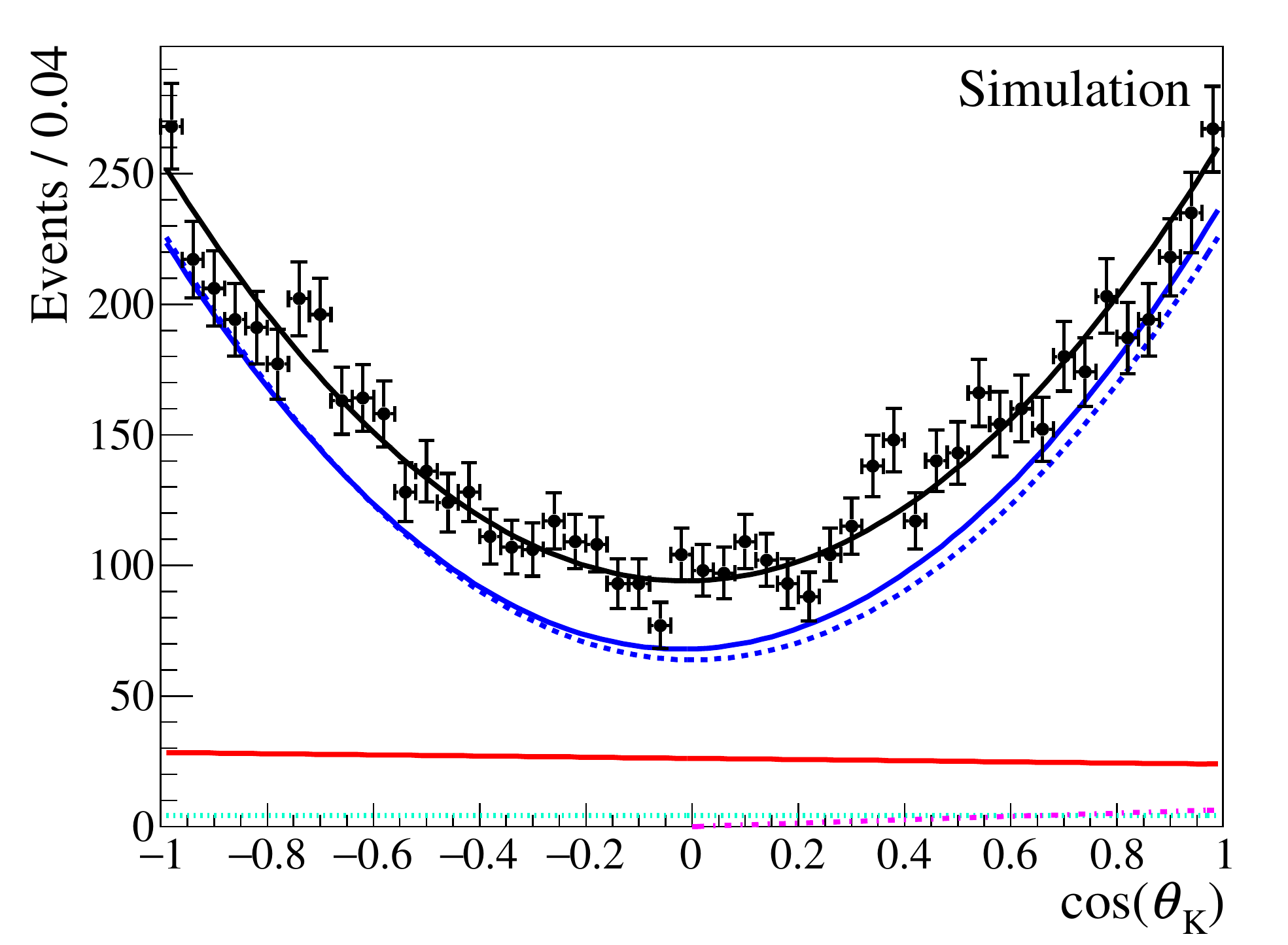}
\includegraphics[width=7cm]{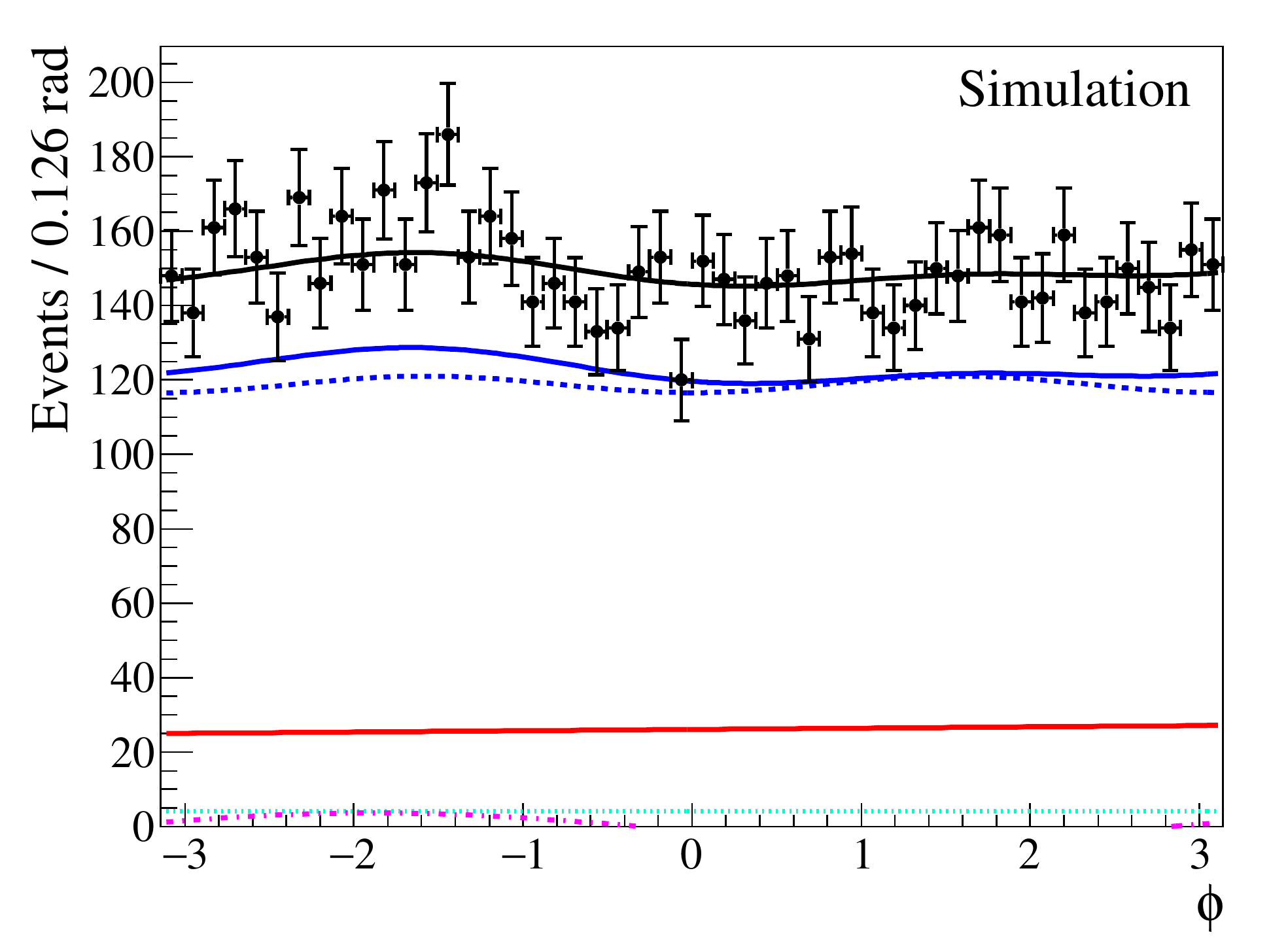}\\
\includegraphics[width=7cm]{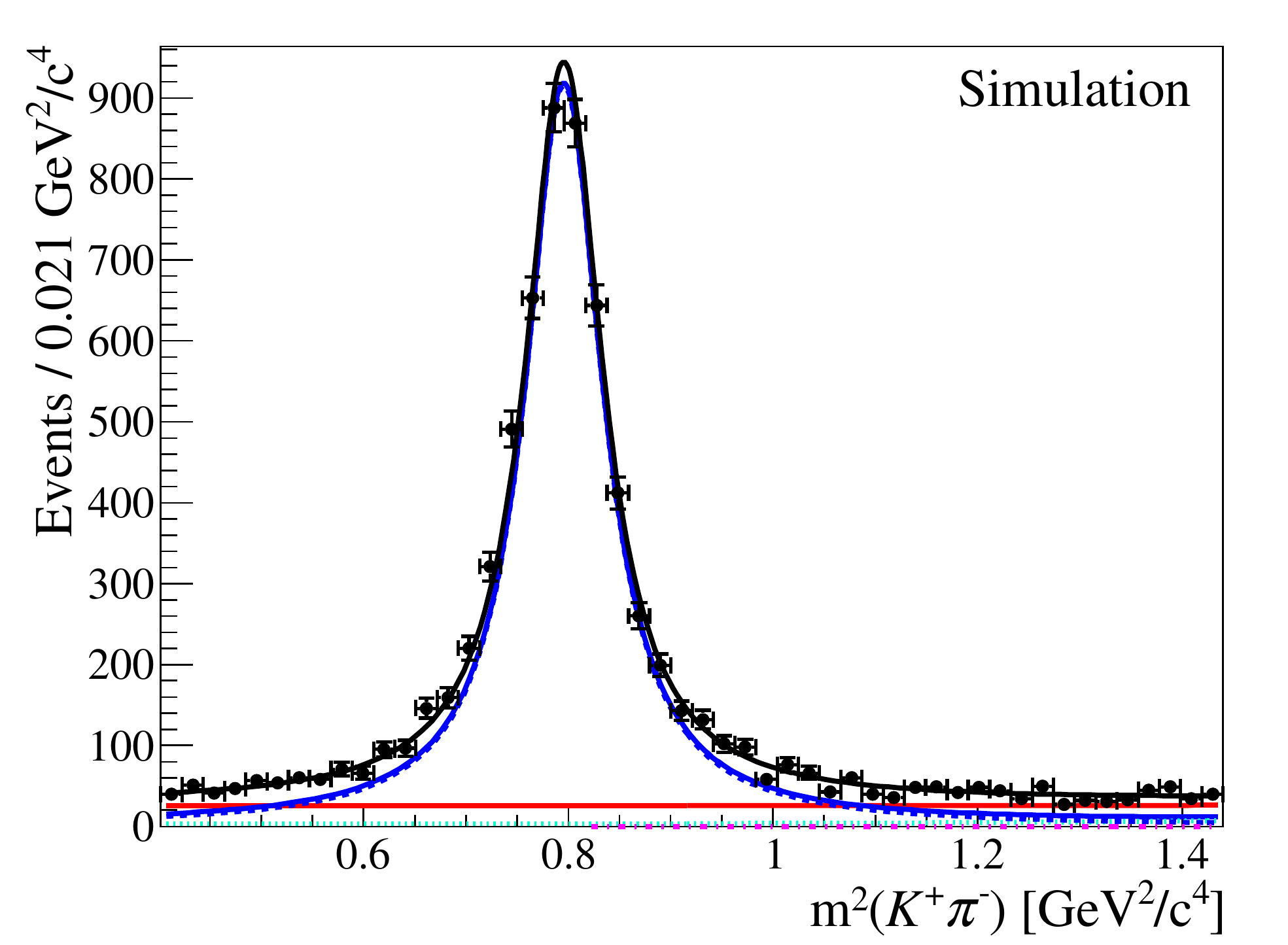}
\caption{
  Results from the fit of a single pseudoexperiment varying the Wilson coefficients ${\rm Re}({\cal C}_7)$ and ${\rm Re}({\cal C}_9)$. 
  Simulated events are overlaid with projections of the fitted PDF on $q^2$, the three decay angles $\cos\thetal$, $\cos\thetak$ and $\phi$ and $m_{K\pi}^2$ in the $q^2$ range $0.1<q^2<8.0\gevgevcccc$. 
  The simulated events and projections are shown for the signal region $\pm 50\mevcc$ around the $\Bd$ mass to enhance the signal fraction. 
  The black solid line denotes the full PDF, the blue solid line the signal component. 
  The blue dashed line gives the P-wave and the teal dotted line the S-wave part. 
  The magenta dash-dotted line finally gives the P-wave/S-wave interference and the red line the background contribution. 
  \label{fig:fit_projections_lowq2}}
\end{figure}

\begin{figure}
  \centering
\includegraphics[width=7cm]{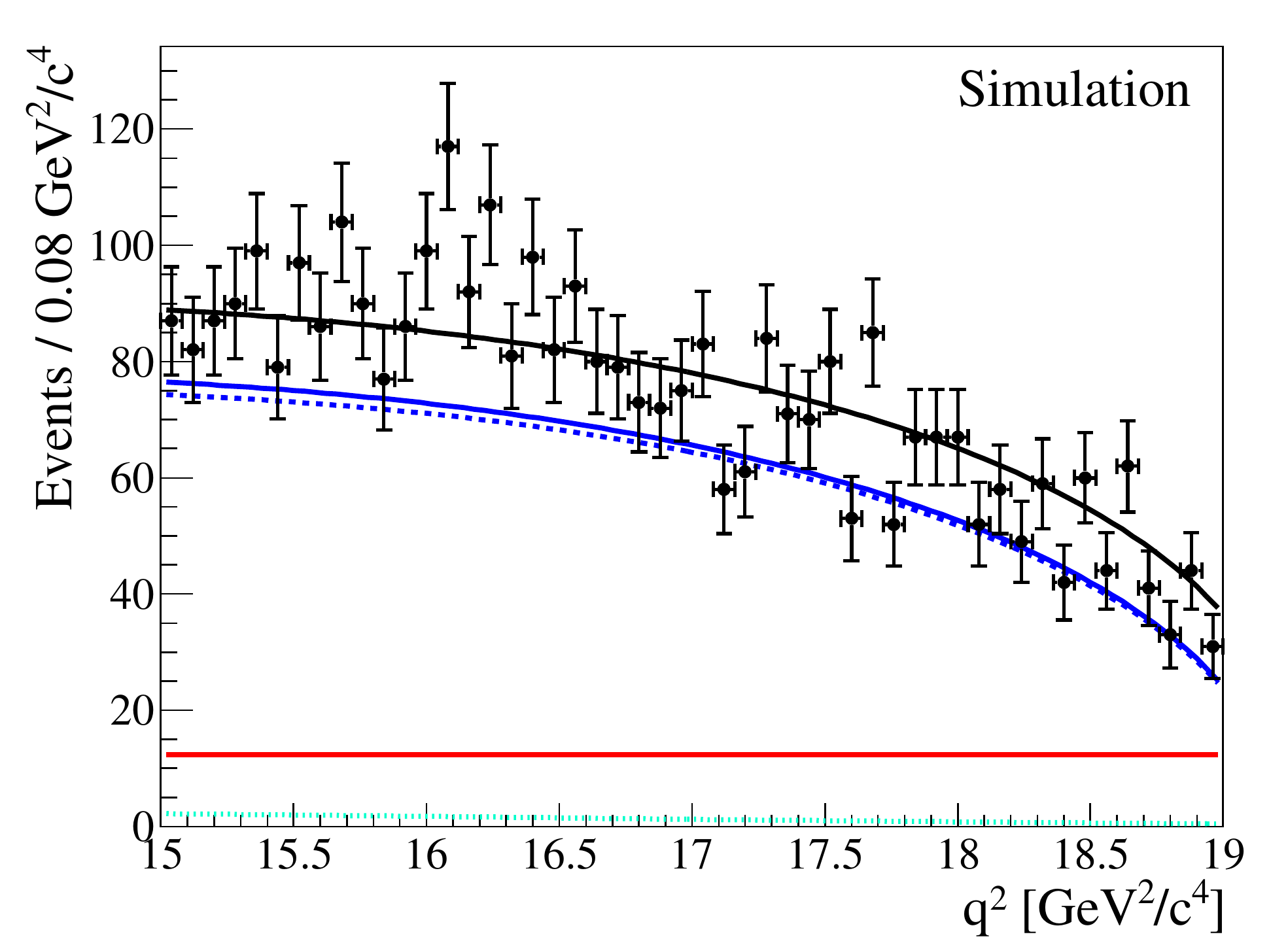}
\includegraphics[width=7cm]{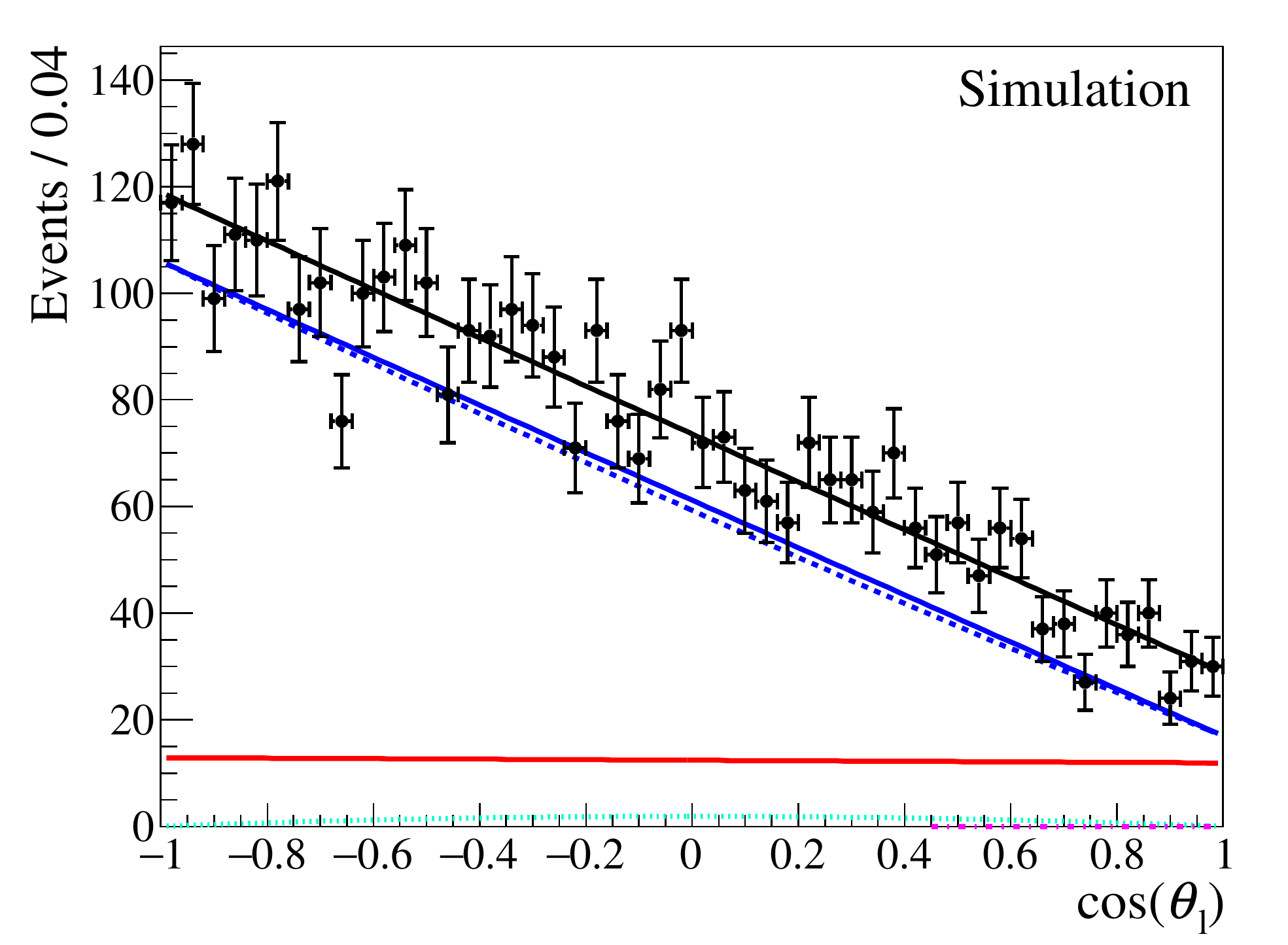}\\
\includegraphics[width=7cm]{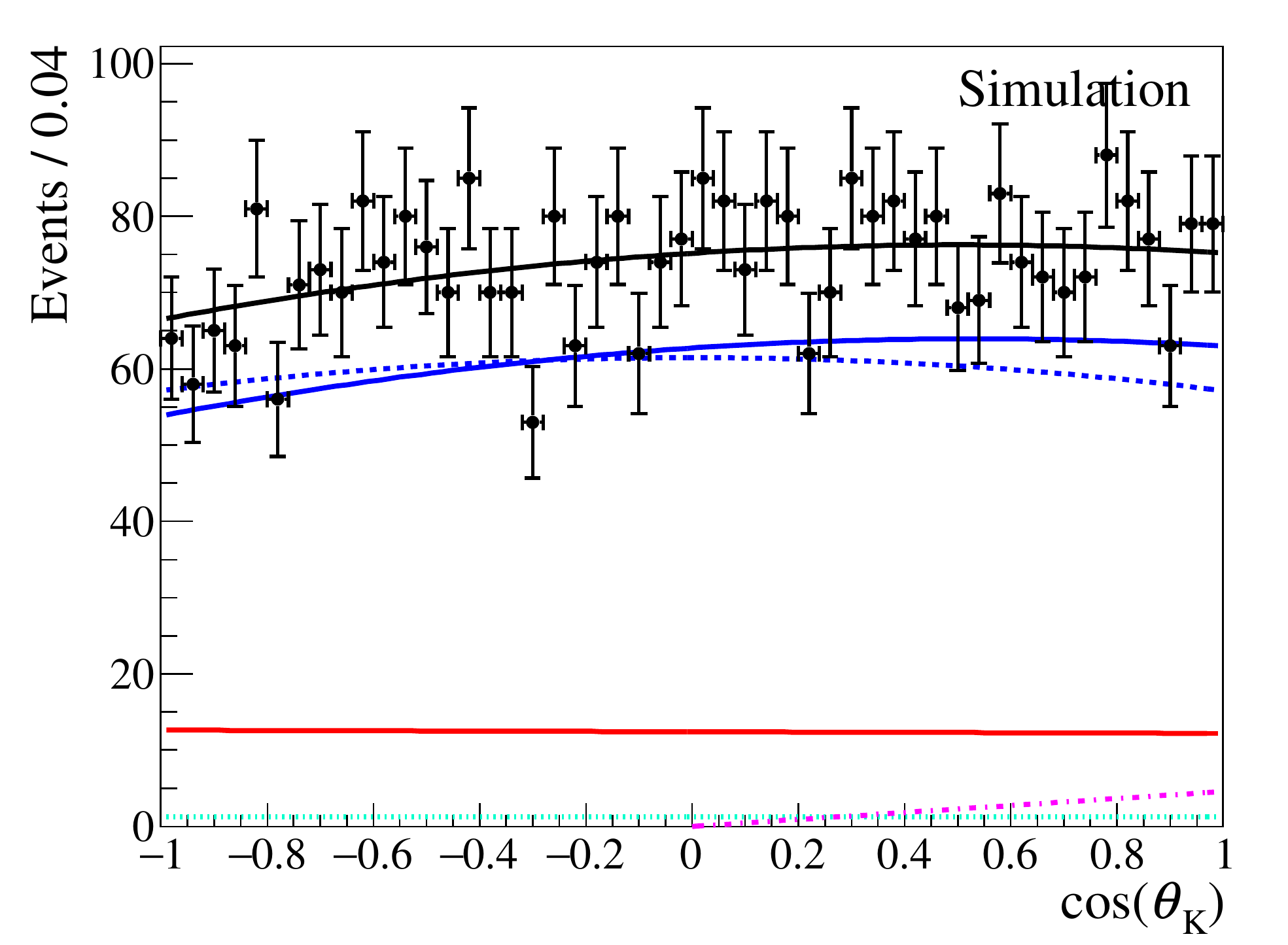}
\includegraphics[width=7cm]{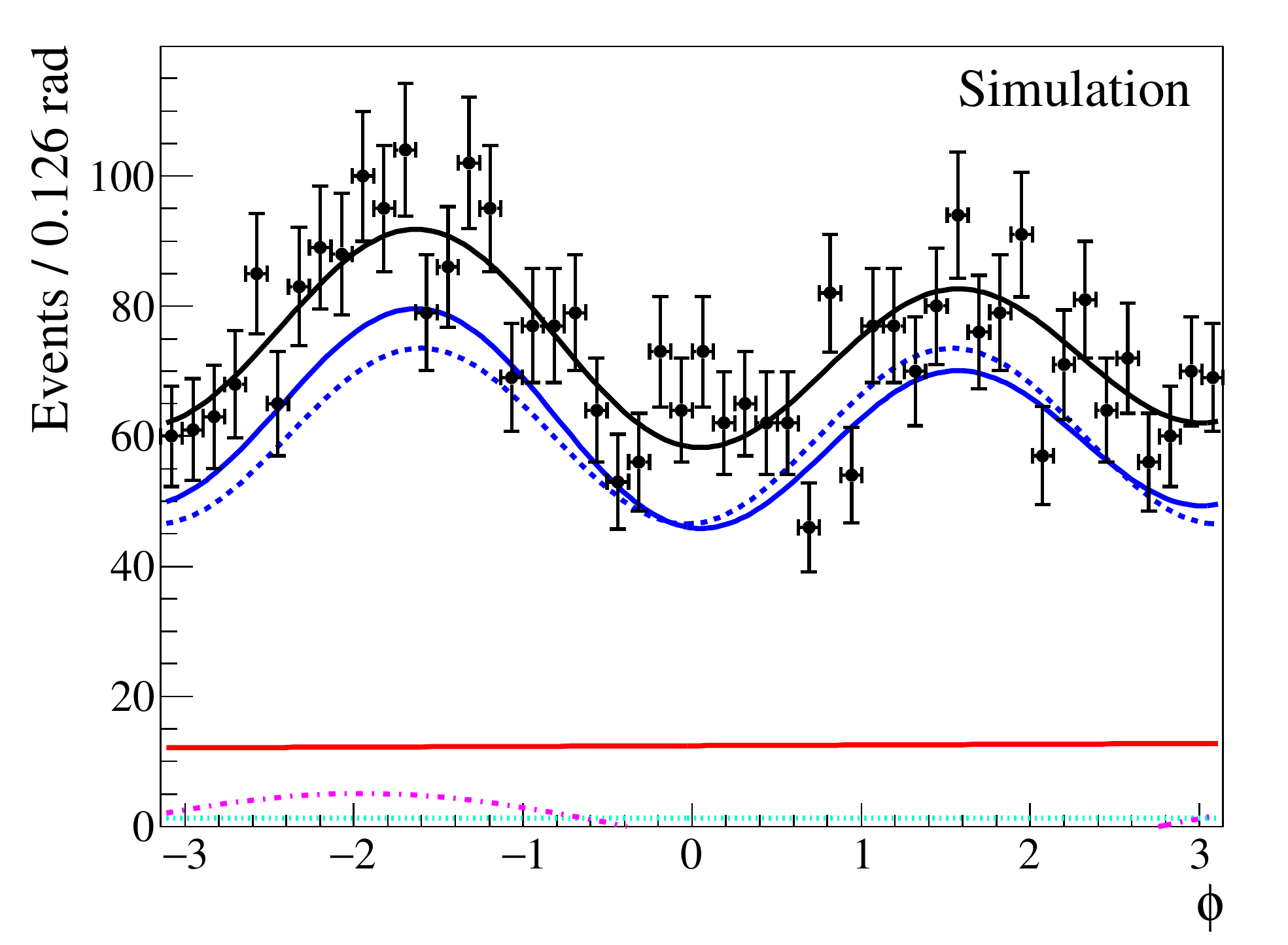}\\
\includegraphics[width=7cm]{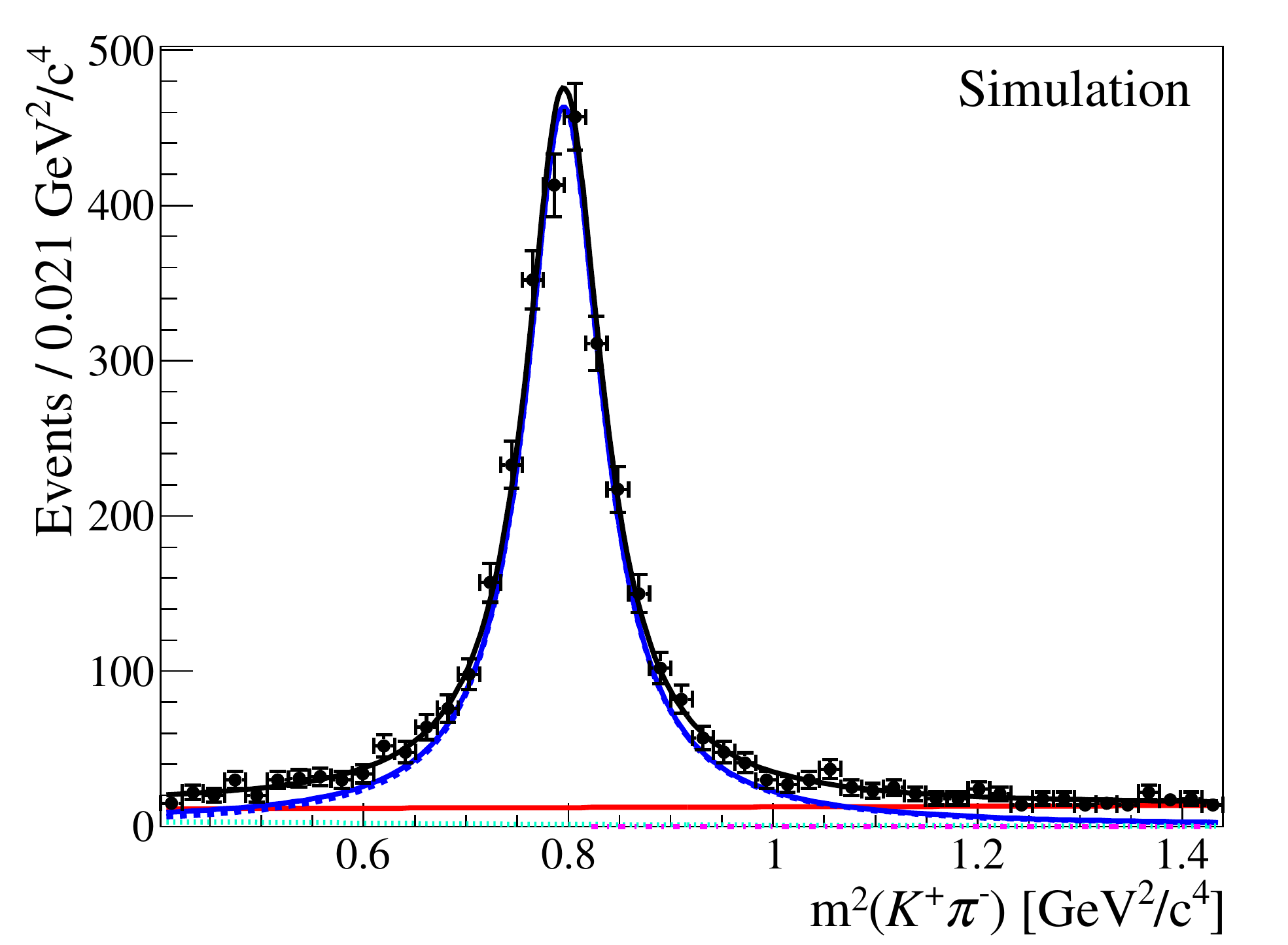}
\caption{
  Results from the fit of a single pseudoexperiment varying the Wilson coefficients ${\rm Re}({\cal C}_7)$ and ${\rm Re}({\cal C}_9)$. 
  Simulated events are overlaid with projections of the fitted PDF on $q^2$, the three decay angles $\cos\thetal$, $\cos\thetak$ and $\phi$ and $m_{K\pi}^2$ in the $q^2$ range $15.0<q^2<19.0\gevgevcccc$. 
  The simulated events and projections are shown for the signal region $\pm 50\mevcc$ around the $\Bd$ mass to enhance the signal fraction. 
  The black solid line denotes the full PDF, the blue solid line the signal component. 
  The blue dashed line gives the P-wave and the teal dotted line the S-wave part. 
  The magenta dash-dotted line finally gives the P-wave/S-wave interference and the red line the background contribution.
  We note again that the $q^2$ distribution in the high $q^2$ region is only shown for illustration, as it is not used in the direct fit method. 
  \label{fig:fit_projections_highq2}}
\end{figure}

\begin{figure}
  \centering
\subfloat[\label{fig:lh_c7c9}]{
  \includegraphics[width=8.5cm,clip=true,trim=0mm 5mm 0mm 0mm]{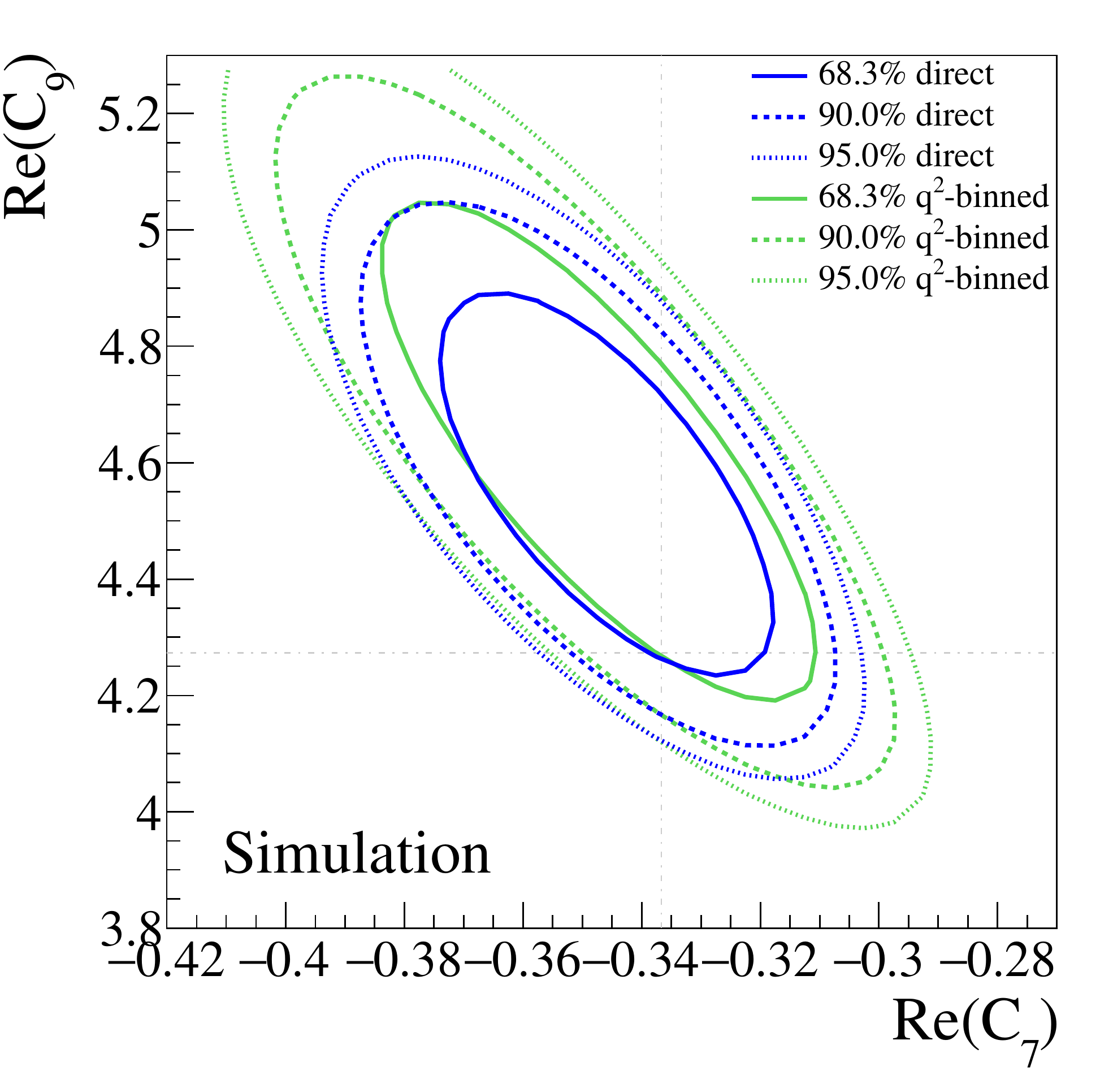}
}\\
\subfloat[\label{fig:lh_c9c10}]{
  \includegraphics[width=8.5cm,clip=true,trim=0mm 5mm 0mm 0mm]{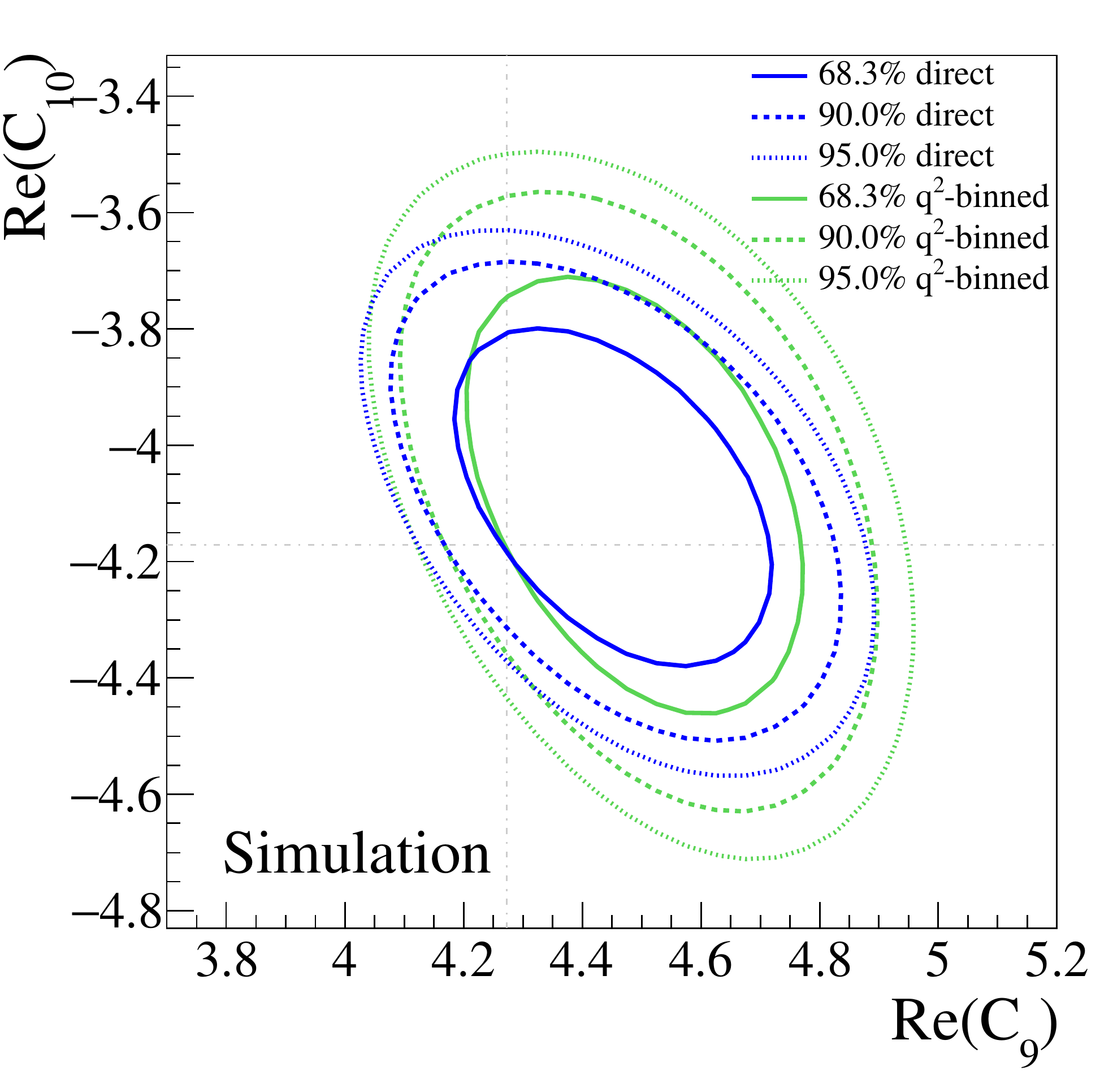}
  }
  \caption{
    Confidence regions for \protect\subref{fig:lh_c7c9} ${\rm Re}({\cal C}_7)$ and ${\rm Re}({\cal C}_9)$ and \protect\subref{fig:lh_c9c10} ${\rm Re}({\cal C}_9)$ and ${\rm Re}({\cal C}_{10})$ for a single pseudoexperiment. 
    The results from the direct fit method are given by the blue contours, the results from the determination using $q^2$-binned angular observables are given by the green contours.
    The contours correspond to confidence levels of $68.3\%$, $90\%$ and $95\%$ for the solid, dashed and dotted lines, respectively. 
    The SM values for the Wilson coefficients used in the generation are indicated by the grey dash-dotted lines. 
    \label{fig:lhprofiles}}
\end{figure}

Results for the fit of a single pseudoexperiment, determining the Wilson coefficients ${\rm Re}({\cal C}_7)$ and ${\rm Re}({\cal C}_9)$ while fixing ${\rm Re}({\cal C}_{10})$, are given in Tab.~\ref{tab:dir_fit} in App.~\ref{sec:singleresults}. 
Projections of the fitted PDF on $q^2$, the decay angles and $m_{K\pi}$ are given in Figs.~\ref{fig:fit_projections_lowq2} and~\ref{fig:fit_projections_highq2}.
Good agreement between simulated events and the PDF projections is observed. 
In Fig.~\ref{fig:lh_c7c9} we show the confidence regions for the two Wilson coefficients resulting from the profile likelihood. 
The solid, dashed and dotted lines correspond to the $68.3\%$, $90\%$ and $95\%$ confidence regions, respectively. 
The SM values for the Wilson coefficients used in the generation are indicated by the dash-dotted grey lines.
They lie within the the $1\,\sigma$ confidence region. 
Furthermore, we perform a fit of ${\rm Re}({\cal C}_9)$ and ${\rm Re}({\cal C}_{10})$ while fixing ${\rm Re}({\cal C}_{7})$. 
The results are given in Tab.~\ref{tab:dir_fit} and the fit projections are shown in Figs.~\ref{fig:fit_c9c10_projections_lowq2} and~\ref{fig:fit_c9c10_projections_highq2} in App.~\ref{sec:projections}. 
The confidence regions for the Wilson coefficients are shown in Fig.~\ref{fig:lh_c9c10}. 
In summary, a good behaviour of the direct fit method is observed for a single pseudoexperiment.
To get a quantitative estimate of the performance of the method and to validate it, it is however necessary to study the full ensemble of pseudoexperiments, 
as detailed in Sec.~\ref{sec:wilsons} below. 

\subsection{Sensitivity on Wilson coefficients and nuisance parameters}
\label{sec:wilsons}
To validate the direct fit method and estimate its performance 
the pseudoexperiments are fit using several different configurations for the Wilson coefficients. 
First, fits of single Wilson coefficients are performed, while all other Wilson coefficients are fixed to their SM values. 
Detailed results for fits of ${\rm Re}({\cal C}_7)$, ${\rm Re}({\cal C}_9)$ and ${\rm Re}({\cal C}_{10})$ are given in Tab.~\ref{tab:dir_cseven}-\ref{tab:dir_cten} in App.~\ref{sec:dir_results}, where background parameters are omitted for brevity. 
The first column of the tables gives the sensitivity to the parameters, determined from the width of the distribution of fitted parameter values. 
The uncertainty on the given sensitivity is the statistical error due to the limited number of pseudoexperiments. 
Furthermore, the mean values and widths of the pull distributions are given for every parameter. 
The physics parameters have pull distributions that are centered around zero with a width of one. 
The direct fit method is thus unbiased and the parameter uncertainties are correctly estimated. 
The nuisance parameters are also determined without bias and with correct uncertainties, 
with the exceptions of the two S-wave parameters $|g_\kappa|$ and $\arg(g_\kappa)$. 
These parameters show non-Gaussian behaviour due to the small size of the S-wave contribution, however this does not negatively affect the other parameters. 
The sensitivities to the Wilson coefficients are, as summarised in Tab.~\ref{tab:dir_summary}, $\sigma_{{\rm Re}({\cal C}_7)}=0.0139 \pm 0.0004$, $\sigma_{{\rm Re}({\cal C}_9)}=0.1534 \pm 0.0049$ and $\sigma_{{\rm Re}({\cal C}_{10})}=0.1833 \pm 0.0058$. 

Furthermore, it is instructive to see how well the direct method performs when separating contributions from different Wilson coefficients. 
We therefore perform studies in which we vary two Wilson coefficients simultaneously. 
First, we vary both ${\rm Re}({\cal C}_7)$ and ${\rm Re}({\cal C}_{9})$, while keeping ${\rm Re}({\cal C}_{10})$ fixed. 
The results are given in Tab.~\ref{tab:dir_csevencnine} in App.~\ref{sec:dir_results}. 
The simultaneous fit has a sensitivity of $\sigma_{{\rm Re}({\cal C}_7)}=0.0193 \pm 0.0006$ and $\sigma_{{\rm Re}({\cal C}_{10})}=0.2130 \pm 0.0068$. 
In a second study we allow both ${\rm Re}({\cal C}_{9})$ and ${\rm Re}({\cal C}_{10})$ to vary, while fixing ${\rm Re}({\cal C}_{7})$. 
Tab.~\ref{tab:dir_cninecten} gives expected uncertainties of $\sigma_{{\rm Re}({\cal C}_9)}=0.1715 \pm 0.0054$ and $\sigma_{{\rm Re}({\cal C}_{10})}=0.2054 \pm 0.0065$ in this case. 
In both cases, the pull distributions show that the fit is unbiased and the uncertainties are correctly estimated. 

Besides the Wilson coefficients, it is interesting to also study the expected sensitivities of the fit to the nuisance parameters. 
Theoretical constraints on theory nuisance parameters are included in the fit using Gaussian constraints as discussed in Sec.~\ref{sec:nuisances} and detailed in Tab.~\ref{tab:nuisances}. 
If the expected uncertainty from the fit is significantly smaller than the Gaussian constraint this shows that the data is able to further constrain these parameters. 
This is particularly visible for the form factor parameters. 
Their uncertainties are significantly reduced through the $q^2$-unbinned fit to the data.  
For the subleading corrections we observe a reduction in uncertainty at high $q^2$ compared to the Gaussian constraints. 
We however do not observe a significant reduction of uncertainty at low $q^2$, 
which is due to the fact that the parameterisation of the subleading corrections affects only a part of the decay amplitudes as discussed in Sec.~\ref{sec:nuisances}. 

\begin{table}
\centering
\subfloat[\label{tab:dir_summary}]{
\begin{tabular}{lrrrr} \hline
    \multicolumn{5}{c}{Direct fit method}\\ \hline\hline
    & sensitivity & rel.\ sens.\ [\%] & pull mean & pull width \\ 
\hline\multicolumn{5}{c}{Single Wilson coefficients}\\\hline
${\rm Re}(C_{7})$ & $0.0139 \pm 0.0004$ & $4.14 \pm 0.13$ & $0.03 \pm 0.04$ & $0.98 \pm 0.03$\\
${\rm Re}(C_{9})$ & $0.1534 \pm 0.0049$ & $3.59 \pm 0.11$ & $-0.02 \pm 0.04$ & $0.99 \pm 0.03$\\
${\rm Re}(C_{10})$ & $0.1833 \pm 0.0058$ & $4.39 \pm 0.14$ & $-0.00 \pm 0.05$ & $1.03 \pm 0.03$\\
\hline\multicolumn{5}{c}{Pairs of Wilson coefficients}\\\hline
${\rm Re}(C_{7})$ & $0.0193 \pm 0.0006$ & $5.74 \pm 0.18$ & $0.03 \pm 0.05$ & $1.02 \pm 0.03$\\
${\rm Re}(C_{9})$ & $0.2130 \pm 0.0068$ & $4.98 \pm 0.16$ & $-0.04 \pm 0.05$ & $1.02 \pm 0.03$\\\hline
${\rm Re}(C_{9})$ & $0.1715 \pm 0.0054$ & $4.01 \pm 0.13$ & $-0.01 \pm 0.04$ & $0.97 \pm 0.03$\\
${\rm Re}(C_{10})$ & $0.2054 \pm 0.0065$ & $4.92 \pm 0.16$ & $0.03 \pm 0.05$ & $1.01 \pm 0.03$\\
\hline \end{tabular}}
\vspace{1em}
\subfloat[\label{tab:obs_summary}]{
  \begin{tabular}{lrrrr} \hline
    \multicolumn{5}{c}{$q^2$-binned observables}\\ \hline\hline
    & sensitivity & rel.\ sens.\ [\%] & pull mean & pull width \\ 
\hline\multicolumn{5}{c}{Single Wilson coefficients}\\\hline
${\rm Re}(C_{7})$ & $0.0159 \pm 0.0005$ & $4.73 \pm 0.15$ & $0.08 \pm 0.05$ & $1.04 \pm 0.03$\\
${\rm Re}(C_{9})$ & $0.1610 \pm 0.0051$ & $3.77 \pm 0.12$ & $0.01 \pm 0.05$ & $1.02 \pm 0.03$\\
${\rm Re}(C_{10})$ & $0.2278 \pm 0.0072$ & $5.46 \pm 0.17$ & $-0.01 \pm 0.05$ & $1.06 \pm 0.03$\\
\hline\multicolumn{5}{c}{Pairs of Wilson coefficients}\\\hline
${\rm Re}(C_{7})$ & $0.0252 \pm 0.0008$ & $7.49 \pm 0.24$ & $0.06 \pm 0.05$ & $1.07 \pm 0.03$\\
${\rm Re}(C_{9})$ & $0.2555 \pm 0.0081$ & $5.98 \pm 0.19$ & $-0.04 \pm 0.05$ & $1.05 \pm 0.03$\\\hline
$Re(C_{9})$ & $0.1869 \pm 0.0059$ & $4.37 \pm 0.14$ & $0.03 \pm 0.05$ & $1.02 \pm 0.03$\\
$Re(C_{10})$ & $0.2663 \pm 0.0085$ & $6.38 \pm 0.20$ & $0.00 \pm 0.05$ & $1.06 \pm 0.03$\\
\hline \end{tabular}}
  \caption{Summary of the sensitivity to the Wilson coefficients (absolute and relative to the SM value of the Wilson coefficient) as well as the means and widths of the pull distributions for \protect\subref{tab:dir_summary} the direct fit method and \protect\subref{tab:obs_summary} the conventional $q^2$-binned approach.\label{tab:wilson_comparison}}  
\end{table}

\subsection{Comparison with the $q^2$-binned method}
\label{sec:comparison}
To compare the performance of the direct fit with the conventional $q^2$-binned approach the same 500 pseudoexperiments are split into bins of $q^2$ and maximum likelihood fits similar to Ref.~\cite{LHCb-PAPER-2015-051} are performed to determine the angular observables in bins of $q^2$.  
For the binned angular observables the $P_i$ basis is used, consisting of $F_{\rm L}$, $P_{1,2,3}$ and $P_{4,5,6,8}^\prime$. 
The $S_i$ basis consisting of $F_{\rm L}$, $A_{\rm FB}$ and $S_{3,4,5,7,8,9}$ gives consistent results.
The $q^2$ binning is analogous to the binning used for the likelihood fit in Ref.~\cite{LHCb-PAPER-2015-051}. 
To constrain the S-wave contribution, the $m_{K\pi}$ distribution is used in the angular fit. 

After the angular observables and their correlations are determined in bins of $q^2$, 
a $\chi^2$-minimisation is performed, using the binned angular observables and their correlations as input. 
The EOS software package is used to provide the binned predictions from theory. 
We perform fits of the pseudoexperiment studied using the direct fit method in Sec.~\ref{sec:single} determining pairs of Wilson coefficients. 
The resulting confidence regions are given by the green contours in Fig.~\ref{fig:lhprofiles}, 
indicating that the direct fit method allows a more precise determination of the Wilson coefficients than the $q^2$-binned approach. 

Subsequently, we study the full ensemble of pseudoexperiments and perform fits 
of the single Wilson coefficients ${\rm Re}({\cal C}_7)$, ${\rm Re}({\cal C}_9)$ and ${\rm Re}({\cal C}_{10})$, while all other coefficients are fixed to their SM values. 
Detailed results are given in Tab.~\ref{tab:dir_cseven}-\ref{tab:dir_cten} in App.~\ref{sec:dir_results}. 
The $q^2$-binned fit is shown to be unbiased and the parameter uncertainties are determined correctly. 
The sensitivities of the binned fit of single Wilson coefficients 
to ${\rm Re}({\cal C}_7)$, ${\rm Re}({\cal C}_9)$ and ${\rm Re}({\cal C}_{10})$ are found to be
$\sigma_{{\rm Re}({\cal C}_7)}=0.0159 \pm 0.0005$, $\sigma_{{\rm Re}({\cal C}_9)}=0.1610 \pm 0.0051$ and $\sigma_{{\rm Re}({\cal C}_{10})}=0.2278 \pm 0.0072$, as summarised in Tab.~\ref{tab:obs_summary}. 
The expected uncertainties from the $q^2$-binned approach are therefore significantly larger than from the proposed direct fit method. 

We furthermore perform simultaneous fits of two Wilson coefficients using the $q^2$-binned approach.
The results of a simultaneous fit of ${\rm Re}({\cal C}_7)$ and ${\rm Re}({\cal C}_{9})$ are given in Tab.~\ref{tab:dir_csevencnine} in App.~\ref{sec:dir_results}. 
The sensitivities found for the simultaneous fit of the Wilson coefficients are $\sigma_{{\rm Re}({\cal C}_7)}=0.0252 \pm 0.0008$ and $\sigma_{{\rm Re}({\cal C}_{9})}=0.2555 \pm 0.0081$. 
Results for the simultaneous fit of ${\rm Re}({\cal C}_7)$ and ${\rm Re}({\cal C}_{9})$ are given in Tab.~\ref{tab:dir_cninecten}. 
The expected uncertainties for the Wilson coefficients in this case are $\sigma_{{\rm Re}({\cal C}_9)}=0.1869 \pm 0.0059$ and $\sigma_{{\rm Re}({\cal C}_{10})}=0.2663 \pm 0.0085$. 
As for the $q^2$-unbinned approach, the binned fit method is unbiased and the uncertainties are correctly estimated.
However, the uncertainties are significantly larger than the uncertainties achievable using the proposed direct fit method detailed in Sec.~\ref{sec:wilsons}. 

The binned method also allows to constrain nuisance parameters. 
Comparing the expected uncertainties of the form factor parameters with the Gaussian constraints listed in Tab.~\ref{tab:nuisances} shows a reduction of the uncertainties.
However, the binned approach is less powerful in reducing the uncertainties than the direct fit of Wilson coefficients.

\section{Conclusions} 
\label{sec:conclusions} 
We present a method to determine the Wilson coefficients directly from a $q^2$-unbinned fit of $\decay{\Bd}{\Kstarz\mumu}$ decays. 
This direct fit method uses all available experimental data on the decay, namely $q^2$, the decay angles, the $B^0$ decay flavour, and $m_{K\pi}$ in a more efficient way than the conventional $q^2$-binned approach. 
The method is validated and shown to be unbiased and to correctly determine the parameter uncertainties. 
In comparison with the conventional $q^2$-binned method the direct fit method gives increased sensitivity to the physics parameters of interest, the Wilson coefficients. 
The $q^2$-unbinned direct fit method is particularly useful for the simultaneous determination of pairs of Wilson coefficients where the different $q^2$ dependencies of the contributions can be exploited. 
The statistical uncertainties for the determination of pairs of Wilson coefficients in the direct fit approach are reduced by $8-23\%$ compared to the $q^2$-binned method, corresponding to an increase in signal yield of around $20-70\%$. 
In addition, theory nuisance parameters can be better constrained through the more efficient use of the data. 
An example are the form factor parameters that can be further constrained in the direct fit. 
We note that the parameterisation of non-factorisable contributions (including leading and power corrections) based on analyticity properties which was very recently proposed in Ref.~\cite{Bobeth:2017vxj}
is possible within the direct fit approach and should be explored in future work. 
Furthermore, the direct fit method can also be applied to other $b\to s\ell\ell$ decays like $\decay{\Bs}{\phi\mumu}$. 

In light of the advantages of the direct fit method we would encourage its use in future analyses of the decay $\decay{\Bd}{\Kstarz\mumu}$ at LHCb and Belle~II. 
Publishing background subtracted and efficiency corrected data samples should also be discussed. 
A drawback of the direct fit method is that it depends on the calculation and assumptions for nuisance parameters used in the fit.
It is thus crucial to continue to also determine and publish the $q^2$-binned observables that are independent of theory considerations. 

\section{Acknowledgements} 
C.\,L.\ gratefully acknowledges support by the Emmy Noether programme of the Deutsche Forschungsgemeinschaft (DFG), grant identifier LA 3937/1-1. 

\setboolean{inbibliography}{true}
\bibliographystyle{LHCb}
\bibliography{main}

\clearpage

\appendix

\section{Fit results for a single pseudoexperiment}\label{sec:singleresults}
\noindent\begin{minipage}[t]{\textwidth}
\centering
\scalebox{0.66}{
  \begin{tabular}{lrr}\hline
Parameter & Result & Pull $[\sigma]$ \\\hline\hline
${\rm Re}(C_{7})$ & $-0.345\pm0.019$ & $-0.4$\\
${\rm Re}(C_{9})$ & $4.55\pm0.22$ & $1.3$\\
\hline\multicolumn{3}{c}{CKM parameters}\\\hline
$A_\mathrm{CKM}$ & $0.824\pm0.020$ & $0.8$\\
$\lambda_\mathrm{CKM}$ & $0.22613\pm0.00064$ & $1.2$\\
$\bar{\rho}_\mathrm{CKM}$ & $0.073\pm0.053$ & $-1.0$\\
$\bar{\eta}_\mathrm{CKM}$ & $0.369\pm0.059$ & $-0.1$\\
\hline\multicolumn{3}{c}{Quark masses}\\\hline
$m_{c}$ & $1.282\pm0.024$ & $0.3$\\
$m_{b}$ & $4.185\pm0.029$ & $0.2$\\
\hline\multicolumn{3}{c}{Form factor parameters}\\\hline
$\alpha^{A_{0}}_{0}$ & $0.389\pm0.018$ & $1.1$\\
$\alpha^{A_{0}}_{1}$ & $-1.14\pm0.22$ & $1.1$\\
$\alpha^{A_{0}}_{2}$ & $0.6\pm1.4$ & $0.4$\\
$\alpha^{A_{1}}_{0}$ & $0.309\pm0.017$ & $0.7$\\
$\alpha^{A_{1}}_{1}$ & $0.57\pm0.16$ & $1.2$\\
$\alpha^{A_{1}}_{2}$ & $1.77\pm0.95$ & $0.6$\\
$\alpha^{A_{12}}_{1}$ & $0.725\pm0.100$ & $1.9$\\
$\alpha^{A_{12}}_{2}$ & $1.23\pm0.59$ & $1.3$\\
$\alpha^{V}_{0}$ & $0.384\pm0.022$ & $0.3$\\
$\alpha^{V}_{1}$ & $-0.97\pm0.21$ & $0.9$\\
$\alpha^{V}_{2}$ & $1.1\pm1.3$ & $-1.0$\\
$\alpha^{T_{1}}_{0}$ & $0.334\pm0.019$ & $1.1$\\
$\alpha^{T_{1}}_{1}$ & $-0.77\pm0.15$ & $1.5$\\
$\alpha^{T_{1}}_{2}$ & $0.9\pm1.5$ & $-0.4$\\
$\alpha^{T_{2}}_{1}$ & $0.67\pm0.14$ & $1.2$\\
$\alpha^{T_{2}}_{2}$ & $2.06\pm0.78$ & $0.6$\\
$\alpha^{T_{23}}_{0}$ & $0.737\pm0.056$ & $1.2$\\
$\alpha^{T_{23}}_{1}$ & $1.25\pm0.21$ & $-0.3$\\
$\alpha^{T_{23}}_{2}$ & $1.9\pm2.1$ & $-0.9$\\
\hline\multicolumn{3}{c}{Subleading corrections}\\\hline
${\rm Re}(a_{0}^{sl})$ & $-0.008\pm0.100$ & $-0.1$\\
${\rm Im}(a_{0}^{sl})$ & $-0.095\pm0.100$ & $-1.0$\\
${\rm Re}(b_{0}^{sl})$ & $-0.16\pm0.24$ & $-0.7$\\
${\rm Im}(b_{0}^{sl})$ & $0.32\pm0.24$ & $1.3$\\
${\rm Re}(a_{\perp}^{sl})$ & $-0.088\pm0.097$ & $-0.9$\\
${\rm Im}(a_{\perp}^{sl})$ & $-0.085\pm0.097$ & $-0.9$\\
${\rm Re}(b_{\perp}^{sl})$ & $-0.05\pm0.24$ & $-0.2$\\
${\rm Im}(b_{\perp}^{sl})$ & $0.17\pm0.24$ & $0.7$\\
${\rm Re}(a_{\parallel}^{sl})$ & $0.039\pm0.097$ & $0.4$\\
${\rm Im}(a_{\parallel}^{sl})$ & $-0.096\pm0.095$ & $-1.0$\\
${\rm Re}(b_{\parallel}^{sl})$ & $0.08\pm0.24$ & $0.3$\\
${\rm Im}(b_{\parallel}^{sl})$ & $-0.35\pm0.24$ & $-1.5$\\
${\rm Re}(c_{0}^{sl})$ & $0.007\pm0.059$ & $0.1$\\
${\rm Im}(c_{0}^{sl})$ & $0.046\pm0.059$ & $0.8$\\
${\rm Re}(c_{\perp}^{sl})$ & $0.073\pm0.072$ & $1.0$\\
${\rm Im}(c_{\perp}^{sl})$ & $-0.026\pm0.069$ & $-0.4$\\
${\rm Re}(c_{\parallel}^{sl})$ & $0.077\pm0.061$ & $1.3$\\
${\rm Im}(c_{\parallel}^{sl})$ & $0.040\pm0.061$ & $0.7$\\
\hline\multicolumn{3}{c}{S-wave parameters}\\\hline
$S(\xi_{\parallel})$ & $0.951\pm0.090$ & $-0.5$\\
$\delta_{S}$ & $3.31\pm0.18$ & $0.9$\\
$|g_{\kappa}|$ & $0.19\pm0.17$ & $0.5$\\
$arg(g_{\kappa})$ & $1.09\pm0.64$ & $-0.8$\\
\hline\end{tabular}}
\hspace{0.1\linewidth}
\scalebox{0.66}{
\begin{tabular}{lrr}\hline
Parameter & Result & Pull $[\sigma]$ \\\hline\hline
${\rm Re}(C_{9})$ & $4.45\pm0.18$ & $1.0$\\
${\rm Re}(C_{10})$ & $-4.08\pm0.19$ & $0.5$\\
\hline\multicolumn{3}{c}{CKM parameters}\\\hline
$A_\mathrm{CKM}$ & $0.824\pm0.020$ & $0.8$\\
$\lambda_\mathrm{CKM}$ & $0.22613\pm0.00064$ & $1.2$\\
$\bar{\rho}_\mathrm{CKM}$ & $0.073\pm0.053$ & $-1.0$\\
$\bar{\eta}_\mathrm{CKM}$ & $0.369\pm0.059$ & $-0.1$\\
\hline\multicolumn{3}{c}{Quark masses}\\\hline
$m_{c}$ & $1.283\pm0.024$ & $0.3$\\
$m_{b}$ & $4.185\pm0.029$ & $0.2$\\
\hline\multicolumn{3}{c}{Form factor parameters}\\\hline
$\alpha^{A_{0}}_{0}$ & $0.389\pm0.018$ & $1.1$\\
$\alpha^{A_{0}}_{1}$ & $-1.14\pm0.22$ & $1.1$\\
$\alpha^{A_{0}}_{2}$ & $0.6\pm1.4$ & $0.4$\\
$\alpha^{A_{1}}_{0}$ & $0.309\pm0.017$ & $0.7$\\
$\alpha^{A_{1}}_{1}$ & $0.57\pm0.16$ & $1.2$\\
$\alpha^{A_{1}}_{2}$ & $1.77\pm0.95$ & $0.6$\\
$\alpha^{A_{12}}_{1}$ & $0.726\pm0.100$ & $1.9$\\
$\alpha^{A_{12}}_{2}$ & $1.23\pm0.59$ & $1.3$\\
$\alpha^{V}_{0}$ & $0.384\pm0.022$ & $0.3$\\
$\alpha^{V}_{1}$ & $-0.98\pm0.21$ & $0.9$\\
$\alpha^{V}_{2}$ & $1.1\pm1.3$ & $-1.0$\\
$\alpha^{T_{1}}_{0}$ & $0.334\pm0.019$ & $1.1$\\
$\alpha^{T_{1}}_{1}$ & $-0.77\pm0.15$ & $1.5$\\
$\alpha^{T_{1}}_{2}$ & $0.9\pm1.5$ & $-0.4$\\
$\alpha^{T_{2}}_{1}$ & $0.66\pm0.14$ & $1.2$\\
$\alpha^{T_{2}}_{2}$ & $2.06\pm0.78$ & $0.6$\\
$\alpha^{T_{23}}_{0}$ & $0.737\pm0.056$ & $1.2$\\
$\alpha^{T_{23}}_{1}$ & $1.25\pm0.21$ & $-0.3$\\
$\alpha^{T_{23}}_{2}$ & $1.9\pm2.1$ & $-0.9$\\
\hline\multicolumn{3}{c}{Subleading corrections}\\\hline
${\rm Re}(a_{0}^{sl})$ & $-0.009\pm0.100$ & $-0.1$\\
${\rm Im}(a_{0}^{sl})$ & $-0.095\pm0.100$ & $-1.0$\\
${\rm Re}(b_{0}^{sl})$ & $-0.16\pm0.24$ & $-0.7$\\
${\rm Im}(b_{0}^{sl})$ & $0.32\pm0.24$ & $1.3$\\
${\rm Re}(a_{\perp}^{sl})$ & $-0.088\pm0.097$ & $-0.9$\\
${\rm Im}(a_{\perp}^{sl})$ & $-0.085\pm0.097$ & $-0.9$\\
${\rm Re}(b_{\perp}^{sl})$ & $-0.05\pm0.24$ & $-0.2$\\
${\rm Im}(b_{\perp}^{sl})$ & $0.17\pm0.24$ & $0.7$\\
${\rm Re}(a_{\parallel}^{sl})$ & $0.038\pm0.097$ & $0.4$\\
${\rm Im}(a_{\parallel}^{sl})$ & $-0.096\pm0.094$ & $-1.0$\\
${\rm Re}(b_{\parallel}^{sl})$ & $0.08\pm0.24$ & $0.3$\\
${\rm Im}(b_{\parallel}^{sl})$ & $-0.35\pm0.24$ & $-1.5$\\
${\rm Re}(c_{0}^{sl})$ & $0.007\pm0.059$ & $0.1$\\
${\rm Im}(c_{0}^{sl})$ & $0.046\pm0.059$ & $0.8$\\
${\rm Re}(c_{\perp}^{sl})$ & $0.073\pm0.072$ & $1.0$\\
${\rm Im}(c_{\perp}^{sl})$ & $-0.026\pm0.069$ & $-0.4$\\
${\rm Re}(c_{\parallel}^{sl})$ & $0.077\pm0.061$ & $1.3$\\
${\rm Im}(c_{\parallel}^{sl})$ & $0.040\pm0.061$ & $0.7$\\
\hline\multicolumn{3}{c}{S-wave parameters}\\\hline
$S(\xi_{\parallel})$ & $0.950\pm0.090$ & $-0.6$\\
$\delta_{S}$ & $3.31\pm0.18$ & $0.9$\\
$|g_{\kappa}|$ & $0.19\pm0.17$ & $0.5$\\
$arg(g_{\kappa})$ & $1.08\pm0.64$ & $-0.8$\\
\hline\end{tabular}}
\captionof{table}{Results from the direct fit method for a single pseudoexperiment varying (left) the Wilson coefficients ${\rm Re}({\cal C}_7)$ and ${\rm Re}({\cal C}_9)$ and (right) the Wilson coefficients ${\rm Re}({\cal C}_9)$ and ${\rm Re}({\cal C}_{10})$. Background parameters are omitted for brevity.\label{tab:dir_fit}}
\end{minipage}

\section{Projections of the probability density function}\label{sec:projections}
\noindent\begin{minipage}{\textwidth}
\centering
\includegraphics[width=6.9cm]{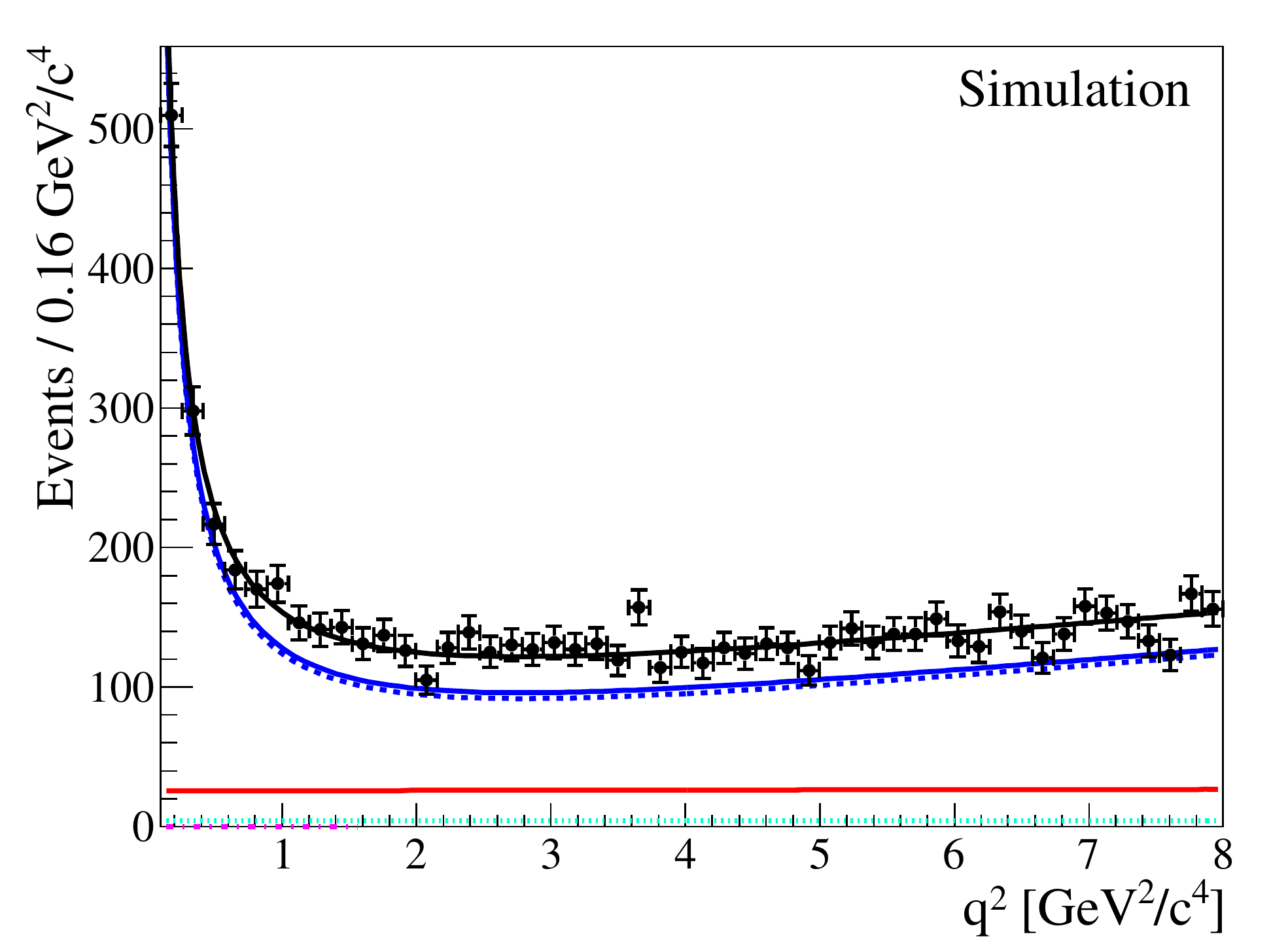}
\includegraphics[width=6.9cm]{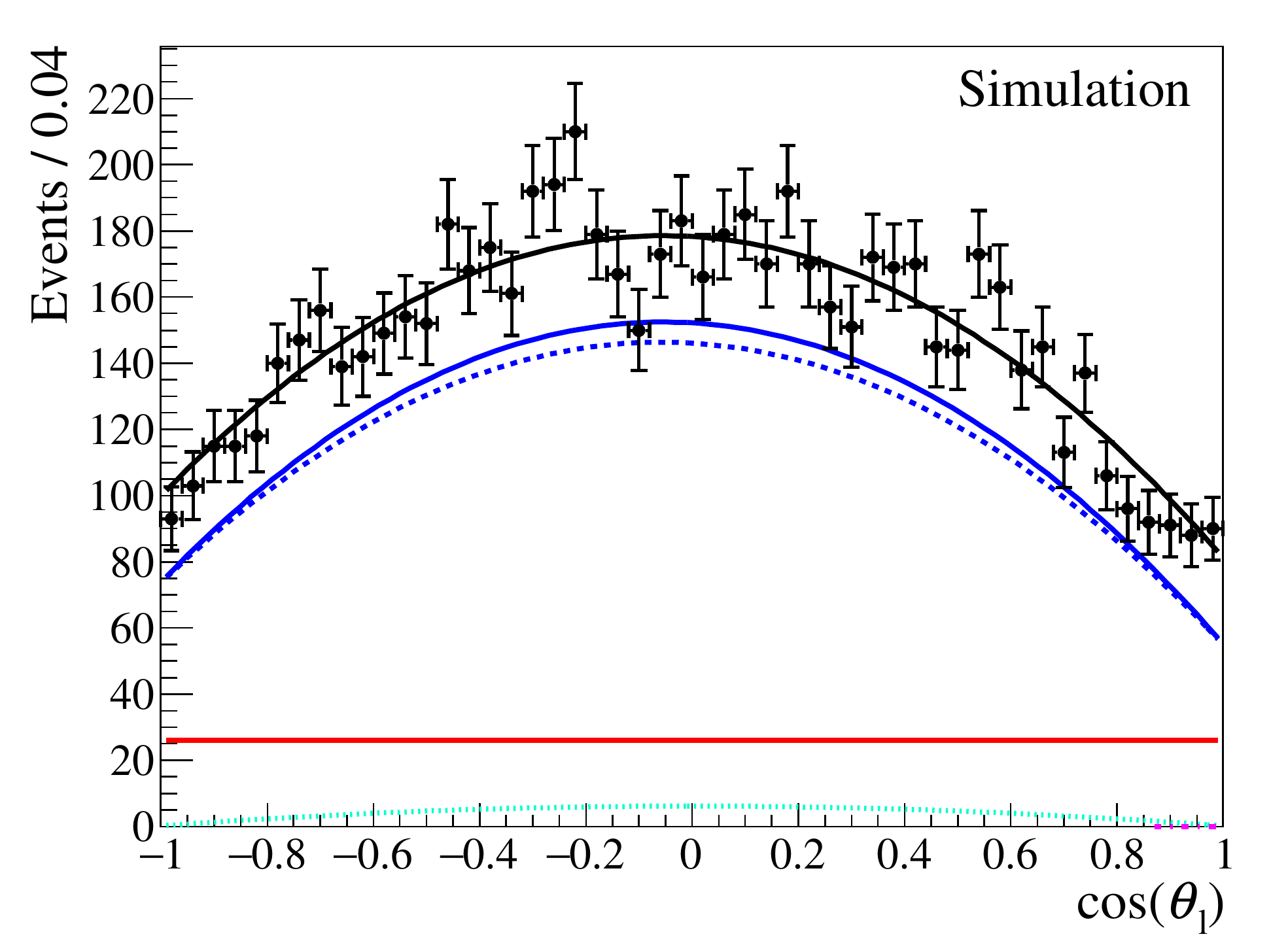}\\
\includegraphics[width=6.9cm]{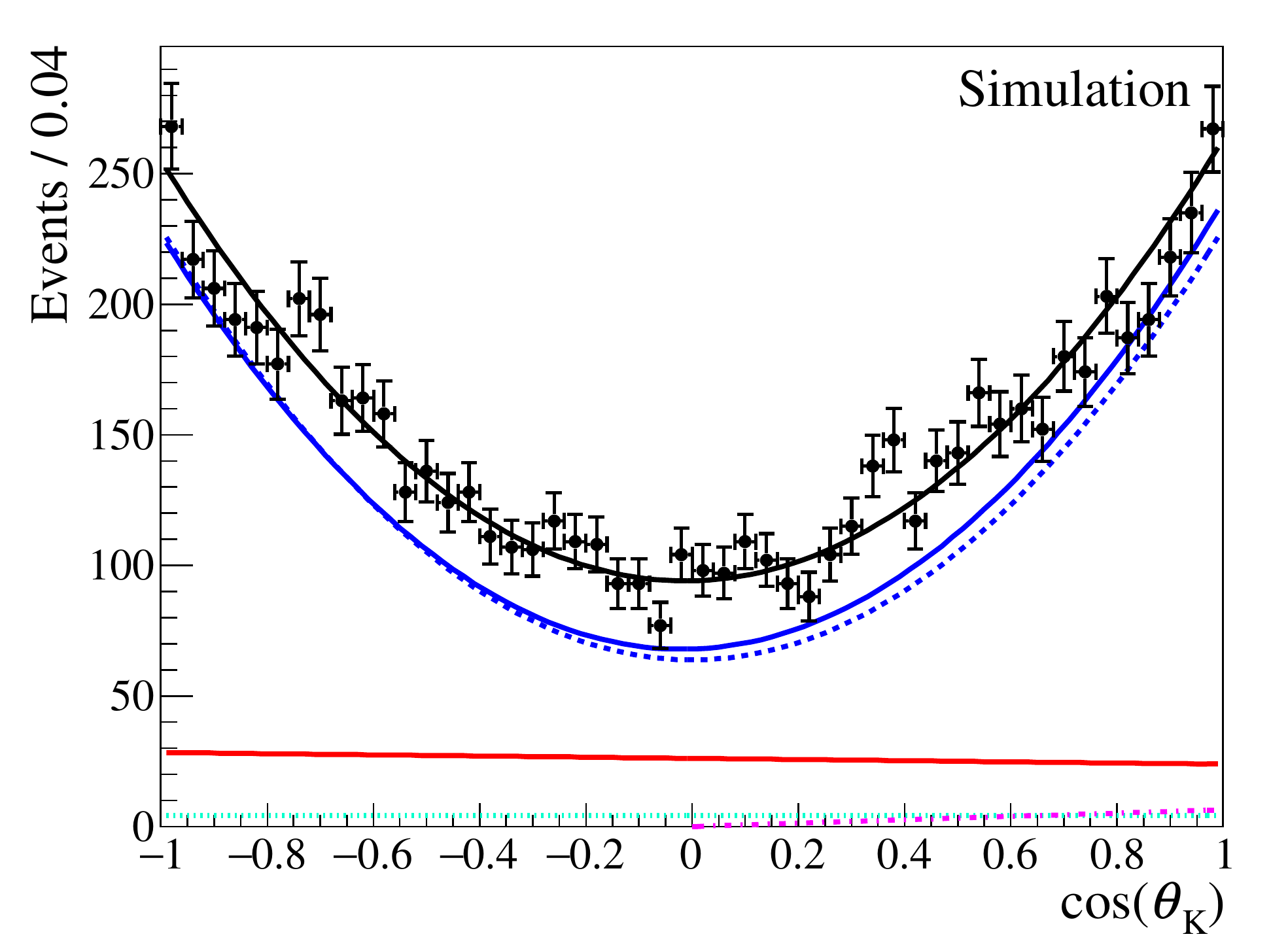}
\includegraphics[width=6.9cm]{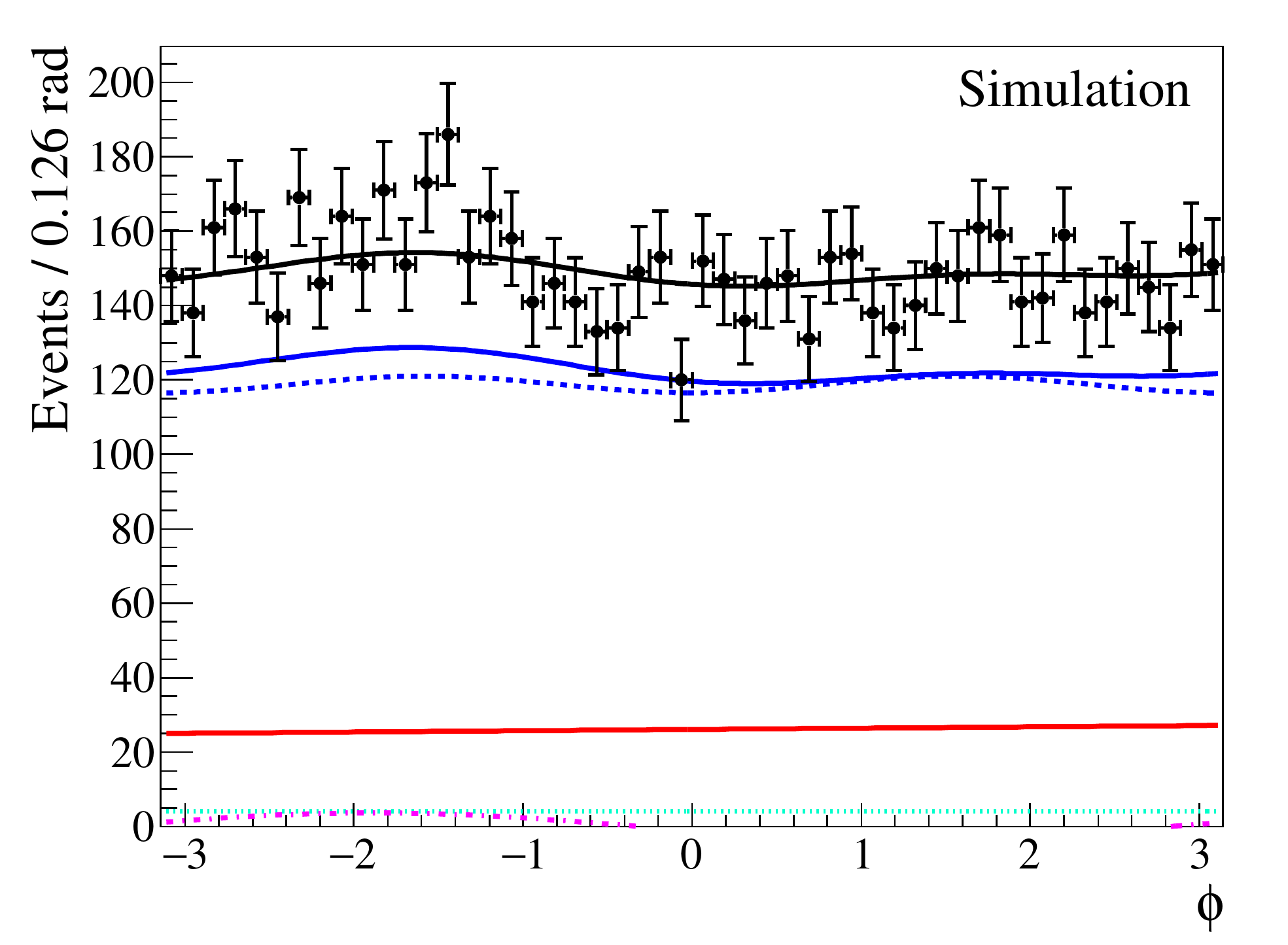}\\
\includegraphics[width=6.9cm]{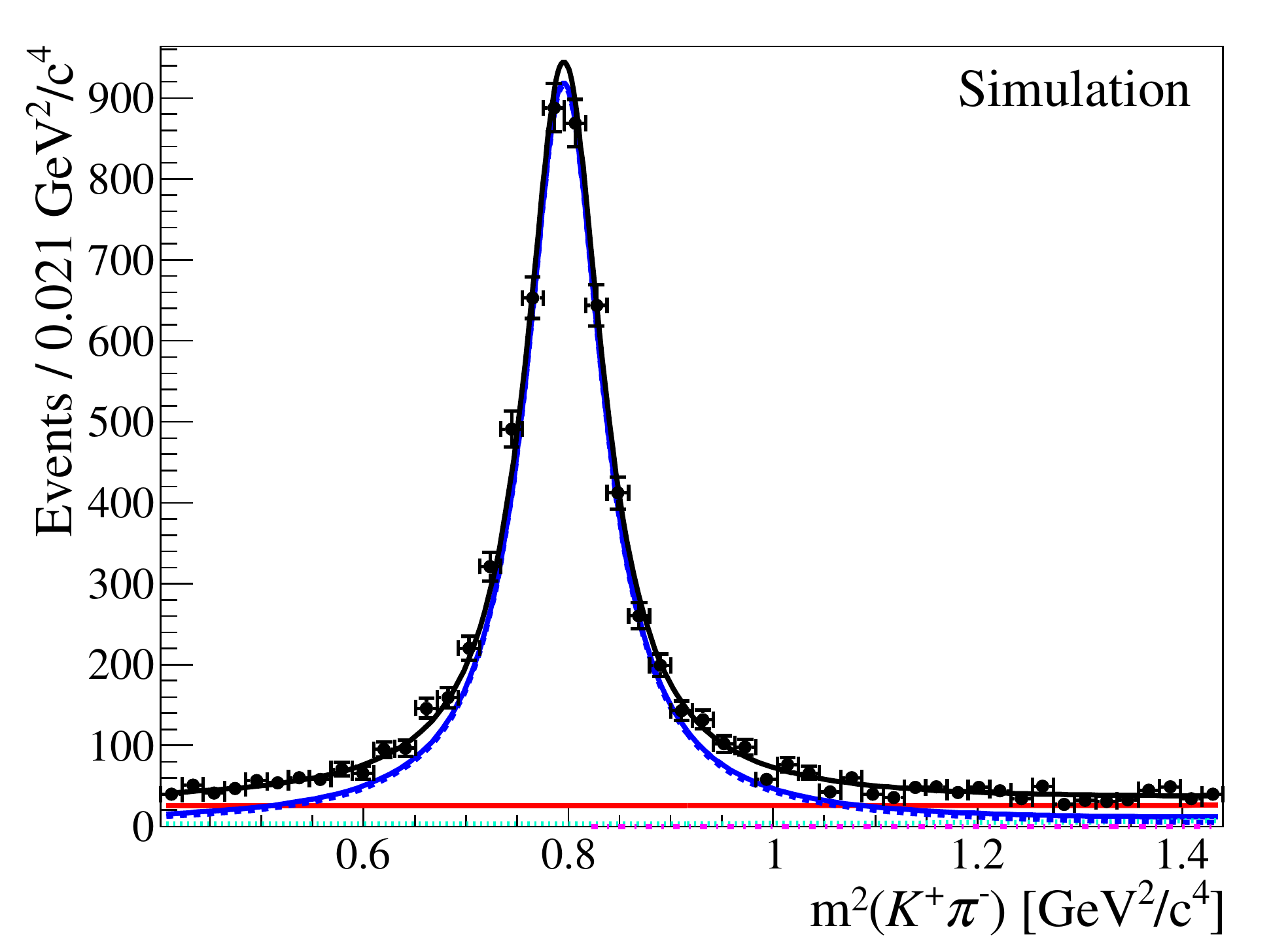}
\captionof{figure}{
  Results from the fit of a single pseudoexperiment varying the Wilson coefficients ${\rm Re}({\cal C}_9)$ and ${\rm Re}({\cal C}_{10})$. 
  Simulated events are overlaid with projections of the fitted PDF on $q^2$, the three decay angles $\cos\thetal$, $\cos\thetak$ and $\phi$ and $m_{K\pi}^2$ in the $q^2$ range $0.1<q^2<8.0\gevgevcccc$. 
  The simulated events and projections are shown for the signal region $\pm 50\mevcc$ around the $\Bd$ mass to enhance the signal fraction. 
  The black solid line denotes the full PDF, the blue solid line the signal component. 
  The blue dashed line gives the P-wave and the teal dotted line the S-wave part. 
  The magenta dash-dotted line finally gives the P-wave/S-wave interference and the red line the background contribution. 
  \label{fig:fit_c9c10_projections_lowq2}}
\end{minipage}

\clearpage

\begin{figure}
  \centering
\includegraphics[width=6.9cm]{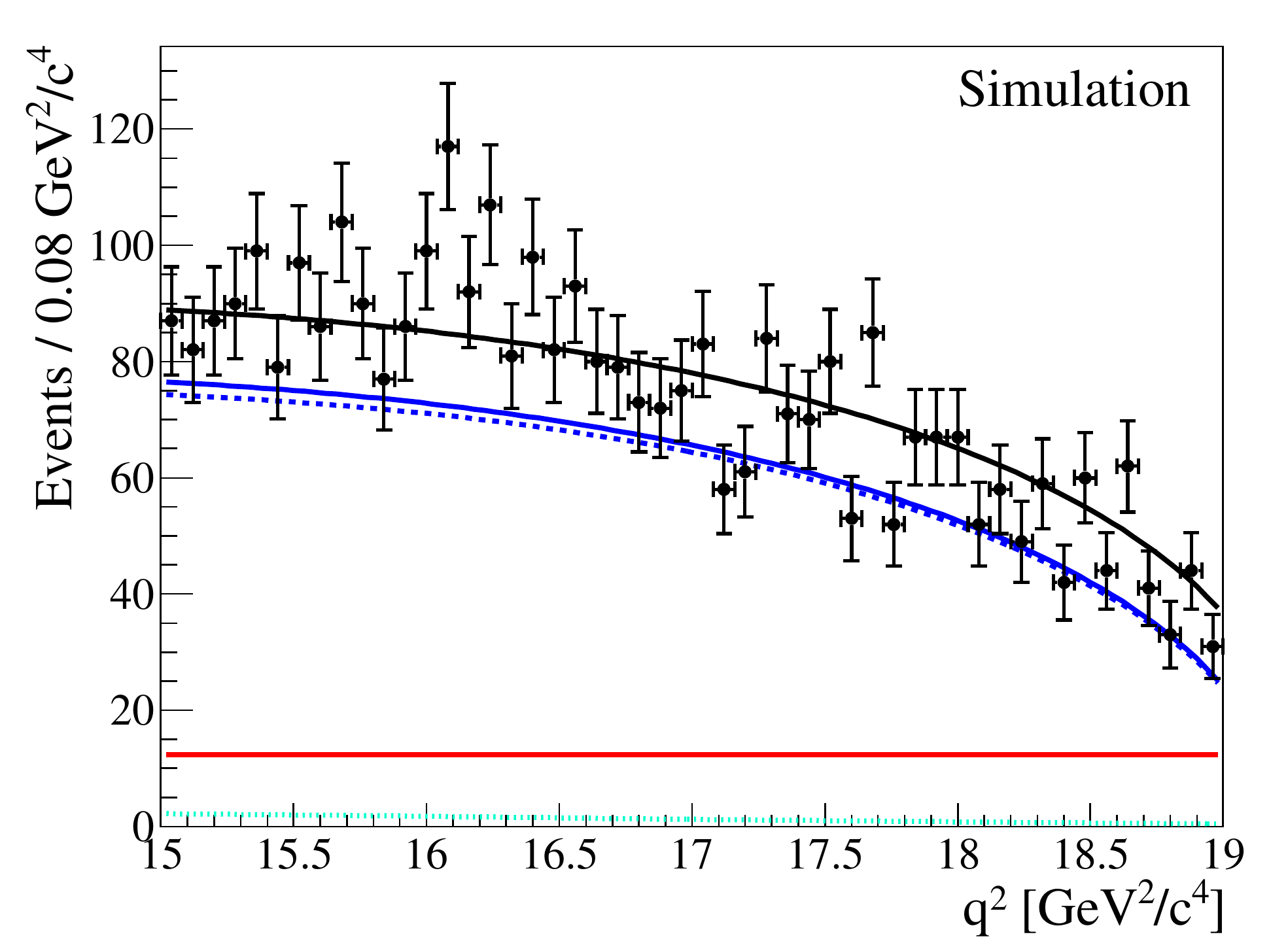}
\includegraphics[width=6.9cm]{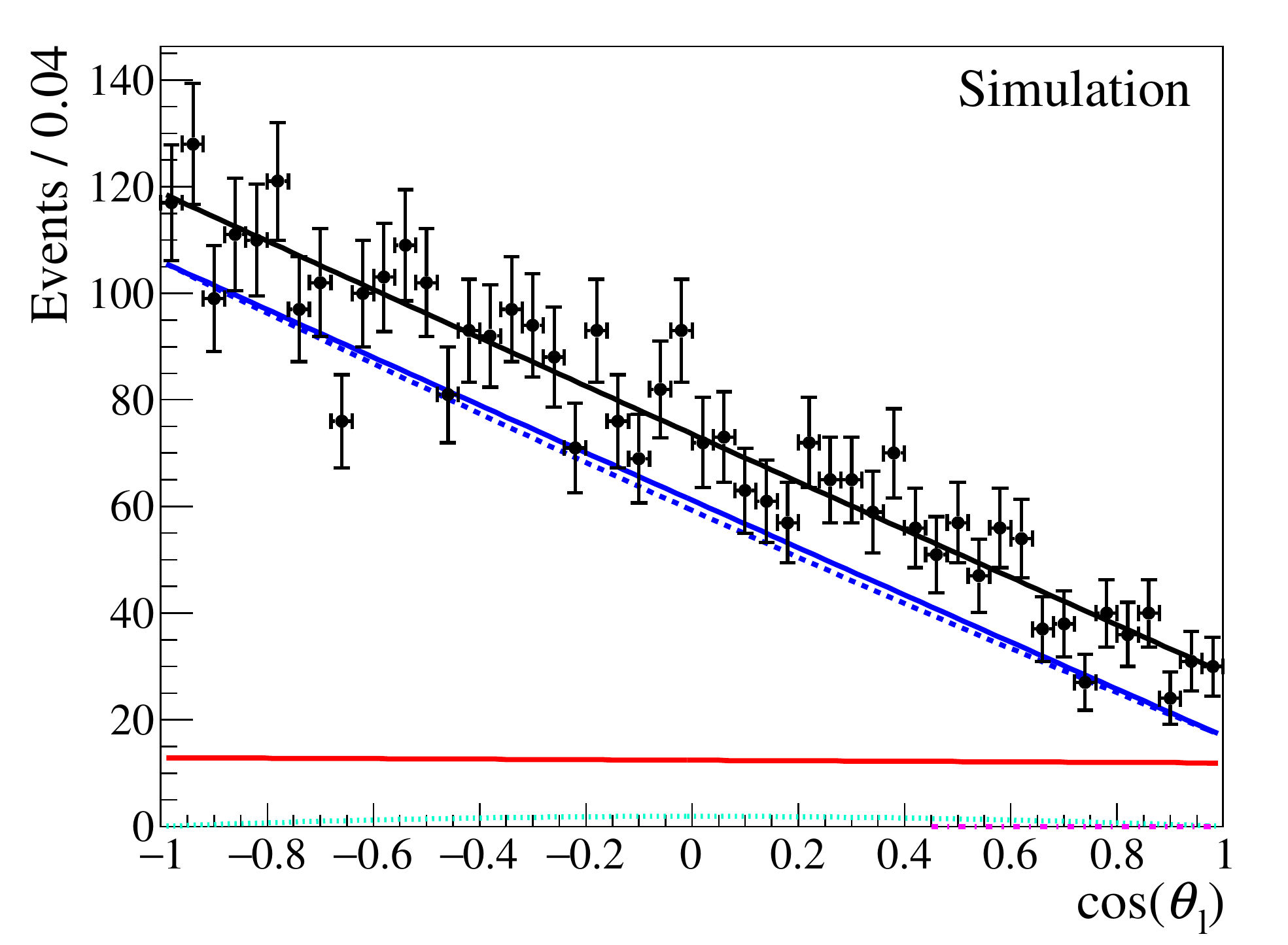}\\
\includegraphics[width=6.9cm]{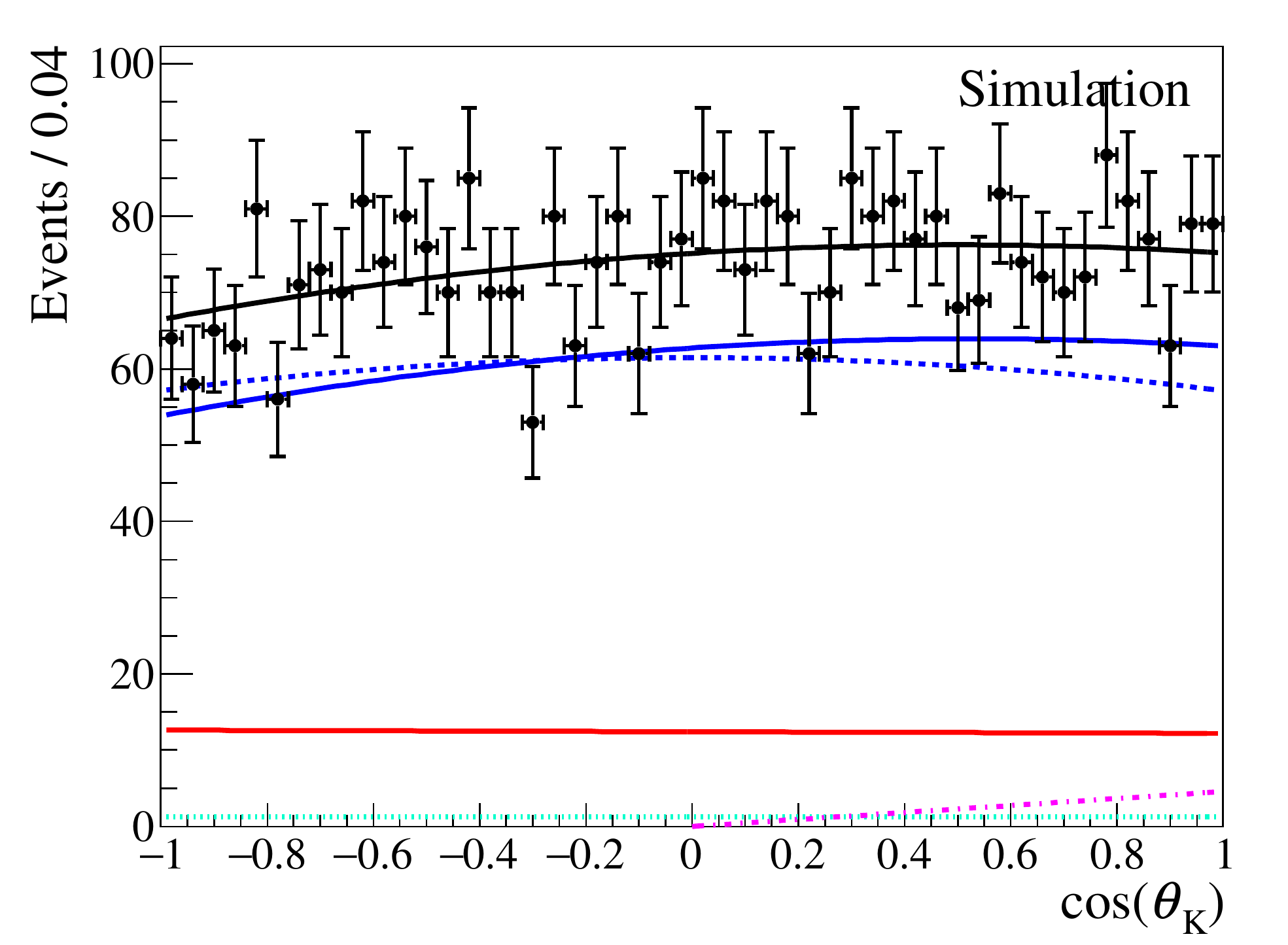}
\includegraphics[width=6.9cm]{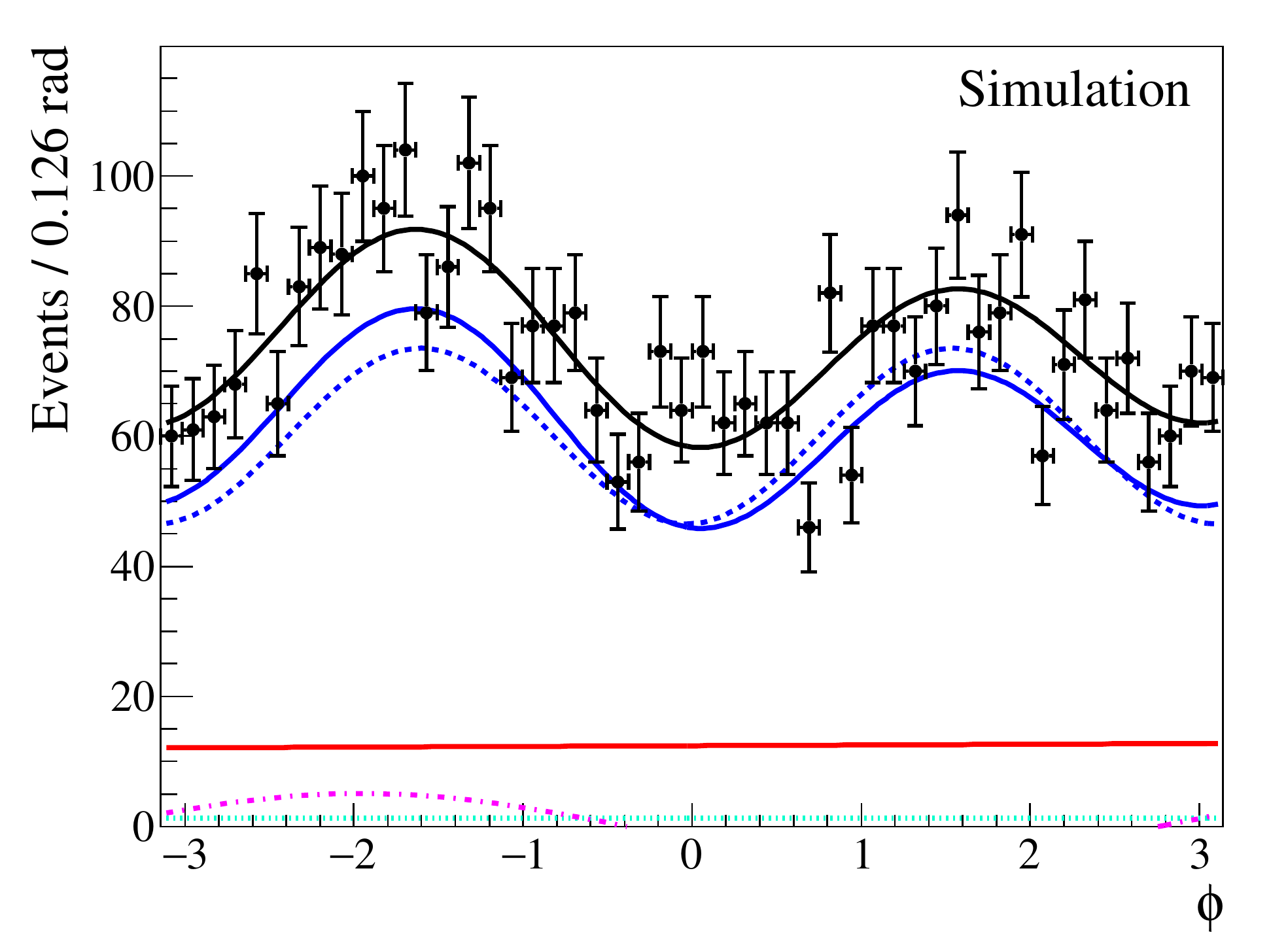}\\
\includegraphics[width=6.9cm]{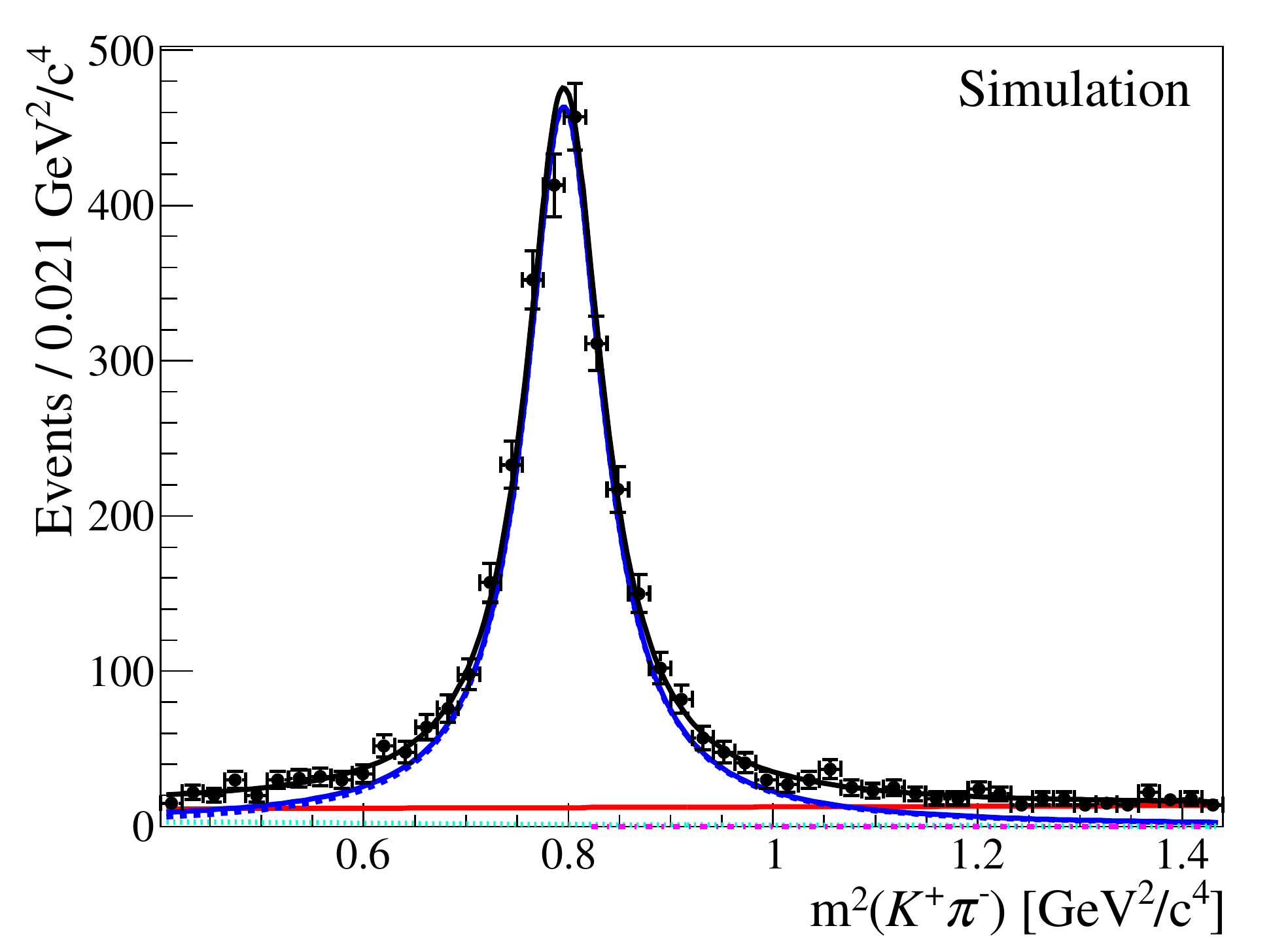}
\caption{
  Results from the fit of a single pseudoexperiment varying the Wilson coefficients ${\rm Re}({\cal C}_9)$ and ${\rm Re}({\cal C}_{10})$. 
  Simulated events are overlaid with projections of the fitted PDF on $q^2$, the three decay angles $\cos\thetal$, $\cos\thetak$ and $\phi$ and $m_{K\pi}^2$ in the $q^2$ range $15.0<q^2<19.0\gevgevcccc$. 
  The simulated events and projections are shown for the signal region $\pm 50\mevcc$ around the $\Bd$ mass to enhance the signal fraction. 
  The black solid line denotes the full PDF, the blue solid line the signal component. 
  The blue dashed line gives the P-wave and the teal dotted line the S-wave part. 
  The magenta dash-dotted line finally gives the P-wave/S-wave interference and the red line the background contribution.
  We note again that the $q^2$ distribution in the high $q^2$ region is only shown for illustration, as it is not used in the direct fit method. 
  \label{fig:fit_c9c10_projections_highq2}}
\end{figure}

\clearpage

\section{Detailed comparison of the direct fit method with the $q^2$-binned approach}\label{sec:dir_results}
\noindent\begin{minipage}[t]{\textwidth}
\centering
\scalebox{0.66}{
  \begin{tabular}[t]{lrrr} \hline
 & sensitivity & pull mean & pull width \\ \hline\hline
${\rm Re}(C_{7})$ & $0.0139 \pm 0.0004$ & $0.03 \pm 0.04$ & $0.98 \pm 0.03$\\
\hline\multicolumn{4}{c}{CKM parameters}\\\hline
$A_\mathrm{CKM}$ & $0.0206 \pm 0.0007$ & $-0.00 \pm 0.05$ & $1.01 \pm 0.03$\\
$\lambda_\mathrm{CKM}$ & $0.0007 \pm 0.0000$ & $-0.09 \pm 0.05$ & $1.03 \pm 0.03$\\
$\bar{\rho}_\mathrm{CKM}$ & $0.0529 \pm 0.0017$ & $0.04 \pm 0.04$ & $0.99 \pm 0.03$\\
$\bar{\eta}_\mathrm{CKM}$ & $0.0589 \pm 0.0019$ & $0.00 \pm 0.04$ & $0.98 \pm 0.03$\\
\hline\multicolumn{4}{c}{Quark masses}\\\hline
$m_{c}$ & $0.0235 \pm 0.0007$ & $-0.04 \pm 0.04$ & $0.98 \pm 0.03$\\
$m_{b}$ & $0.0298 \pm 0.0009$ & $-0.04 \pm 0.05$ & $1.02 \pm 0.03$\\
\hline\multicolumn{4}{c}{Form factor parameters}\\\hline
$\alpha^{A_{0}}_{0}$ & $0.0179 \pm 0.0006$ & $-0.03 \pm 0.05$ & $1.04 \pm 0.03$\\
$\alpha^{A_{0}}_{1}$ & $0.2115 \pm 0.0067$ & $-0.01 \pm 0.04$ & $0.99 \pm 0.03$\\
$\alpha^{A_{0}}_{2}$ & $1.3459 \pm 0.0428$ & $-0.05 \pm 0.04$ & $0.95 \pm 0.03$\\
$\alpha^{A_{1}}_{0}$ & $0.0172 \pm 0.0005$ & $-0.04 \pm 0.05$ & $1.01 \pm 0.03$\\
$\alpha^{A_{1}}_{1}$ & $0.1499 \pm 0.0048$ & $-0.05 \pm 0.04$ & $0.97 \pm 0.03$\\
$\alpha^{A_{1}}_{2}$ & $0.9163 \pm 0.0291$ & $-0.05 \pm 0.04$ & $0.96 \pm 0.03$\\
$\alpha^{A_{12}}_{1}$ & $0.0976 \pm 0.0031$ & $-0.02 \pm 0.04$ & $0.99 \pm 0.03$\\
$\alpha^{A_{12}}_{2}$ & $0.5869 \pm 0.0187$ & $-0.02 \pm 0.04$ & $1.00 \pm 0.03$\\
$\alpha^{V}_{0}$ & $0.0222 \pm 0.0007$ & $-0.03 \pm 0.04$ & $1.00 \pm 0.03$\\
$\alpha^{V}_{1}$ & $0.2130 \pm 0.0068$ & $-0.03 \pm 0.05$ & $1.02 \pm 0.03$\\
$\alpha^{V}_{2}$ & $1.3687 \pm 0.0435$ & $-0.02 \pm 0.05$ & $1.01 \pm 0.03$\\
$\alpha^{T_{1}}_{0}$ & $0.0194 \pm 0.0006$ & $-0.02 \pm 0.05$ & $1.00 \pm 0.03$\\
$\alpha^{T_{1}}_{1}$ & $0.1585 \pm 0.0050$ & $-0.03 \pm 0.05$ & $1.05 \pm 0.03$\\
$\alpha^{T_{1}}_{2}$ & $1.4405 \pm 0.0458$ & $-0.00 \pm 0.04$ & $0.93 \pm 0.03$\\
$\alpha^{T_{2}}_{1}$ & $0.1357 \pm 0.0043$ & $-0.03 \pm 0.04$ & $1.00 \pm 0.03$\\
$\alpha^{T_{2}}_{2}$ & $0.7237 \pm 0.0230$ & $-0.02 \pm 0.04$ & $0.93 \pm 0.03$\\
$\alpha^{T_{23}}_{0}$ & $0.0570 \pm 0.0018$ & $-0.02 \pm 0.05$ & $1.02 \pm 0.03$\\
$\alpha^{T_{23}}_{1}$ & $0.1905 \pm 0.0061$ & $-0.05 \pm 0.04$ & $0.93 \pm 0.03$\\
$\alpha^{T_{23}}_{2}$ & $1.9632 \pm 0.0624$ & $-0.00 \pm 0.04$ & $0.96 \pm 0.03$\\
\hline\multicolumn{4}{c}{Subleading corrections}\\\hline
${\rm Re}(a_{0}^{sl})$ & $0.0982 \pm 0.0031$ & $-0.01 \pm 0.04$ & $0.99 \pm 0.03$\\
${\rm Im}(a_{0}^{sl})$ & $0.1040 \pm 0.0033$ & $0.04 \pm 0.05$ & $1.04 \pm 0.03$\\
${\rm Re}(b_{0}^{sl})$ & $0.2407 \pm 0.0077$ & $-0.02 \pm 0.05$ & $1.01 \pm 0.03$\\
${\rm Im}(b_{0}^{sl})$ & $0.2465 \pm 0.0078$ & $-0.08 \pm 0.05$ & $1.02 \pm 0.03$\\
${\rm Re}(a_{\perp}^{sl})$ & $0.0983 \pm 0.0031$ & $0.01 \pm 0.05$ & $1.02 \pm 0.03$\\
${\rm Im}(a_{\perp}^{sl})$ & $0.0906 \pm 0.0029$ & $-0.01 \pm 0.04$ & $0.94 \pm 0.03$\\
${\rm Re}(b_{\perp}^{sl})$ & $0.2485 \pm 0.0079$ & $-0.06 \pm 0.05$ & $1.06 \pm 0.03$\\
${\rm Im}(b_{\perp}^{sl})$ & $0.2499 \pm 0.0079$ & $0.04 \pm 0.05$ & $1.05 \pm 0.03$\\
${\rm Re}(a_{\parallel}^{sl})$ & $0.0939 \pm 0.0030$ & $-0.01 \pm 0.04$ & $0.99 \pm 0.03$\\
${\rm Im}(a_{\parallel}^{sl})$ & $0.0943 \pm 0.0030$ & $0.06 \pm 0.05$ & $1.01 \pm 0.03$\\
${\rm Re}(b_{\parallel}^{sl})$ & $0.2309 \pm 0.0073$ & $0.04 \pm 0.04$ & $0.99 \pm 0.03$\\
${\rm Im}(b_{\parallel}^{sl})$ & $0.2434 \pm 0.0077$ & $0.11 \pm 0.05$ & $1.04 \pm 0.03$\\
${\rm Re}(c_{0}^{sl})$ & $0.0590 \pm 0.0019$ & $-0.06 \pm 0.04$ & $0.97 \pm 0.03$\\
${\rm Im}(c_{0}^{sl})$ & $0.0581 \pm 0.0018$ & $0.00 \pm 0.04$ & $0.97 \pm 0.03$\\
${\rm Re}(c_{\perp}^{sl})$ & $0.0692 \pm 0.0022$ & $-0.05 \pm 0.05$ & $1.01 \pm 0.03$\\
${\rm Im}(c_{\perp}^{sl})$ & $0.0659 \pm 0.0021$ & $0.09 \pm 0.04$ & $0.99 \pm 0.03$\\
${\rm Re}(c_{\parallel}^{sl})$ & $0.0569 \pm 0.0018$ & $-0.03 \pm 0.04$ & $0.97 \pm 0.03$\\
${\rm Im}(c_{\parallel}^{sl})$ & $0.0566 \pm 0.0018$ & $0.01 \pm 0.04$ & $0.96 \pm 0.03$\\
\hline\multicolumn{4}{c}{S-wave parameters}\\\hline
$S(\xi_{\parallel})$ & $0.0861 \pm 0.0027$ & $0.05 \pm 0.05$ & $1.04 \pm 0.03$\\
$\delta_S$ & $0.1776 \pm 0.0056$ & $0.10 \pm 0.05$ & $1.13 \pm 0.04$\\
$|g_{\kappa}|$ & $0.0527 \pm 0.0017$ & $0.27 \pm 0.04$ & $0.91 \pm 0.03$\\
${\rm arg}(g_{\kappa})$ & $0.8666 \pm 0.0275$ & $0.01 \pm 0.07$ & $1.64 \pm 0.05$\\
\hline \end{tabular}}
\hspace*{0.25cm}
\scalebox{0.66}{
  \begin{tabular}[t]{lrrr} \hline
    & sensitivity & pull mean & pull width \\ \hline\hline
${\rm Re}(C_{7})$ & $0.0159 \pm 0.0005$ & $0.08 \pm 0.05$ & $1.04 \pm 0.03$\\
\hline\multicolumn{4}{c}{CKM parameters}\\\hline
$A_\mathrm{CKM}$ & $0.0205 \pm 0.0007$ & $-0.00 \pm 0.05$ & $1.00 \pm 0.03$\\
$\lambda_\mathrm{CKM}$ & $0.0007 \pm 0.0000$ & $-0.09 \pm 0.05$ & $1.03 \pm 0.03$\\
$\bar{\rho}_\mathrm{CKM}$ & $0.0532 \pm 0.0017$ & $0.04 \pm 0.04$ & $0.99 \pm 0.03$\\
$\bar{\eta}_\mathrm{CKM}$ & $0.0589 \pm 0.0019$ & $0.00 \pm 0.04$ & $0.98 \pm 0.03$\\
\hline\multicolumn{4}{c}{Quark masses}\\\hline
$m_{c}$ & $0.0237 \pm 0.0008$ & $-0.04 \pm 0.04$ & $0.98 \pm 0.03$\\
$m_{b}$ & $0.0298 \pm 0.0009$ & $-0.05 \pm 0.05$ & $1.02 \pm 0.03$\\
\hline\multicolumn{4}{c}{Form factor parameters}\\\hline
$\alpha^{A_{0}}_{0}$ & $0.0197 \pm 0.0006$ & $-0.14 \pm 0.05$ & $1.06 \pm 0.03$\\
$\alpha^{A_{0}}_{1}$ & $0.2407 \pm 0.0076$ & $-0.04 \pm 0.05$ & $1.02 \pm 0.03$\\
$\alpha^{A_{0}}_{2}$ & $1.4745 \pm 0.0469$ & $0.01 \pm 0.05$ & $1.01 \pm 0.03$\\
$\alpha^{A_{1}}_{0}$ & $0.0188 \pm 0.0006$ & $0.07 \pm 0.05$ & $1.04 \pm 0.03$\\
$\alpha^{A_{1}}_{1}$ & $0.1636 \pm 0.0052$ & $0.02 \pm 0.05$ & $1.00 \pm 0.03$\\
$\alpha^{A_{1}}_{2}$ & $0.9523 \pm 0.0303$ & $-0.03 \pm 0.04$ & $0.98 \pm 0.03$\\
$\alpha^{A_{12}}_{1}$ & $0.1148 \pm 0.0037$ & $-0.06 \pm 0.05$ & $1.00 \pm 0.03$\\
$\alpha^{A_{12}}_{2}$ & $0.6432 \pm 0.0205$ & $-0.00 \pm 0.05$ & $1.01 \pm 0.03$\\
$\alpha^{V}_{0}$ & $0.0241 \pm 0.0008$ & $0.09 \pm 0.05$ & $1.01 \pm 0.03$\\
$\alpha^{V}_{1}$ & $0.2385 \pm 0.0076$ & $0.05 \pm 0.05$ & $1.05 \pm 0.03$\\
$\alpha^{V}_{2}$ & $1.3483 \pm 0.0429$ & $-0.06 \pm 0.04$ & $0.99 \pm 0.03$\\
$\alpha^{T_{1}}_{0}$ & $0.0209 \pm 0.0007$ & $0.10 \pm 0.05$ & $1.03 \pm 0.03$\\
$\alpha^{T_{1}}_{1}$ & $0.1735 \pm 0.0055$ & $0.06 \pm 0.05$ & $1.08 \pm 0.03$\\
$\alpha^{T_{1}}_{2}$ & $1.4726 \pm 0.0468$ & $-0.03 \pm 0.04$ & $0.94 \pm 0.03$\\
$\alpha^{T_{2}}_{1}$ & $0.1484 \pm 0.0047$ & $0.04 \pm 0.05$ & $1.02 \pm 0.03$\\
$\alpha^{T_{2}}_{2}$ & $0.7400 \pm 0.0235$ & $-0.03 \pm 0.04$ & $0.94 \pm 0.03$\\
$\alpha^{T_{23}}_{0}$ & $0.0615 \pm 0.0020$ & $-0.00 \pm 0.05$ & $1.06 \pm 0.03$\\
$\alpha^{T_{23}}_{1}$ & $0.2085 \pm 0.0066$ & $-0.03 \pm 0.04$ & $0.96 \pm 0.03$\\
$\alpha^{T_{23}}_{2}$ & $2.0483 \pm 0.0652$ & $-0.01 \pm 0.04$ & $0.99 \pm 0.03$\\
\hline\multicolumn{4}{c}{Subleading corrections}\\\hline
${\rm Re}(a_{0}^{sl})$ & $0.1030 \pm 0.0033$ & $-0.04 \pm 0.05$ & $1.03 \pm 0.03$\\
${\rm Im}(a_{0}^{sl})$ & $0.0986 \pm 0.0031$ & $-0.01 \pm 0.04$ & $0.99 \pm 0.03$\\
${\rm Re}(b_{0}^{sl})$ & $0.2605 \pm 0.0083$ & $0.07 \pm 0.05$ & $1.07 \pm 0.03$\\
${\rm Im}(b_{0}^{sl})$ & $0.2419 \pm 0.0077$ & $-0.03 \pm 0.04$ & $1.00 \pm 0.03$\\
${\rm Re}(a_{\perp}^{sl})$ & $0.0997 \pm 0.0032$ & $-0.06 \pm 0.05$ & $1.03 \pm 0.03$\\
${\rm Im}(a_{\perp}^{sl})$ & $0.0952 \pm 0.0030$ & $-0.00 \pm 0.04$ & $0.99 \pm 0.03$\\
${\rm Re}(b_{\perp}^{sl})$ & $0.2260 \pm 0.0072$ & $0.01 \pm 0.04$ & $0.95 \pm 0.03$\\
${\rm Im}(b_{\perp}^{sl})$ & $0.2553 \pm 0.0081$ & $-0.05 \pm 0.05$ & $1.07 \pm 0.03$\\
${\rm Re}(a_{\parallel}^{sl})$ & $0.0994 \pm 0.0032$ & $0.03 \pm 0.05$ & $1.04 \pm 0.03$\\
${\rm Im}(a_{\parallel}^{sl})$ & $0.0936 \pm 0.0030$ & $-0.02 \pm 0.05$ & $1.00 \pm 0.03$\\
${\rm Re}(b_{\parallel}^{sl})$ & $0.2373 \pm 0.0075$ & $0.08 \pm 0.05$ & $1.01 \pm 0.03$\\
${\rm Im}(b_{\parallel}^{sl})$ & $0.2309 \pm 0.0073$ & $0.03 \pm 0.04$ & $0.98 \pm 0.03$\\
${\rm Re}(c_{0}^{sl})$ & $0.0654 \pm 0.0021$ & $0.11 \pm 0.05$ & $1.00 \pm 0.03$\\
${\rm Im}(c_{0}^{sl})$ & $0.0611 \pm 0.0019$ & $-0.04 \pm 0.04$ & $0.96 \pm 0.03$\\
${\rm Re}(c_{\perp}^{sl})$ & $0.0716 \pm 0.0023$ & $0.05 \pm 0.05$ & $1.01 \pm 0.03$\\
${\rm Im}(c_{\perp}^{sl})$ & $0.0668 \pm 0.0021$ & $0.04 \pm 0.04$ & $0.97 \pm 0.03$\\
${\rm Re}(c_{\parallel}^{sl})$ & $0.0638 \pm 0.0020$ & $0.05 \pm 0.05$ & $1.02 \pm 0.03$\\
${\rm Im}(c_{\parallel}^{sl})$ & $0.0630 \pm 0.0020$ & $-0.05 \pm 0.05$ & $1.02 \pm 0.03$\\
\hline \end{tabular}}
\captionof{table}{
  Results from pseudoexperiments determining ${\rm Re}({\cal C}_7)$ using (left) the proposed direct fit method and (right) the $q^2$-binned approach.\label{tab:dir_cseven}
}  
\end{minipage}

\clearpage

\begin{table}
  \begin{minipage}[t]{\textwidth}
  \centering
\scalebox{0.7}{
  \begin{tabular}[t]{lrrr} \hline
 & sensitivity & pull mean & pull width \\ \hline\hline
${\rm Re}(C_{9})$ & $0.1534 \pm 0.0049$ & $-0.02 \pm 0.04$ & $0.99 \pm 0.03$\\
\hline\multicolumn{4}{c}{CKM parameters}\\\hline
$A_\mathrm{CKM}$ & $0.0206 \pm 0.0007$ & $-0.01 \pm 0.05$ & $1.01 \pm 0.03$\\
$\lambda_\mathrm{CKM}$ & $0.0007 \pm 0.0000$ & $-0.09 \pm 0.05$ & $1.03 \pm 0.03$\\
$\bar{\rho}_\mathrm{CKM}$ & $0.0531 \pm 0.0017$ & $0.04 \pm 0.04$ & $0.99 \pm 0.03$\\
$\bar{\eta}_\mathrm{CKM}$ & $0.0587 \pm 0.0019$ & $-0.00 \pm 0.04$ & $0.97 \pm 0.03$\\
\hline\multicolumn{4}{c}{Quark masses}\\\hline
$m_{c}$ & $0.0235 \pm 0.0007$ & $-0.04 \pm 0.04$ & $0.97 \pm 0.03$\\
$m_{b}$ & $0.0296 \pm 0.0009$ & $-0.04 \pm 0.05$ & $1.01 \pm 0.03$\\
\hline\multicolumn{4}{c}{Form factor parameters}\\\hline
$\alpha^{A_{0}}_{0}$ & $0.0179 \pm 0.0006$ & $-0.04 \pm 0.05$ & $1.04 \pm 0.03$\\
$\alpha^{A_{0}}_{1}$ & $0.2130 \pm 0.0068$ & $-0.02 \pm 0.04$ & $1.00 \pm 0.03$\\
$\alpha^{A_{0}}_{2}$ & $1.3281 \pm 0.0422$ & $-0.04 \pm 0.04$ & $0.95 \pm 0.03$\\
$\alpha^{A_{1}}_{0}$ & $0.0171 \pm 0.0005$ & $-0.05 \pm 0.05$ & $1.01 \pm 0.03$\\
$\alpha^{A_{1}}_{1}$ & $0.1471 \pm 0.0047$ & $-0.04 \pm 0.04$ & $0.96 \pm 0.03$\\
$\alpha^{A_{1}}_{2}$ & $0.9074 \pm 0.0288$ & $-0.04 \pm 0.04$ & $0.96 \pm 0.03$\\
$\alpha^{A_{12}}_{1}$ & $0.0971 \pm 0.0031$ & $-0.01 \pm 0.04$ & $0.99 \pm 0.03$\\
$\alpha^{A_{12}}_{2}$ & $0.5825 \pm 0.0185$ & $-0.01 \pm 0.04$ & $0.99 \pm 0.03$\\
$\alpha^{V}_{0}$ & $0.0221 \pm 0.0007$ & $-0.04 \pm 0.05$ & $1.00 \pm 0.03$\\
$\alpha^{V}_{1}$ & $0.2050 \pm 0.0065$ & $-0.02 \pm 0.05$ & $1.01 \pm 0.03$\\
$\alpha^{V}_{2}$ & $1.3732 \pm 0.0436$ & $-0.01 \pm 0.05$ & $1.01 \pm 0.03$\\
$\alpha^{T_{1}}_{0}$ & $0.0181 \pm 0.0006$ & $-0.04 \pm 0.05$ & $1.03 \pm 0.03$\\
$\alpha^{T_{1}}_{1}$ & $0.1482 \pm 0.0047$ & $-0.02 \pm 0.05$ & $1.03 \pm 0.03$\\
$\alpha^{T_{1}}_{2}$ & $1.4485 \pm 0.0460$ & $0.01 \pm 0.04$ & $0.94 \pm 0.03$\\
$\alpha^{T_{2}}_{1}$ & $0.1295 \pm 0.0041$ & $-0.03 \pm 0.04$ & $0.99 \pm 0.03$\\
$\alpha^{T_{2}}_{2}$ & $0.7279 \pm 0.0231$ & $0.00 \pm 0.04$ & $0.94 \pm 0.03$\\
$\alpha^{T_{23}}_{0}$ & $0.0566 \pm 0.0018$ & $-0.02 \pm 0.05$ & $1.02 \pm 0.03$\\
$\alpha^{T_{23}}_{1}$ & $0.1927 \pm 0.0061$ & $-0.05 \pm 0.04$ & $0.94 \pm 0.03$\\
$\alpha^{T_{23}}_{2}$ & $1.9581 \pm 0.0622$ & $0.01 \pm 0.04$ & $0.95 \pm 0.03$\\
\hline\multicolumn{4}{c}{Subleading corrections}\\\hline
${\rm Re}(a_{0}^{sl})$ & $0.0977 \pm 0.0031$ & $-0.01 \pm 0.04$ & $0.98 \pm 0.03$\\
${\rm Im}(a_{0}^{sl})$ & $0.1044 \pm 0.0033$ & $0.04 \pm 0.05$ & $1.05 \pm 0.03$\\
${\rm Re}(b_{0}^{sl})$ & $0.2404 \pm 0.0076$ & $-0.03 \pm 0.05$ & $1.01 \pm 0.03$\\
${\rm Im}(b_{0}^{sl})$ & $0.2475 \pm 0.0079$ & $-0.08 \pm 0.05$ & $1.02 \pm 0.03$\\
${\rm Re}(a_{\perp}^{sl})$ & $0.0963 \pm 0.0031$ & $0.01 \pm 0.05$ & $1.02 \pm 0.03$\\
${\rm Im}(a_{\perp}^{sl})$ & $0.0900 \pm 0.0029$ & $-0.00 \pm 0.04$ & $0.94 \pm 0.03$\\
${\rm Re}(b_{\perp}^{sl})$ & $0.2530 \pm 0.0080$ & $-0.07 \pm 0.05$ & $1.06 \pm 0.03$\\
${\rm Im}(b_{\perp}^{sl})$ & $0.2553 \pm 0.0081$ & $0.04 \pm 0.05$ & $1.06 \pm 0.03$\\
${\rm Re}(a_{\parallel}^{sl})$ & $0.0921 \pm 0.0029$ & $-0.01 \pm 0.04$ & $0.99 \pm 0.03$\\
${\rm Im}(a_{\parallel}^{sl})$ & $0.0935 \pm 0.0030$ & $0.07 \pm 0.05$ & $1.01 \pm 0.03$\\
${\rm Re}(b_{\parallel}^{sl})$ & $0.2317 \pm 0.0074$ & $0.04 \pm 0.04$ & $0.99 \pm 0.03$\\
${\rm Im}(b_{\parallel}^{sl})$ & $0.2438 \pm 0.0077$ & $0.10 \pm 0.05$ & $1.04 \pm 0.03$\\
${\rm Re}(c_{0}^{sl})$ & $0.0588 \pm 0.0019$ & $-0.06 \pm 0.04$ & $0.97 \pm 0.03$\\
${\rm Im}(c_{0}^{sl})$ & $0.0582 \pm 0.0018$ & $-0.00 \pm 0.04$ & $0.97 \pm 0.03$\\
${\rm Re}(c_{\perp}^{sl})$ & $0.0693 \pm 0.0022$ & $-0.05 \pm 0.05$ & $1.01 \pm 0.03$\\
${\rm Im}(c_{\perp}^{sl})$ & $0.0656 \pm 0.0021$ & $0.09 \pm 0.04$ & $0.98 \pm 0.03$\\
${\rm Re}(c_{\parallel}^{sl})$ & $0.0566 \pm 0.0018$ & $-0.05 \pm 0.04$ & $0.97 \pm 0.03$\\
${\rm Im}(c_{\parallel}^{sl})$ & $0.0566 \pm 0.0018$ & $0.00 \pm 0.04$ & $0.96 \pm 0.03$\\
\hline\multicolumn{4}{c}{S-wave parameters}\\\hline
$S(\xi_{\parallel})$ & $0.0864 \pm 0.0027$ & $0.04 \pm 0.05$ & $1.05 \pm 0.03$\\
$\delta_S$ & $0.1779 \pm 0.0057$ & $0.09 \pm 0.05$ & $1.12 \pm 0.04$\\
$|g_{\kappa}|$ & $0.0525 \pm 0.0017$ & $0.28 \pm 0.04$ & $0.91 \pm 0.03$\\
${\rm arg}(g_{\kappa})$ & $0.8531 \pm 0.0271$ & $0.09 \pm 0.09$ & $2.02 \pm 0.06$\\
\hline \end{tabular}}
\hspace*{0.25cm}
\scalebox{0.7}{
\begin{tabular}[t]{lrrr} \hline
 & sensitivity & pull mean & pull width \\ \hline\hline
${\rm Re}(C_{9})$ & $0.1610 \pm 0.0051$ & $0.01 \pm 0.05$ & $1.02 \pm 0.03$\\
\hline\multicolumn{4}{c}{CKM parameters}\\\hline
$A_\mathrm{CKM}$ & $0.0205 \pm 0.0007$ & $-0.02 \pm 0.05$ & $1.01 \pm 0.03$\\
$\lambda_\mathrm{CKM}$ & $0.0007 \pm 0.0000$ & $-0.08 \pm 0.05$ & $1.03 \pm 0.03$\\
$\bar{\rho}_\mathrm{CKM}$ & $0.0528 \pm 0.0017$ & $0.04 \pm 0.04$ & $0.98 \pm 0.03$\\
$\bar{\eta}_\mathrm{CKM}$ & $0.0589 \pm 0.0019$ & $0.01 \pm 0.04$ & $0.98 \pm 0.03$\\
\hline\multicolumn{4}{c}{Quark masses}\\\hline
$m_{c}$ & $0.0234 \pm 0.0007$ & $-0.02 \pm 0.04$ & $0.96 \pm 0.03$\\
$m_{b}$ & $0.0298 \pm 0.0010$ & $-0.04 \pm 0.05$ & $1.01 \pm 0.03$\\
\hline\multicolumn{4}{c}{Form factor parameters}\\\hline
$\alpha^{A_{0}}_{0}$ & $0.0197 \pm 0.0006$ & $-0.14 \pm 0.05$ & $1.06 \pm 0.03$\\
$\alpha^{A_{0}}_{1}$ & $0.2400 \pm 0.0076$ & $-0.03 \pm 0.05$ & $1.02 \pm 0.03$\\
$\alpha^{A_{0}}_{2}$ & $1.4643 \pm 0.0467$ & $0.00 \pm 0.05$ & $1.00 \pm 0.03$\\
$\alpha^{A_{1}}_{0}$ & $0.0188 \pm 0.0006$ & $0.07 \pm 0.05$ & $1.04 \pm 0.03$\\
$\alpha^{A_{1}}_{1}$ & $0.1624 \pm 0.0052$ & $0.02 \pm 0.04$ & $0.99 \pm 0.03$\\
$\alpha^{A_{1}}_{2}$ & $0.9478 \pm 0.0302$ & $-0.04 \pm 0.04$ & $0.98 \pm 0.03$\\
$\alpha^{A_{12}}_{1}$ & $0.1145 \pm 0.0037$ & $-0.06 \pm 0.05$ & $1.00 \pm 0.03$\\
$\alpha^{A_{12}}_{2}$ & $0.6453 \pm 0.0206$ & $-0.00 \pm 0.05$ & $1.02 \pm 0.03$\\
$\alpha^{V}_{0}$ & $0.0240 \pm 0.0008$ & $0.09 \pm 0.05$ & $1.01 \pm 0.03$\\
$\alpha^{V}_{1}$ & $0.2371 \pm 0.0076$ & $0.04 \pm 0.05$ & $1.05 \pm 0.03$\\
$\alpha^{V}_{2}$ & $1.3559 \pm 0.0432$ & $-0.07 \pm 0.04$ & $0.99 \pm 0.03$\\
$\alpha^{T_{1}}_{0}$ & $0.0202 \pm 0.0006$ & $0.08 \pm 0.05$ & $1.02 \pm 0.03$\\
$\alpha^{T_{1}}_{1}$ & $0.1698 \pm 0.0054$ & $0.05 \pm 0.05$ & $1.06 \pm 0.03$\\
$\alpha^{T_{1}}_{2}$ & $1.4814 \pm 0.0472$ & $-0.03 \pm 0.04$ & $0.95 \pm 0.03$\\
$\alpha^{T_{2}}_{1}$ & $0.1464 \pm 0.0047$ & $0.03 \pm 0.05$ & $1.01 \pm 0.03$\\
$\alpha^{T_{2}}_{2}$ & $0.7476 \pm 0.0238$ & $-0.04 \pm 0.04$ & $0.95 \pm 0.03$\\
$\alpha^{T_{23}}_{0}$ & $0.0614 \pm 0.0020$ & $0.00 \pm 0.05$ & $1.06 \pm 0.03$\\
$\alpha^{T_{23}}_{1}$ & $0.2078 \pm 0.0066$ & $-0.03 \pm 0.04$ & $0.96 \pm 0.03$\\
$\alpha^{T_{23}}_{2}$ & $2.0454 \pm 0.0652$ & $-0.01 \pm 0.04$ & $0.99 \pm 0.03$\\
\hline\multicolumn{4}{c}{Subleading corrections}\\\hline
${\rm Re}(a_{0}^{sl})$ & $0.1034 \pm 0.0033$ & $-0.03 \pm 0.05$ & $1.04 \pm 0.03$\\
${\rm Im}(a_{0}^{sl})$ & $0.0982 \pm 0.0031$ & $-0.01 \pm 0.04$ & $0.99 \pm 0.03$\\
${\rm Re}(b_{0}^{sl})$ & $0.2600 \pm 0.0083$ & $0.07 \pm 0.05$ & $1.06 \pm 0.03$\\
${\rm Im}(b_{0}^{sl})$ & $0.2411 \pm 0.0077$ & $-0.03 \pm 0.04$ & $0.99 \pm 0.03$\\
${\rm Re}(a_{\perp}^{sl})$ & $0.0989 \pm 0.0032$ & $-0.08 \pm 0.05$ & $1.04 \pm 0.03$\\
${\rm Im}(a_{\perp}^{sl})$ & $0.0953 \pm 0.0030$ & $0.00 \pm 0.04$ & $0.99 \pm 0.03$\\
${\rm Re}(b_{\perp}^{sl})$ & $0.2284 \pm 0.0073$ & $0.02 \pm 0.04$ & $0.95 \pm 0.03$\\
${\rm Im}(b_{\perp}^{sl})$ & $0.2565 \pm 0.0082$ & $-0.05 \pm 0.05$ & $1.06 \pm 0.03$\\
${\rm Re}(a_{\parallel}^{sl})$ & $0.0990 \pm 0.0032$ & $0.03 \pm 0.05$ & $1.05 \pm 0.03$\\
${\rm Im}(a_{\parallel}^{sl})$ & $0.0932 \pm 0.0030$ & $-0.01 \pm 0.05$ & $1.00 \pm 0.03$\\
${\rm Re}(b_{\parallel}^{sl})$ & $0.2358 \pm 0.0075$ & $0.09 \pm 0.05$ & $1.00 \pm 0.03$\\
${\rm Im}(b_{\parallel}^{sl})$ & $0.2310 \pm 0.0074$ & $0.03 \pm 0.04$ & $0.98 \pm 0.03$\\
${\rm Re}(c_{0}^{sl})$ & $0.0655 \pm 0.0021$ & $0.11 \pm 0.05$ & $1.01 \pm 0.03$\\
${\rm Im}(c_{0}^{sl})$ & $0.0611 \pm 0.0019$ & $-0.05 \pm 0.04$ & $0.96 \pm 0.03$\\
${\rm Re}(c_{\perp}^{sl})$ & $0.0720 \pm 0.0023$ & $0.06 \pm 0.05$ & $1.01 \pm 0.03$\\
${\rm Im}(c_{\perp}^{sl})$ & $0.0668 \pm 0.0021$ & $0.04 \pm 0.04$ & $0.97 \pm 0.03$\\
${\rm Re}(c_{\parallel}^{sl})$ & $0.0636 \pm 0.0020$ & $0.04 \pm 0.05$ & $1.01 \pm 0.03$\\
${\rm Im}(c_{\parallel}^{sl})$ & $0.0630 \pm 0.0020$ & $-0.05 \pm 0.05$ & $1.02 \pm 0.03$\\
\hline \end{tabular}
}
\end{minipage}
  \caption{Results from pseudoexperiments determining ${\rm Re}({\cal C}_9)$ using (left) the proposed direct fit method and (right) the $q^2$-binned approach.\label{tab:dir_cnine}
  }  
\end{table}

\begin{table}
  \begin{minipage}[t]{\textwidth}
  \centering
\scalebox{0.7}{
  \begin{tabular}[t]{lrrr} \hline
 & sensitivity & pull mean & pull width \\ \hline\hline
${\rm Re}(C_{10})$ & $0.1833 \pm 0.0058$ & $-0.00 \pm 0.05$ & $1.03 \pm 0.03$\\
  \hline\multicolumn{4}{c}{CKM parameters}\\\hline
$A_\mathrm{CKM}$ & $0.0206 \pm 0.0007$ & $-0.01 \pm 0.05$ & $1.01 \pm 0.03$\\
$\lambda_\mathrm{CKM}$ & $0.0007 \pm 0.0000$ & $-0.09 \pm 0.05$ & $1.04 \pm 0.03$\\
$\bar{\rho}_\mathrm{CKM}$ & $0.0530 \pm 0.0017$ & $0.05 \pm 0.04$ & $0.99 \pm 0.03$\\
$\bar{\eta}_\mathrm{CKM}$ & $0.0588 \pm 0.0019$ & $-0.00 \pm 0.04$ & $0.98 \pm 0.03$\\
\hline\multicolumn{4}{c}{Quark masses}\\\hline
$m_{c}$ & $0.0237 \pm 0.0008$ & $-0.04 \pm 0.04$ & $0.98 \pm 0.03$\\
$m_{b}$ & $0.0296 \pm 0.0009$ & $-0.04 \pm 0.05$ & $1.01 \pm 0.03$\\
\hline\multicolumn{4}{c}{Form factor parameters}\\\hline
$\alpha^{A_{0}}_{0}$ & $0.0178 \pm 0.0006$ & $-0.05 \pm 0.05$ & $1.03 \pm 0.03$\\
$\alpha^{A_{0}}_{1}$ & $0.2117 \pm 0.0067$ & $-0.02 \pm 0.04$ & $0.99 \pm 0.03$\\
$\alpha^{A_{0}}_{2}$ & $1.3447 \pm 0.0427$ & $-0.04 \pm 0.04$ & $0.95 \pm 0.03$\\
$\alpha^{A_{1}}_{0}$ & $0.0170 \pm 0.0005$ & $-0.06 \pm 0.05$ & $1.01 \pm 0.03$\\
$\alpha^{A_{1}}_{1}$ & $0.1497 \pm 0.0048$ & $-0.05 \pm 0.04$ & $0.97 \pm 0.03$\\
$\alpha^{A_{1}}_{2}$ & $0.9260 \pm 0.0294$ & $-0.04 \pm 0.04$ & $0.97 \pm 0.03$\\
$\alpha^{A_{12}}_{1}$ & $0.0973 \pm 0.0031$ & $-0.02 \pm 0.04$ & $0.99 \pm 0.03$\\
$\alpha^{A_{12}}_{2}$ & $0.5855 \pm 0.0186$ & $-0.02 \pm 0.04$ & $1.00 \pm 0.03$\\
$\alpha^{V}_{0}$ & $0.0223 \pm 0.0007$ & $-0.04 \pm 0.05$ & $1.01 \pm 0.03$\\
$\alpha^{V}_{1}$ & $0.2095 \pm 0.0067$ & $-0.04 \pm 0.05$ & $1.02 \pm 0.03$\\
$\alpha^{V}_{2}$ & $1.3687 \pm 0.0435$ & $-0.02 \pm 0.05$ & $1.01 \pm 0.03$\\
$\alpha^{T_{1}}_{0}$ & $0.0180 \pm 0.0006$ & $-0.04 \pm 0.05$ & $1.04 \pm 0.03$\\
$\alpha^{T_{1}}_{1}$ & $0.1529 \pm 0.0049$ & $-0.03 \pm 0.05$ & $1.04 \pm 0.03$\\
$\alpha^{T_{1}}_{2}$ & $1.4414 \pm 0.0458$ & $-0.00 \pm 0.04$ & $0.94 \pm 0.03$\\
$\alpha^{T_{2}}_{1}$ & $0.1340 \pm 0.0043$ & $-0.04 \pm 0.05$ & $1.01 \pm 0.03$\\
$\alpha^{T_{2}}_{2}$ & $0.7302 \pm 0.0232$ & $-0.01 \pm 0.04$ & $0.94 \pm 0.03$\\
$\alpha^{T_{23}}_{0}$ & $0.0571 \pm 0.0018$ & $-0.03 \pm 0.05$ & $1.02 \pm 0.03$\\
$\alpha^{T_{23}}_{1}$ & $0.1921 \pm 0.0061$ & $-0.06 \pm 0.04$ & $0.94 \pm 0.03$\\
$\alpha^{T_{23}}_{2}$ & $1.9618 \pm 0.0624$ & $-0.00 \pm 0.04$ & $0.96 \pm 0.03$\\
\hline\multicolumn{4}{c}{Subleading corrections}\\\hline
${\rm Re}(a_{0}^{sl})$ & $0.0979 \pm 0.0031$ & $-0.02 \pm 0.04$ & $0.98 \pm 0.03$\\
${\rm Im}(a_{0}^{sl})$ & $0.1047 \pm 0.0033$ & $0.03 \pm 0.05$ & $1.05 \pm 0.03$\\
${\rm Re}(b_{0}^{sl})$ & $0.2414 \pm 0.0077$ & $-0.03 \pm 0.05$ & $1.01 \pm 0.03$\\
${\rm Im}(b_{0}^{sl})$ & $0.2479 \pm 0.0079$ & $-0.08 \pm 0.05$ & $1.02 \pm 0.03$\\
${\rm Re}(a_{\perp}^{sl})$ & $0.0962 \pm 0.0031$ & $0.02 \pm 0.05$ & $1.02 \pm 0.03$\\
${\rm Im}(a_{\perp}^{sl})$ & $0.0890 \pm 0.0028$ & $0.00 \pm 0.04$ & $0.93 \pm 0.03$\\
${\rm Re}(b_{\perp}^{sl})$ & $0.2480 \pm 0.0079$ & $-0.06 \pm 0.05$ & $1.06 \pm 0.03$\\
${\rm Im}(b_{\perp}^{sl})$ & $0.2482 \pm 0.0079$ & $0.04 \pm 0.05$ & $1.05 \pm 0.03$\\
${\rm Re}(a_{\parallel}^{sl})$ & $0.0946 \pm 0.0030$ & $-0.01 \pm 0.04$ & $1.00 \pm 0.03$\\
${\rm Im}(a_{\parallel}^{sl})$ & $0.0940 \pm 0.0030$ & $0.06 \pm 0.05$ & $1.01 \pm 0.03$\\
${\rm Re}(b_{\parallel}^{sl})$ & $0.2311 \pm 0.0073$ & $0.04 \pm 0.04$ & $0.99 \pm 0.03$\\
${\rm Im}(b_{\parallel}^{sl})$ & $0.2418 \pm 0.0077$ & $0.11 \pm 0.05$ & $1.04 \pm 0.03$\\
${\rm Re}(c_{0}^{sl})$ & $0.0592 \pm 0.0019$ & $-0.06 \pm 0.04$ & $0.97 \pm 0.03$\\
${\rm Im}(c_{0}^{sl})$ & $0.0582 \pm 0.0018$ & $-0.01 \pm 0.04$ & $0.97 \pm 0.03$\\
${\rm Re}(c_{\perp}^{sl})$ & $0.0691 \pm 0.0022$ & $-0.05 \pm 0.05$ & $1.01 \pm 0.03$\\
${\rm Im}(c_{\perp}^{sl})$ & $0.0657 \pm 0.0021$ & $0.09 \pm 0.04$ & $0.98 \pm 0.03$\\
${\rm Re}(c_{\parallel}^{sl})$ & $0.0569 \pm 0.0018$ & $-0.04 \pm 0.04$ & $0.97 \pm 0.03$\\
${\rm Im}(c_{\parallel}^{sl})$ & $0.0565 \pm 0.0018$ & $-0.01 \pm 0.04$ & $0.96 \pm 0.03$\\
\hline\multicolumn{4}{c}{S-wave parameters}\\\hline
$S(\xi_{\parallel})$ & $0.0858 \pm 0.0027$ & $0.04 \pm 0.05$ & $1.05 \pm 0.03$\\
$\delta_S$ & $0.1785 \pm 0.0057$ & $0.10 \pm 0.05$ & $1.14 \pm 0.04$\\
$|g_{\kappa}|$ & $0.0533 \pm 0.0017$ & $0.26 \pm 0.04$ & $0.97 \pm 0.03$\\
${\rm arg}(g_{\kappa})$ & $0.8466 \pm 0.0269$ & $0.02 \pm 0.07$ & $1.58 \pm 0.05$\\
\hline \end{tabular}}
\hspace*{0.25cm}
\scalebox{0.7}{
\begin{tabular}[t]{lrrr} \hline
 & sensitivity & pull mean & pull width \\ \hline\hline
${\rm Re}(C_{10})$ & $0.2278 \pm 0.0072$ & $-0.01 \pm 0.05$ & $1.06 \pm 0.03$\\
\hline\multicolumn{4}{c}{CKM parameters}\\\hline
$A_\mathrm{CKM}$ & $0.0206 \pm 0.0007$ & $-0.01 \pm 0.05$ & $1.01 \pm 0.03$\\
$\lambda_\mathrm{CKM}$ & $0.0007 \pm 0.0000$ & $-0.09 \pm 0.05$ & $1.03 \pm 0.03$\\
$\bar{\rho}_\mathrm{CKM}$ & $0.0530 \pm 0.0017$ & $0.05 \pm 0.04$ & $0.98 \pm 0.03$\\
$\bar{\eta}_\mathrm{CKM}$ & $0.0588 \pm 0.0019$ & $0.00 \pm 0.04$ & $0.98 \pm 0.03$\\
\hline\multicolumn{4}{c}{Quark masses}\\\hline
$m_{c}$ & $0.0237 \pm 0.0008$ & $-0.04 \pm 0.04$ & $0.98 \pm 0.03$\\
$m_{b}$ & $0.0296 \pm 0.0009$ & $-0.04 \pm 0.04$ & $1.00 \pm 0.03$\\
\hline\multicolumn{4}{c}{Form factor parameters}\\\hline
$\alpha^{A_{0}}_{0}$ & $0.0197 \pm 0.0006$ & $-0.14 \pm 0.05$ & $1.06 \pm 0.03$\\
$\alpha^{A_{0}}_{1}$ & $0.2408 \pm 0.0076$ & $-0.04 \pm 0.05$ & $1.02 \pm 0.03$\\
$\alpha^{A_{0}}_{2}$ & $1.4634 \pm 0.0464$ & $0.01 \pm 0.04$ & $1.00 \pm 0.03$\\
$\alpha^{A_{1}}_{0}$ & $0.0188 \pm 0.0006$ & $0.06 \pm 0.05$ & $1.04 \pm 0.03$\\
$\alpha^{A_{1}}_{1}$ & $0.1622 \pm 0.0051$ & $0.01 \pm 0.04$ & $1.00 \pm 0.03$\\
$\alpha^{A_{1}}_{2}$ & $0.9500 \pm 0.0301$ & $-0.04 \pm 0.04$ & $0.98 \pm 0.03$\\
$\alpha^{A_{12}}_{1}$ & $0.1146 \pm 0.0036$ & $-0.07 \pm 0.04$ & $1.00 \pm 0.03$\\
$\alpha^{A_{12}}_{2}$ & $0.6426 \pm 0.0204$ & $-0.01 \pm 0.05$ & $1.01 \pm 0.03$\\
$\alpha^{V}_{0}$ & $0.0242 \pm 0.0008$ & $0.09 \pm 0.05$ & $1.02 \pm 0.03$\\
$\alpha^{V}_{1}$ & $0.2358 \pm 0.0075$ & $0.04 \pm 0.05$ & $1.05 \pm 0.03$\\
$\alpha^{V}_{2}$ & $1.3573 \pm 0.0431$ & $-0.07 \pm 0.04$ & $0.99 \pm 0.03$\\
$\alpha^{T_{1}}_{0}$ & $0.0195 \pm 0.0006$ & $0.07 \pm 0.05$ & $1.03 \pm 0.03$\\
$\alpha^{T_{1}}_{1}$ & $0.1692 \pm 0.0054$ & $0.04 \pm 0.05$ & $1.06 \pm 0.03$\\
$\alpha^{T_{1}}_{2}$ & $1.4759 \pm 0.0468$ & $-0.02 \pm 0.04$ & $0.96 \pm 0.03$\\
$\alpha^{T_{2}}_{1}$ & $0.1462 \pm 0.0046$ & $0.03 \pm 0.05$ & $1.02 \pm 0.03$\\
$\alpha^{T_{2}}_{2}$ & $0.7442 \pm 0.0236$ & $-0.03 \pm 0.04$ & $0.95 \pm 0.03$\\
$\alpha^{T_{23}}_{0}$ & $0.0617 \pm 0.0020$ & $-0.00 \pm 0.05$ & $1.06 \pm 0.03$\\
$\alpha^{T_{23}}_{1}$ & $0.2082 \pm 0.0066$ & $-0.04 \pm 0.04$ & $0.96 \pm 0.03$\\
$\alpha^{T_{23}}_{2}$ & $2.0536 \pm 0.0651$ & $-0.01 \pm 0.04$ & $0.99 \pm 0.03$\\
\hline\multicolumn{4}{c}{Subleading corrections}\\\hline
${\rm Re}(a_{0}^{sl})$ & $0.1030 \pm 0.0033$ & $-0.04 \pm 0.05$ & $1.03 \pm 0.03$\\
${\rm Im}(a_{0}^{sl})$ & $0.0984 \pm 0.0031$ & $-0.00 \pm 0.04$ & $0.99 \pm 0.03$\\
${\rm Re}(b_{0}^{sl})$ & $0.2597 \pm 0.0082$ & $0.07 \pm 0.05$ & $1.06 \pm 0.03$\\
${\rm Im}(b_{0}^{sl})$ & $0.2408 \pm 0.0076$ & $-0.03 \pm 0.04$ & $0.99 \pm 0.03$\\
${\rm Re}(a_{\perp}^{sl})$ & $0.0970 \pm 0.0031$ & $-0.08 \pm 0.05$ & $1.03 \pm 0.03$\\
${\rm Im}(a_{\perp}^{sl})$ & $0.0944 \pm 0.0030$ & $0.01 \pm 0.04$ & $0.98 \pm 0.03$\\
${\rm Re}(b_{\perp}^{sl})$ & $0.2279 \pm 0.0072$ & $0.00 \pm 0.04$ & $0.97 \pm 0.03$\\
${\rm Im}(b_{\perp}^{sl})$ & $0.2500 \pm 0.0079$ & $-0.04 \pm 0.05$ & $1.05 \pm 0.03$\\
${\rm Re}(a_{\parallel}^{sl})$ & $0.1000 \pm 0.0032$ & $0.03 \pm 0.05$ & $1.05 \pm 0.03$\\
${\rm Im}(a_{\parallel}^{sl})$ & $0.0930 \pm 0.0029$ & $-0.02 \pm 0.04$ & $1.00 \pm 0.03$\\
${\rm Re}(b_{\parallel}^{sl})$ & $0.2356 \pm 0.0075$ & $0.08 \pm 0.05$ & $1.01 \pm 0.03$\\
${\rm Im}(b_{\parallel}^{sl})$ & $0.2279 \pm 0.0072$ & $0.04 \pm 0.04$ & $0.97 \pm 0.03$\\
${\rm Re}(c_{0}^{sl})$ & $0.0656 \pm 0.0021$ & $0.11 \pm 0.05$ & $1.01 \pm 0.03$\\
${\rm Im}(c_{0}^{sl})$ & $0.0613 \pm 0.0019$ & $-0.05 \pm 0.04$ & $0.96 \pm 0.03$\\
${\rm Re}(c_{\perp}^{sl})$ & $0.0719 \pm 0.0023$ & $0.06 \pm 0.05$ & $1.01 \pm 0.03$\\
${\rm Im}(c_{\perp}^{sl})$ & $0.0667 \pm 0.0021$ & $0.04 \pm 0.04$ & $0.97 \pm 0.03$\\
${\rm Re}(c_{\parallel}^{sl})$ & $0.0637 \pm 0.0020$ & $0.04 \pm 0.05$ & $1.02 \pm 0.03$\\
${\rm Im}(c_{\parallel}^{sl})$ & $0.0629 \pm 0.0020$ & $-0.05 \pm 0.05$ & $1.02 \pm 0.03$\\
\hline \end{tabular}
}
\end{minipage}
\caption{Results from pseudoexperiments determining ${\rm Re}({\cal C}_{10})$ using (left) the proposed direct fit method and (right) the $q^2$-binned approach.\label{tab:dir_cten}}  
\end{table}

\begin{table}
\begin{minipage}[t]{\textwidth}
\centering
\scalebox{0.7}{
  \begin{tabular}[t]{lrrr} \hline
 & sensitivity & pull mean & pull width \\ \hline\hline
${\rm Re}(C_{7})$ & $0.0193 \pm 0.0006$ & $0.03 \pm 0.05$ & $1.02 \pm 0.03$\\
${\rm Re}(C_{9})$ & $0.2130 \pm 0.0068$ & $-0.04 \pm 0.05$ & $1.02 \pm 0.03$\\
\hline\multicolumn{4}{c}{CKM parameters}\\\hline
$A_\mathrm{CKM}$ & $0.0204 \pm 0.0006$ & $-0.00 \pm 0.05$ & $1.01 \pm 0.03$\\
$\lambda_\mathrm{CKM}$ & $0.0007 \pm 0.0000$ & $-0.10 \pm 0.05$ & $1.03 \pm 0.03$\\
$\bar{\rho}_\mathrm{CKM}$ & $0.0529 \pm 0.0017$ & $0.04 \pm 0.04$ & $0.99 \pm 0.03$\\
$\bar{\eta}_\mathrm{CKM}$ & $0.0588 \pm 0.0019$ & $0.00 \pm 0.04$ & $0.98 \pm 0.03$\\
\hline\multicolumn{4}{c}{Quark masses}\\\hline
$m_{c}$ & $0.0236 \pm 0.0008$ & $-0.04 \pm 0.04$ & $0.98 \pm 0.03$\\
$m_{b}$ & $0.0298 \pm 0.0009$ & $-0.04 \pm 0.05$ & $1.02 \pm 0.03$\\
\hline\multicolumn{4}{c}{Form factor parameters}\\\hline
$\alpha^{A_{0}}_{0}$ & $0.0179 \pm 0.0006$ & $-0.05 \pm 0.05$ & $1.04 \pm 0.03$\\
$\alpha^{A_{0}}_{1}$ & $0.2126 \pm 0.0068$ & $-0.03 \pm 0.04$ & $0.99 \pm 0.03$\\
$\alpha^{A_{0}}_{2}$ & $1.3563 \pm 0.0431$ & $-0.05 \pm 0.04$ & $0.96 \pm 0.03$\\
$\alpha^{A_{1}}_{0}$ & $0.0171 \pm 0.0005$ & $-0.06 \pm 0.05$ & $1.01 \pm 0.03$\\
$\alpha^{A_{1}}_{1}$ & $0.1515 \pm 0.0048$ & $-0.05 \pm 0.04$ & $0.97 \pm 0.03$\\
$\alpha^{A_{1}}_{2}$ & $0.9288 \pm 0.0295$ & $-0.04 \pm 0.04$ & $0.97 \pm 0.03$\\
$\alpha^{A_{12}}_{1}$ & $0.0974 \pm 0.0031$ & $-0.02 \pm 0.04$ & $0.99 \pm 0.03$\\
$\alpha^{A_{12}}_{2}$ & $0.5866 \pm 0.0186$ & $-0.02 \pm 0.04$ & $0.99 \pm 0.03$\\
$\alpha^{V}_{0}$ & $0.0225 \pm 0.0007$ & $-0.05 \pm 0.05$ & $1.01 \pm 0.03$\\
$\alpha^{V}_{1}$ & $0.2166 \pm 0.0069$ & $-0.03 \pm 0.05$ & $1.03 \pm 0.03$\\
$\alpha^{V}_{2}$ & $1.3718 \pm 0.0436$ & $-0.02 \pm 0.05$ & $1.01 \pm 0.03$\\
$\alpha^{T_{1}}_{0}$ & $0.0196 \pm 0.0006$ & $-0.03 \pm 0.05$ & $1.01 \pm 0.03$\\
$\alpha^{T_{1}}_{1}$ & $0.1614 \pm 0.0051$ & $-0.02 \pm 0.05$ & $1.06 \pm 0.03$\\
$\alpha^{T_{1}}_{2}$ & $1.4383 \pm 0.0457$ & $0.01 \pm 0.04$ & $0.93 \pm 0.03$\\
$\alpha^{T_{2}}_{1}$ & $0.1385 \pm 0.0044$ & $-0.03 \pm 0.05$ & $1.01 \pm 0.03$\\
$\alpha^{T_{2}}_{2}$ & $0.7301 \pm 0.0232$ & $-0.01 \pm 0.04$ & $0.94 \pm 0.03$\\
$\alpha^{T_{23}}_{0}$ & $0.0575 \pm 0.0018$ & $-0.03 \pm 0.05$ & $1.02 \pm 0.03$\\
$\alpha^{T_{23}}_{1}$ & $0.1914 \pm 0.0061$ & $-0.06 \pm 0.04$ & $0.93 \pm 0.03$\\
$\alpha^{T_{23}}_{2}$ & $1.9653 \pm 0.0625$ & $-0.00 \pm 0.04$ & $0.96 \pm 0.03$\\
\hline\multicolumn{4}{c}{Subleading corrections}\\\hline
${\rm Re}(a_{0}^{sl})$ & $0.0981 \pm 0.0031$ & $-0.00 \pm 0.04$ & $0.99 \pm 0.03$\\
${\rm Im}(a_{0}^{sl})$ & $0.1040 \pm 0.0033$ & $0.04 \pm 0.05$ & $1.04 \pm 0.03$\\
${\rm Re}(b_{0}^{sl})$ & $0.2408 \pm 0.0077$ & $-0.02 \pm 0.05$ & $1.01 \pm 0.03$\\
${\rm Im}(b_{0}^{sl})$ & $0.2474 \pm 0.0079$ & $-0.08 \pm 0.05$ & $1.02 \pm 0.03$\\
${\rm Re}(a_{\perp}^{sl})$ & $0.0982 \pm 0.0031$ & $0.02 \pm 0.05$ & $1.02 \pm 0.03$\\
${\rm Im}(a_{\perp}^{sl})$ & $0.0904 \pm 0.0029$ & $-0.01 \pm 0.04$ & $0.94 \pm 0.03$\\
${\rm Re}(b_{\perp}^{sl})$ & $0.2536 \pm 0.0081$ & $-0.06 \pm 0.05$ & $1.06 \pm 0.03$\\
${\rm Im}(b_{\perp}^{sl})$ & $0.2519 \pm 0.0080$ & $0.05 \pm 0.05$ & $1.05 \pm 0.03$\\
${\rm Re}(a_{\parallel}^{sl})$ & $0.0951 \pm 0.0030$ & $-0.01 \pm 0.04$ & $0.99 \pm 0.03$\\
${\rm Im}(a_{\parallel}^{sl})$ & $0.0942 \pm 0.0030$ & $0.06 \pm 0.05$ & $1.01 \pm 0.03$\\
${\rm Re}(b_{\parallel}^{sl})$ & $0.2315 \pm 0.0074$ & $0.04 \pm 0.04$ & $0.99 \pm 0.03$\\
${\rm Im}(b_{\parallel}^{sl})$ & $0.2417 \pm 0.0077$ & $0.10 \pm 0.05$ & $1.03 \pm 0.03$\\
${\rm Re}(c_{0}^{sl})$ & $0.0592 \pm 0.0019$ & $-0.06 \pm 0.04$ & $0.97 \pm 0.03$\\
${\rm Im}(c_{0}^{sl})$ & $0.0582 \pm 0.0018$ & $-0.00 \pm 0.04$ & $0.97 \pm 0.03$\\
${\rm Re}(c_{\perp}^{sl})$ & $0.0696 \pm 0.0022$ & $-0.05 \pm 0.05$ & $1.01 \pm 0.03$\\
${\rm Im}(c_{\perp}^{sl})$ & $0.0657 \pm 0.0021$ & $0.10 \pm 0.04$ & $0.98 \pm 0.03$\\
${\rm Re}(c_{\parallel}^{sl})$ & $0.0571 \pm 0.0018$ & $-0.05 \pm 0.04$ & $0.97 \pm 0.03$\\
${\rm Im}(c_{\parallel}^{sl})$ & $0.0567 \pm 0.0018$ & $0.00 \pm 0.04$ & $0.96 \pm 0.03$\\
\hline\multicolumn{4}{c}{S-wave parameters}\\\hline
$S(\xi_{\parallel})$ & $0.0869 \pm 0.0028$ & $0.04 \pm 0.05$ & $1.06 \pm 0.03$\\
$\delta_S$ & $0.1785 \pm 0.0057$ & $0.10 \pm 0.05$ & $1.14 \pm 0.04$\\
$|g_{\kappa}|$ & $0.0529 \pm 0.0017$ & $0.27 \pm 0.04$ & $0.93 \pm 0.03$\\
${\rm arg}(g_{\kappa})$ & $0.8346 \pm 0.0265$ & $0.02 \pm 0.07$ & $1.62 \pm 0.05$\\
\hline \end{tabular}}
\hspace*{0.25cm}
\scalebox{0.7}{
  \begin{tabular}[t]{lrrr} \hline
 & sensitivity & pull mean & pull width \\ \hline\hline
${\rm Re}(C_{7})$ & $0.0252 \pm 0.0008$ & $0.06 \pm 0.05$ & $1.07 \pm 0.03$\\
${\rm Re}(C_{9})$ & $0.2555 \pm 0.0081$ & $-0.04 \pm 0.05$ & $1.05 \pm 0.03$\\
\hline\multicolumn{4}{c}{CKM parameters}\\\hline
$A_\mathrm{CKM}$ & $0.0205 \pm 0.0007$ & $-0.00 \pm 0.05$ & $1.01 \pm 0.03$\\
$\lambda_\mathrm{CKM}$ & $0.0007 \pm 0.0000$ & $-0.09 \pm 0.05$ & $1.03 \pm 0.03$\\
$\bar{\rho}_\mathrm{CKM}$ & $0.0525 \pm 0.0017$ & $0.04 \pm 0.04$ & $0.98 \pm 0.03$\\
$\bar{\eta}_\mathrm{CKM}$ & $0.0587 \pm 0.0019$ & $0.01 \pm 0.04$ & $0.97 \pm 0.03$\\
\hline\multicolumn{4}{c}{Quark masses}\\\hline
$m_{c}$ & $0.0238 \pm 0.0008$ & $-0.04 \pm 0.04$ & $0.99 \pm 0.03$\\
$m_{b}$ & $0.0296 \pm 0.0009$ & $-0.04 \pm 0.05$ & $1.01 \pm 0.03$\\
\hline\multicolumn{4}{c}{Form factor parameters}\\\hline
$\alpha^{A_{0}}_{0}$ & $0.0197 \pm 0.0006$ & $-0.14 \pm 0.05$ & $1.06 \pm 0.03$\\
$\alpha^{A_{0}}_{1}$ & $0.2408 \pm 0.0076$ & $-0.05 \pm 0.05$ & $1.02 \pm 0.03$\\
$\alpha^{A_{0}}_{2}$ & $1.4688 \pm 0.0466$ & $-0.00 \pm 0.04$ & $1.00 \pm 0.03$\\
$\alpha^{A_{1}}_{0}$ & $0.0188 \pm 0.0006$ & $0.07 \pm 0.05$ & $1.04 \pm 0.03$\\
$\alpha^{A_{1}}_{1}$ & $0.1632 \pm 0.0052$ & $0.01 \pm 0.04$ & $1.00 \pm 0.03$\\
$\alpha^{A_{1}}_{2}$ & $0.9549 \pm 0.0303$ & $-0.04 \pm 0.04$ & $0.98 \pm 0.03$\\
$\alpha^{A_{12}}_{1}$ & $0.1147 \pm 0.0036$ & $-0.08 \pm 0.04$ & $1.00 \pm 0.03$\\
$\alpha^{A_{12}}_{2}$ & $0.6430 \pm 0.0204$ & $-0.02 \pm 0.05$ & $1.01 \pm 0.03$\\
$\alpha^{V}_{0}$ & $0.0241 \pm 0.0008$ & $0.09 \pm 0.05$ & $1.01 \pm 0.03$\\
$\alpha^{V}_{1}$ & $0.2377 \pm 0.0075$ & $0.03 \pm 0.05$ & $1.05 \pm 0.03$\\
$\alpha^{V}_{2}$ & $1.3600 \pm 0.0431$ & $-0.07 \pm 0.04$ & $1.00 \pm 0.03$\\
$\alpha^{T_{1}}_{0}$ & $0.0208 \pm 0.0007$ & $0.09 \pm 0.05$ & $1.02 \pm 0.03$\\
$\alpha^{T_{1}}_{1}$ & $0.1730 \pm 0.0055$ & $0.04 \pm 0.05$ & $1.07 \pm 0.03$\\
$\alpha^{T_{1}}_{2}$ & $1.4766 \pm 0.0468$ & $-0.04 \pm 0.04$ & $0.95 \pm 0.03$\\
$\alpha^{T_{2}}_{1}$ & $0.1488 \pm 0.0047$ & $0.03 \pm 0.05$ & $1.02 \pm 0.03$\\
$\alpha^{T_{2}}_{2}$ & $0.7448 \pm 0.0236$ & $-0.04 \pm 0.04$ & $0.94 \pm 0.03$\\
$\alpha^{T_{23}}_{0}$ & $0.0615 \pm 0.0020$ & $-0.00 \pm 0.05$ & $1.06 \pm 0.03$\\
$\alpha^{T_{23}}_{1}$ & $0.2098 \pm 0.0067$ & $-0.04 \pm 0.04$ & $0.97 \pm 0.03$\\
$\alpha^{T_{23}}_{2}$ & $2.0523 \pm 0.0651$ & $-0.02 \pm 0.04$ & $0.99 \pm 0.03$\\
\hline\multicolumn{4}{c}{Subleading corrections}\\\hline
${\rm Re}(a_{0}^{sl})$ & $0.1031 \pm 0.0033$ & $-0.04 \pm 0.05$ & $1.03 \pm 0.03$\\
${\rm Im}(a_{0}^{sl})$ & $0.0983 \pm 0.0031$ & $-0.01 \pm 0.04$ & $0.99 \pm 0.03$\\
${\rm Re}(b_{0}^{sl})$ & $0.2594 \pm 0.0082$ & $0.07 \pm 0.05$ & $1.06 \pm 0.03$\\
${\rm Im}(b_{0}^{sl})$ & $0.2409 \pm 0.0076$ & $-0.03 \pm 0.04$ & $0.99 \pm 0.03$\\
${\rm Re}(a_{\perp}^{sl})$ & $0.0998 \pm 0.0032$ & $-0.07 \pm 0.05$ & $1.03 \pm 0.03$\\
${\rm Im}(a_{\perp}^{sl})$ & $0.0953 \pm 0.0030$ & $0.00 \pm 0.04$ & $0.99 \pm 0.03$\\
${\rm Re}(b_{\perp}^{sl})$ & $0.2290 \pm 0.0073$ & $0.01 \pm 0.04$ & $0.95 \pm 0.03$\\
${\rm Im}(b_{\perp}^{sl})$ & $0.2565 \pm 0.0081$ & $-0.05 \pm 0.05$ & $1.06 \pm 0.03$\\
${\rm Re}(a_{\parallel}^{sl})$ & $0.1004 \pm 0.0032$ & $0.03 \pm 0.05$ & $1.05 \pm 0.03$\\
${\rm Im}(a_{\parallel}^{sl})$ & $0.0948 \pm 0.0030$ & $-0.02 \pm 0.05$ & $1.01 \pm 0.03$\\
${\rm Re}(b_{\parallel}^{sl})$ & $0.2385 \pm 0.0076$ & $0.08 \pm 0.05$ & $1.01 \pm 0.03$\\
${\rm Im}(b_{\parallel}^{sl})$ & $0.2310 \pm 0.0073$ & $0.04 \pm 0.04$ & $0.98 \pm 0.03$\\
${\rm Re}(c_{0}^{sl})$ & $0.0653 \pm 0.0021$ & $0.10 \pm 0.04$ & $1.00 \pm 0.03$\\
${\rm Im}(c_{0}^{sl})$ & $0.0607 \pm 0.0019$ & $-0.04 \pm 0.04$ & $0.96 \pm 0.03$\\
${\rm Re}(c_{\perp}^{sl})$ & $0.0720 \pm 0.0023$ & $0.06 \pm 0.05$ & $1.01 \pm 0.03$\\
${\rm Im}(c_{\perp}^{sl})$ & $0.0674 \pm 0.0021$ & $0.05 \pm 0.04$ & $0.98 \pm 0.03$\\
${\rm Re}(c_{\parallel}^{sl})$ & $0.0635 \pm 0.0020$ & $0.04 \pm 0.05$ & $1.01 \pm 0.03$\\
${\rm Im}(c_{\parallel}^{sl})$ & $0.0628 \pm 0.0020$ & $-0.05 \pm 0.05$ & $1.01 \pm 0.03$\\
    \hline \end{tabular}}
\end{minipage}
\caption{Results from pseudoexperiments determining ${\rm Re}({\cal C}_{7})$ and ${\rm Re}({\cal C}_{9})$ using (left) the proposed direct fit method and (right) the $q^2$-binned approach.\label{tab:dir_csevencnine}}  
\end{table}

\begin{table}
  \begin{minipage}[t]{\textwidth}
  \centering
\scalebox{0.7}{
  \begin{tabular}[t]{lrrr} \hline
 & sensitivity & pull mean & pull width \\ \hline\hline
${\rm Re}(C_{9})$ & $0.1715 \pm 0.0054$ & $-0.01 \pm 0.04$ & $0.97 \pm 0.03$\\
${\rm Re}(C_{10})$ & $0.2054 \pm 0.0065$ & $0.03 \pm 0.05$ & $1.01 \pm 0.03$\\
\hline\multicolumn{4}{c}{CKM parameters}\\\hline
$A_\mathrm{CKM}$ & $0.0205 \pm 0.0007$ & $-0.01 \pm 0.05$ & $1.00 \pm 0.03$\\
$\lambda_\mathrm{CKM}$ & $0.0007 \pm 0.0000$ & $-0.10 \pm 0.05$ & $1.03 \pm 0.03$\\
$\bar{\rho}_\mathrm{CKM}$ & $0.0530 \pm 0.0017$ & $0.04 \pm 0.04$ & $0.99 \pm 0.03$\\
$\bar{\eta}_\mathrm{CKM}$ & $0.0584 \pm 0.0019$ & $-0.00 \pm 0.04$ & $0.97 \pm 0.03$\\
\hline\multicolumn{4}{c}{Quark masses}\\\hline
$m_{c}$ & $0.0237 \pm 0.0008$ & $-0.04 \pm 0.04$ & $0.98 \pm 0.03$\\
$m_{b}$ & $0.0296 \pm 0.0009$ & $-0.03 \pm 0.05$ & $1.01 \pm 0.03$\\
\hline\multicolumn{4}{c}{Form factor parameters}\\\hline
$\alpha^{A_{0}}_{0}$ & $0.0180 \pm 0.0006$ & $-0.05 \pm 0.05$ & $1.04 \pm 0.03$\\
$\alpha^{A_{0}}_{1}$ & $0.2119 \pm 0.0067$ & $-0.02 \pm 0.04$ & $0.99 \pm 0.03$\\
$\alpha^{A_{0}}_{2}$ & $1.3510 \pm 0.0429$ & $-0.04 \pm 0.04$ & $0.95 \pm 0.03$\\
$\alpha^{A_{1}}_{0}$ & $0.0171 \pm 0.0005$ & $-0.05 \pm 0.05$ & $1.00 \pm 0.03$\\
$\alpha^{A_{1}}_{1}$ & $0.1511 \pm 0.0048$ & $-0.05 \pm 0.04$ & $0.97 \pm 0.03$\\
$\alpha^{A_{1}}_{2}$ & $0.9289 \pm 0.0295$ & $-0.04 \pm 0.04$ & $0.97 \pm 0.03$\\
$\alpha^{A_{12}}_{1}$ & $0.0972 \pm 0.0031$ & $-0.02 \pm 0.04$ & $0.99 \pm 0.03$\\
$\alpha^{A_{12}}_{2}$ & $0.5864 \pm 0.0186$ & $-0.02 \pm 0.04$ & $0.99 \pm 0.03$\\
$\alpha^{V}_{0}$ & $0.0224 \pm 0.0007$ & $-0.04 \pm 0.05$ & $1.00 \pm 0.03$\\
$\alpha^{V}_{1}$ & $0.2147 \pm 0.0068$ & $-0.03 \pm 0.05$ & $1.03 \pm 0.03$\\
$\alpha^{V}_{2}$ & $1.3701 \pm 0.0435$ & $-0.02 \pm 0.05$ & $1.01 \pm 0.03$\\
$\alpha^{T_{1}}_{0}$ & $0.0195 \pm 0.0006$ & $-0.03 \pm 0.05$ & $1.00 \pm 0.03$\\
$\alpha^{T_{1}}_{1}$ & $0.1600 \pm 0.0051$ & $-0.02 \pm 0.05$ & $1.05 \pm 0.03$\\
$\alpha^{T_{1}}_{2}$ & $1.4404 \pm 0.0457$ & $0.00 \pm 0.04$ & $0.93 \pm 0.03$\\
$\alpha^{T_{2}}_{1}$ & $0.1372 \pm 0.0044$ & $-0.03 \pm 0.05$ & $1.01 \pm 0.03$\\
$\alpha^{T_{2}}_{2}$ & $0.7294 \pm 0.0232$ & $-0.01 \pm 0.04$ & $0.94 \pm 0.03$\\
$\alpha^{T_{23}}_{0}$ & $0.0572 \pm 0.0018$ & $-0.02 \pm 0.05$ & $1.02 \pm 0.03$\\
$\alpha^{T_{23}}_{1}$ & $0.1923 \pm 0.0061$ & $-0.05 \pm 0.04$ & $0.94 \pm 0.03$\\
$\alpha^{T_{23}}_{2}$ & $1.9576 \pm 0.0622$ & $-0.00 \pm 0.04$ & $0.95 \pm 0.03$\\
\hline\multicolumn{4}{c}{Subleading corrections}\\\hline
${\rm Re}(a_{0}^{sl})$ & $0.0980 \pm 0.0031$ & $-0.01 \pm 0.04$ & $0.98 \pm 0.03$\\
${\rm Im}(a_{0}^{sl})$ & $0.1047 \pm 0.0033$ & $0.04 \pm 0.05$ & $1.05 \pm 0.03$\\
${\rm Re}(b_{0}^{sl})$ & $0.2403 \pm 0.0076$ & $-0.03 \pm 0.05$ & $1.01 \pm 0.03$\\
${\rm Im}(b_{0}^{sl})$ & $0.2479 \pm 0.0079$ & $-0.08 \pm 0.05$ & $1.02 \pm 0.03$\\
${\rm Re}(a_{\perp}^{sl})$ & $0.0983 \pm 0.0031$ & $0.02 \pm 0.05$ & $1.02 \pm 0.03$\\
${\rm Im}(a_{\perp}^{sl})$ & $0.0896 \pm 0.0028$ & $0.01 \pm 0.04$ & $0.93 \pm 0.03$\\
${\rm Re}(b_{\perp}^{sl})$ & $0.2534 \pm 0.0080$ & $-0.06 \pm 0.05$ & $1.06 \pm 0.03$\\
${\rm Im}(b_{\perp}^{sl})$ & $0.2550 \pm 0.0081$ & $0.05 \pm 0.05$ & $1.06 \pm 0.03$\\
${\rm Re}(a_{\parallel}^{sl})$ & $0.0949 \pm 0.0030$ & $-0.02 \pm 0.04$ & $0.99 \pm 0.03$\\
${\rm Im}(a_{\parallel}^{sl})$ & $0.0937 \pm 0.0030$ & $0.07 \pm 0.05$ & $1.01 \pm 0.03$\\
${\rm Re}(b_{\parallel}^{sl})$ & $0.2318 \pm 0.0074$ & $0.04 \pm 0.04$ & $0.99 \pm 0.03$\\
${\rm Im}(b_{\parallel}^{sl})$ & $0.2421 \pm 0.0077$ & $0.10 \pm 0.05$ & $1.03 \pm 0.03$\\
${\rm Re}(c_{0}^{sl})$ & $0.0593 \pm 0.0019$ & $-0.06 \pm 0.04$ & $0.98 \pm 0.03$\\
${\rm Im}(c_{0}^{sl})$ & $0.0584 \pm 0.0019$ & $-0.00 \pm 0.04$ & $0.97 \pm 0.03$\\
${\rm Re}(c_{\perp}^{sl})$ & $0.0696 \pm 0.0022$ & $-0.05 \pm 0.05$ & $1.02 \pm 0.03$\\
${\rm Im}(c_{\perp}^{sl})$ & $0.0652 \pm 0.0021$ & $0.09 \pm 0.04$ & $0.98 \pm 0.03$\\
${\rm Re}(c_{\parallel}^{sl})$ & $0.0571 \pm 0.0018$ & $-0.05 \pm 0.04$ & $0.97 \pm 0.03$\\
${\rm Im}(c_{\parallel}^{sl})$ & $0.0565 \pm 0.0018$ & $0.00 \pm 0.04$ & $0.96 \pm 0.03$\\
\hline\multicolumn{4}{c}{S-wave parameters}\\\hline
$S(\xi_{\parallel})$ & $0.0864 \pm 0.0027$ & $0.04 \pm 0.05$ & $1.05 \pm 0.03$\\
$\delta_S$ & $0.1774 \pm 0.0056$ & $0.10 \pm 0.05$ & $1.13 \pm 0.04$\\
$|g_{\kappa}|$ & $0.0526 \pm 0.0017$ & $0.28 \pm 0.04$ & $0.91 \pm 0.03$\\
${\rm arg}(g_{\kappa})$ & $0.8599 \pm 0.0273$ & $0.00 \pm 0.07$ & $1.63 \pm 0.05$\\
\hline \end{tabular}}
\hspace*{0.25cm}
\scalebox{0.7}{
\begin{tabular}[t]{lrrr} \hline
 & sensitivity & pull mean & pull width \\ \hline\hline
${\rm Re}(C_{9})$ & $0.1869 \pm 0.0059$ & $0.03 \pm 0.05$ & $1.02 \pm 0.03$\\
${\rm Re}(C_{10})$ & $0.2663 \pm 0.0085$ & $0.00 \pm 0.05$ & $1.06 \pm 0.03$\\
\hline\multicolumn{4}{c}{CKM parameters}\\\hline
$A_\mathrm{CKM}$ & $0.0205 \pm 0.0007$ & $-0.01 \pm 0.05$ & $1.01 \pm 0.03$\\
$\lambda_\mathrm{CKM}$ & $0.0007 \pm 0.0000$ & $-0.09 \pm 0.05$ & $1.03 \pm 0.03$\\
$\bar{\rho}_\mathrm{CKM}$ & $0.0526 \pm 0.0017$ & $0.04 \pm 0.04$ & $0.98 \pm 0.03$\\
$\bar{\eta}_\mathrm{CKM}$ & $0.0582 \pm 0.0018$ & $0.01 \pm 0.04$ & $0.97 \pm 0.03$\\
\hline\multicolumn{4}{c}{Quark masses}\\\hline
$m_{c}$ & $0.0239 \pm 0.0008$ & $-0.04 \pm 0.04$ & $0.99 \pm 0.03$\\
$m_{b}$ & $0.0298 \pm 0.0009$ & $-0.05 \pm 0.05$ & $1.02 \pm 0.03$\\
\hline\multicolumn{4}{c}{Form factor parameters}\\\hline
$\alpha^{A_{0}}_{0}$ & $0.0198 \pm 0.0006$ & $-0.14 \pm 0.05$ & $1.06 \pm 0.03$\\
$\alpha^{A_{0}}_{1}$ & $0.2413 \pm 0.0077$ & $-0.04 \pm 0.05$ & $1.02 \pm 0.03$\\
$\alpha^{A_{0}}_{2}$ & $1.4661 \pm 0.0465$ & $0.00 \pm 0.04$ & $1.00 \pm 0.03$\\
$\alpha^{A_{1}}_{0}$ & $0.0188 \pm 0.0006$ & $0.06 \pm 0.05$ & $1.04 \pm 0.03$\\
$\alpha^{A_{1}}_{1}$ & $0.1629 \pm 0.0052$ & $0.02 \pm 0.04$ & $1.00 \pm 0.03$\\
$\alpha^{A_{1}}_{2}$ & $0.9539 \pm 0.0303$ & $-0.04 \pm 0.04$ & $0.98 \pm 0.03$\\
$\alpha^{A_{12}}_{1}$ & $0.1144 \pm 0.0036$ & $-0.07 \pm 0.04$ & $0.99 \pm 0.03$\\
$\alpha^{A_{12}}_{2}$ & $0.6414 \pm 0.0204$ & $-0.01 \pm 0.05$ & $1.01 \pm 0.03$\\
$\alpha^{V}_{0}$ & $0.0242 \pm 0.0008$ & $0.09 \pm 0.05$ & $1.01 \pm 0.03$\\
$\alpha^{V}_{1}$ & $0.2382 \pm 0.0076$ & $0.04 \pm 0.05$ & $1.05 \pm 0.03$\\
$\alpha^{V}_{2}$ & $1.3610 \pm 0.0432$ & $-0.06 \pm 0.04$ & $1.00 \pm 0.03$\\
$\alpha^{T_{1}}_{0}$ & $0.0207 \pm 0.0007$ & $0.09 \pm 0.05$ & $1.02 \pm 0.03$\\
$\alpha^{T_{1}}_{1}$ & $0.1727 \pm 0.0055$ & $0.05 \pm 0.05$ & $1.07 \pm 0.03$\\
$\alpha^{T_{1}}_{2}$ & $1.4731 \pm 0.0468$ & $-0.03 \pm 0.04$ & $0.94 \pm 0.03$\\
$\alpha^{T_{2}}_{1}$ & $0.1478 \pm 0.0047$ & $0.03 \pm 0.05$ & $1.02 \pm 0.03$\\
$\alpha^{T_{2}}_{2}$ & $0.7426 \pm 0.0236$ & $-0.03 \pm 0.04$ & $0.94 \pm 0.03$\\
$\alpha^{T_{23}}_{0}$ & $0.0618 \pm 0.0020$ & $-0.00 \pm 0.05$ & $1.06 \pm 0.03$\\
$\alpha^{T_{23}}_{1}$ & $0.2089 \pm 0.0066$ & $-0.04 \pm 0.04$ & $0.96 \pm 0.03$\\
$\alpha^{T_{23}}_{2}$ & $2.0531 \pm 0.0652$ & $-0.01 \pm 0.04$ & $0.99 \pm 0.03$\\
\hline\multicolumn{4}{c}{Subleading corrections}\\\hline
${\rm Re}(a_{0}^{sl})$ & $0.1032 \pm 0.0033$ & $-0.04 \pm 0.05$ & $1.03 \pm 0.03$\\
${\rm Im}(a_{0}^{sl})$ & $0.0986 \pm 0.0031$ & $-0.01 \pm 0.04$ & $0.99 \pm 0.03$\\
${\rm Re}(b_{0}^{sl})$ & $0.2592 \pm 0.0082$ & $0.07 \pm 0.05$ & $1.06 \pm 0.03$\\
${\rm Im}(b_{0}^{sl})$ & $0.2405 \pm 0.0076$ & $-0.03 \pm 0.04$ & $0.99 \pm 0.03$\\
${\rm Re}(a_{\perp}^{sl})$ & $0.0992 \pm 0.0031$ & $-0.07 \pm 0.05$ & $1.03 \pm 0.03$\\
${\rm Im}(a_{\perp}^{sl})$ & $0.0951 \pm 0.0030$ & $0.00 \pm 0.04$ & $0.98 \pm 0.03$\\
${\rm Re}(b_{\perp}^{sl})$ & $0.2294 \pm 0.0073$ & $0.01 \pm 0.04$ & $0.95 \pm 0.03$\\
${\rm Im}(b_{\perp}^{sl})$ & $0.2547 \pm 0.0081$ & $-0.05 \pm 0.05$ & $1.06 \pm 0.03$\\
${\rm Re}(a_{\parallel}^{sl})$ & $0.1006 \pm 0.0032$ & $0.04 \pm 0.05$ & $1.04 \pm 0.03$\\
${\rm Im}(a_{\parallel}^{sl})$ & $0.0941 \pm 0.0030$ & $-0.02 \pm 0.05$ & $1.01 \pm 0.03$\\
${\rm Re}(b_{\parallel}^{sl})$ & $0.2391 \pm 0.0076$ & $0.09 \pm 0.05$ & $1.02 \pm 0.03$\\
${\rm Im}(b_{\parallel}^{sl})$ & $0.2312 \pm 0.0073$ & $0.03 \pm 0.04$ & $0.98 \pm 0.03$\\
${\rm Re}(c_{0}^{sl})$ & $0.0654 \pm 0.0021$ & $0.10 \pm 0.05$ & $1.00 \pm 0.03$\\
${\rm Im}(c_{0}^{sl})$ & $0.0610 \pm 0.0019$ & $-0.04 \pm 0.04$ & $0.96 \pm 0.03$\\
${\rm Re}(c_{\perp}^{sl})$ & $0.0720 \pm 0.0023$ & $0.06 \pm 0.05$ & $1.01 \pm 0.03$\\
${\rm Im}(c_{\perp}^{sl})$ & $0.0671 \pm 0.0021$ & $0.04 \pm 0.04$ & $0.98 \pm 0.03$\\
${\rm Re}(c_{\parallel}^{sl})$ & $0.0634 \pm 0.0020$ & $0.03 \pm 0.05$ & $1.01 \pm 0.03$\\
${\rm Im}(c_{\parallel}^{sl})$ & $0.0631 \pm 0.0020$ & $-0.05 \pm 0.05$ & $1.02 \pm 0.03$\\
\hline \end{tabular}
}
\end{minipage}
\caption{Results from pseudoexperiments determining ${\rm Re}({\cal C}_{9})$ and ${\rm Re}({\cal C}_{10})$ using (left) the proposed direct fit method and (right) the $q^2$-binned approach.\label{tab:dir_cninecten}}  
\end{table}

\end{document}